\documentclass[longauth]{aa}
\usepackage[colorlinks=true, allcolors=blue]{hyperref}

\usepackage[utf8]{inputenc}
\usepackage{xcolor}
\usepackage{graphicx}
\usepackage[varg]{txfonts}
\usepackage{afterpage}
\usepackage{CJK}
\usepackage{placeins}
\usepackage{longtable}
\usepackage{amsmath} 
\usepackage{footnote}
\usepackage{rotating}
\usepackage{booktabs}
\usepackage{siunitx}
\usepackage{graphicx}
\usepackage{float} 
\usepackage{fancyhdr}
\usepackage{threeparttable}
\usepackage{enumitem}
\usepackage{multirow}
\usepackage{orcidlink}
 
\usepackage{silence}

\newcommand{\ha} {\mbox{H$\alpha$}\,}
\newcommand{\hb} {\mbox{H$\beta$}\,}

\newcommand{\hd} {\mbox{H$\delta$}\,}
\newcommand{\Ha} {\mbox{H$\alpha$}\,}
\newcommand{\Hb} {\mbox{H$\beta$}\,}

\newcommand{\Nai}{\ion{Na}{i}\,}

\newcommand{\Caii} {\ion{Ca}{ii}\,}

\newcommand{\Ciii} {\ion{C}{iii}\,}

\newcommand{\Hei} {\ion{He}{i}\,}
\newcommand{\Heii} {\ion{He}{ii}\,}

\newcommand{\Niii} {\ion{N}{iii}\,}

\newcommand{\msun}{\mbox{M$_{\odot}$\,}}
\newcommand{\msol}{\mbox{M$_{\odot}$\,}}

\newcommand{\kms}{\mbox{$\rm{km}\,s^{-1}$}}

\begin{document}

   \title{SN 2019vxm: A luminous and long-lived Type IIn supernova with early flash-ionisation features}

   \subtitle{}
\author{ 
    Y.-Z. Cai \orcidlink{0000-0002-7714-493X}\inst{\ref{inst7},\ref{inst1},\ref{inst5}}\corrauth{yongzhi.cai@inaf.it (CYZ)} 
    \and A. Pastorello \orcidlink{0000-0002-7259-4624}\inst{\ref{inst5}}
    \and T. J. Moriya \orcidlink{0000-0003-1169-1954}\inst{\ref{inst11}, \ref{inst12}, \ref{inst13}}
    \and X.-F. Wang \orcidlink{0000-0002-7334-2357}\inst{\ref{inst7},\ref{inst8}}\corrauth{wang\_xf@mail.tsinghua.edu.cn (WXF)} 
    \and A.~Reguitti \orcidlink{0000-0003-4254-2724}\inst{\ref{inst5},\ref{inst9}}
    \and A.~V.~Filippenko \orcidlink{0000-0003-3460-0103}\inst{\ref{inst23}}
    \and L.~Tomasella  \orcidlink{0000-0002-3697-2616}\inst{\ref{inst5}}
    \and K. \v{C}otar \orcidlink{0000-0002-3439-0858}\inst{\ref{inst17}} 
    \and A. Siviero \inst{\ref{uni:padua},\ref{inst5}} 
    \and N. Elias-Rosa \orcidlink{0000-0002-1381-9125} \inst{\ref{inst5},\ref{inst6}}
    \and T.~G.~Brink \orcidlink{0000-0001-5955-2502}\inst{\ref{inst23}} 
    \and G. Valerin \orcidlink{0000-0002-3334-4585}\inst{\ref{inst5}}  
    \and S.~Benetti \orcidlink{0000-0002-3256-0016} \inst{\ref{inst5}}
    \and J.-W.~Zhao\orcidlink{0009-0009-8633-8582}\inst{\ref{inst3}}
    \and Z.-H. Peng \orcidlink{0009-0000-7773-553X}\inst{\ref{inst4}}
    \and Z.-Y. Wang \orcidlink{0000-0002-0025-0179} \inst{\ref{UCAS},\ref{NAOC}} 
    \and I. Altunin   \inst{\ref{inst23},\ref{inst:Nevada}} 
    \and R. Baer-Way \orcidlink{0009-0004-7268-7283} \inst{\ref{inst23}, \ref{inst:Virginia}} 
    \and E. Baron \orcidlink{0000-0001-5393-1608} \inst{\ref{inst111},\ref{inst112},\ref{inst113}} 
    \and V.~Chander  \inst{\ref{inst23}} 
    \and M. Chu \orcidlink{0000-0001-5884-911X}  \inst{\ref{inst:UCSD}} 
    \and A. deGraw   \inst{\ref{inst23}}  
    \and J.~M.~DerKacy  \orcidlink{0000-0002-7566-6080}\inst{\ref{inst114}} 
    \and J. Isern \inst{\ref{inst6}, \ref{inst15.5},\ref{inst16}} 
    \and C.~Jennings  \inst{\ref{inst23},\ref{yale}} 
    \and R.~Kotak  \orcidlink{0000-0001-5455-3653} \inst{\ref{inst37}}
    \and L.-P.~Li \orcidlink{0009-0003-3758-0598}\inst{\ref{inst1}} 
    \and P.~Marziani~\orcidlink{0000-0002-6058-4912}\inst{\ref{inst5},\ref{inst:IAA-CSIC}} 
    \and M. May   \inst{\ref{inst23},\ref{Knoxville}}
    \and P.~A.~Mazzali\inst{\ref{inst16.33}, \ref{inst16.66}}
    \and A. Morales-Garoffolo \orcidlink{0000-0001-8830-7063}\inst{\ref{inst14}}
    \and S.~Moran \orcidlink{0000-0001-5221-0243}\inst{\ref{inst10}} 
    \and P.~Ochner \inst{\ref{uni:padua},\ref{inst5}}
    \and I.~Salmaso \inst{\ref{inst35}}
    \and L.~Tartaglia \inst{\ref{inst110}}
    \and M. Turatto\inst{\ref{inst5}} 
    \and H.-Y. Wu \orcidlink{0009-0001-0094-3020}\inst{\ref{inst:Yangtze}} 
    \and D.-F. Xiang \orcidlink{0000-0002-1089-1519}\inst{\ref{inst307}} 
    \and S.-Y. Yan \orcidlink{0009-0004-4256-1209} \inst{\ref{inst7}}
    \and J.-J.~Zhang \orcidlink{0000-0002-8296-2590}\inst{\ref{inst1}}
    \and W.~Zheng \orcidlink{0000-0002-2636-6508}\inst{\ref{inst23}}
}
\institute{
\label{inst7}Department of Physics, Tsinghua University, Beijing 100084, P.R. China     \and
\label{inst1}International Centre of Supernovae (ICESUN), Yunnan Key Laboratory of Supernova Research, Yunnan Observatories, Chinese Academy of Sciences (CAS), Kunming, 650216, China \and
\label{inst5}INAF - Osservatorio Astronomico di Padova, Vicolo dell'Osservatorio 5, 35122 Padova, Italy \and
\label{inst11}National Astronomical Observatory of Japan, National Institutes of Natural Sciences, 2-21-1 Osawa, Mitaka, Tokyo 181-8588, Japan \and
\label{inst12}Graduate Institute for Advanced Studies, SOKENDAI, 2-21-1 Osawa, Mitaka, Tokyo 181-8588, Japan \and
\label{inst13}School of Physics and Astronomy, Monash University, Clayton, VIC 3800, Australia \and
\label{inst8}Purple Mountain Observatory, Chinese Academy of Sciences, Nanjing, 210023, P.R. China     \and
\label{inst9}INAF - Osservatorio Astronomico di Brera, Via E. Bianchi 46, 23807 Merate (LC), Italy    \and
\label{inst23}Department of Astronomy, University of California, Berkeley, CA 94720-3411, USA \and
\label{inst17}Flai d.o.o., Bravni\v{c}arjeva ulica 13, 1000 Ljubljana, Slovenia \and
\label{uni:padua}Universit\`a degli Studi di Padova, Dipartimento di Fisica e Astronomia, Vicolo dell'Osservatorio 3, 35122 Padova, Italy \and
\label{inst6}Institute of Space Sciences (ICE, CSIC), Campus UAB, Carrer de Can Magrans, s/n, E-08193 Barcelona, Spain \and
\label{inst3}South-Western Institute for Astronomy Research, Yunnan Key Laboratory of Survey Science, Yunnan University, Kunming, Yunnan 650500, P.R. China \and
\label{inst4}School of Electronic Science and Engineering, Chongqing University of Posts and Telecommunications, Chongqing 400065, P.R. China \and
\label{UCAS}School of Astronomy and Space Science, University of Chinese Academy of Sciences, Beijing 100049, China \and
\label{NAOC}National Astronomical Observatories, Chinese Academy of Sciences, Beijing 100101, China \and
\label{inst:Nevada}Department of Physics, University of Nevada, Reno, NV 89557, USA  \and
\label{inst:Virginia}Department of Astronomy, University of Virginia, 530 McCormick Road, Charlottesville, VA 22904, USA  \and
\label{inst111}Planetary Science Institute, 1700 East Fort Lowell Road, Suite 106, Tucson, AZ 85719-2395 USA  \and
\label{inst112}Hamburger Sternwarte, Gojenbergsweg 112, 21029 Hamburg, Germany  \and
\label{inst113}Homer L.~Dodge Department of Physics and Astronomy, University of Oklahoma, Rm 100 440 W. Brooks, Norman, OK 73019-2061, USA  \and
\label{inst:UCSD}Department of Physics, University of California, San Diego, CA 92093, USA \and
\label{inst114}Space Telescope Science Institute, 3700 San Martin Drive, Baltimore, MD 21218-2410, USA \and
\label{inst15.5}Fabra Observatory, Royal Academy of Sciences and Arts of Barcelona (RACAB), 08001 Barcelona, Spain   \and
\label{inst16}Institute for Space Studies of Catalonia (IEEC), Campus UPC, 08860 Castelldefels (Barcelona), Spain \and
\label{yale}Department of Astronomy, Yale University, New Haven, CT 06520-8101, USA \and
\label{inst37}Department of Physics and Astronomy, FI-20014, University of Turku, Finland \and
\label{inst:IAA-CSIC}Instituto de Astrof\'\i sica de Andaluc\'\i a, CSIC, Glorieta de Astronom\'\i a s/n, ES18008, Granada, Spain \and
\label{Knoxville}Department of Nuclear Engineering, University of Tennessee, Knoxville TN 37996-2300, USA \and 
\label{inst16.33}Astrophysics Research Institute, Liverpool John Moores University, IC2, 146 Brownlow Hill, Liverpool L3 5RF, UK \and
\label{inst16.66}Max-Planck-Institut f\"ur Astrophysik, Karl-Schwarzschild Str. 1, D-85741 Garching, Germany \and
\label{inst14}Department of Applied Physics, School of Engineering, University of C\'{a}diz, Campus of Puerto Real, E-11519 C\'{a}diz, Spain \and
\label{inst10}School of Physics and Astronomy, University of Leicester, University Road, Leicester LE1 7RH, UK   \and
\label{inst35}INAF - Osservatorio Astronomico di Capodimonte, Salita Moiariello 16, 80131 Napoli, Italy   \and
\label{inst110}INAF - Osservatorio Astronomico d'Abruzzo, Via M. Maggini snc, 64100 Teramo, Italy    \and
\label{inst:Yangtze}School of Physics and Optoelectronic, Yangtze University, Jingzhou 434023, P.R. China \and
\label{inst307}Beijing Planetarium, Beijing Academy of Sciences and Technology, 100044, P.R. China 
}

   \authorrunning{Y.-Z. Cai et al.} 
   \titlerunning{SN 2019vxm}
   
   \date{Received 2026; accepted 2026}
 
  \abstract
  {
  We present the photometric and spectroscopic analysis of the luminous and long-lasting Type IIn supernova (SN) 2019vxm. The SN reaches a peak $V$-band absolute magnitude of $M_V = -20.01 \pm 0.13$~mag in 35.0 days, and displays slow evolution in both the light curves and spectra, resembling that of long-lived SNe IIn. 
  A mid-infrared (MIR) excess is detected starting from seven months after maximum brightness, suggesting a few $10^{-3}$ \msun of dust are newly formed at $\gtrsim 210$ days (and up to 0.01 \msol at $+4.5$ yr).
  The spectra are dominated by a blue continuum at early stages, with narrow, symmetric Balmer lines and flash-ionisation emission lines of C {\sc iii}, N {\sc iii}, and He {\sc ii}. 
  Comparing our flash-ionised spectrum with early interacting SN spectral models, we estimate a lower limit for the mass-loss rate of the progenitor of $\dot{M} \gtrsim 10^{-2}\,\mathrm{M_{\odot}\,yr^{-1}}$. 
  A weak P~Cygni absorption feature is detected in the \Hb\ profile of the high-resolution Echelle spectrum at $+$19.7 d, suggesting the presence of slow-moving ($60\pm10$ \kms), unshocked circumstellar material (CSM) arising from the pre-SN wind of the progenitor. 
  The \Ha\ and \Hb\ profiles gradually evolve and become broader and asymmetric, showing a progressively increasing blueshift, with a clear flux deficit in the red wings of the broad velocity component after $+$102 days. 
  Our observed bolometric light curve before $\sim$100 days can be well fitted by a power-law function ($L(t)=2\times~10^{44}(t/{\rm day})^{-0.49}$ erg s$^{-1}$), which is very similar to SN\,2010jl. The bolometric luminosity evolution indicates that the progenitor mass-loss rate was $\sim 0.05~\mathrm{M_\odot~yr^{-1}}$ if we assume $\rho_\mathrm{CSM}\propto r^{-2}$. 
  These findings are consistent with an asymmetric, dense circumstellar environment, indicative of a massive, compact progenitor such as a luminous blue variable (LBV)  embedded in a clumpy, asymmetric medium. However, we cannot entirely rule out an extreme red supergiant (RSG) with superwinds or a transitional post-RSG/LBV star as the potential progenitor.
  }
 

   \maketitle
   \nolinenumbers
   
\section{Introduction}

Type IIn supernovae (SNe IIn) were first introduced by \citet[][]{Schlegel1990MNRAS.244..269S} \citep[but see also][for earlier examples]{Filippenko1989AJ.....97..726F} to describe a subclass of hydrogen-rich SNe characterised by relatively narrow hydrogen emission lines, ranging from $\sim$100 to $\sim$1000 \kms, superposed on a blue continuum in their early-time spectra \citep[see][for a review]{Filippenko1997ARA&A..35..309F}. These emission lines indicate the presence of a slowly expanding circumstellar medium (CSM). 
The interaction between the rapidly expanding SN ejecta and the CSM generates a forward shock that propagates into the CSM and a reverse shock that moves into the ejecta \citep[see, e.g.,][]{Chevalier1994ApJ...420..268C, Chugai1997Ap&SS.252..225C}. The presence of ejecta-CSM interaction is revealed by typical  observables such as strong X-ray and radio emission, a blue spectral continuum, and multicomponent line profiles, sometimes with boxy components \citep[see, e.g.,][]{Fraser2020RSOS....700467F,Dessart2022A&A...660L...9D,Dessart2024arXiv240504259D,Gangopadhyay2024arXiv241104107G}.

Type IIn SNe are relatively rare, representing around 10\% of all core-collapse (CC) SNe in the local Universe \citep[][]{Smartt2009ARA&A..47...63S,Smith2011MNRAS.412.1522S,Li2011MNRAS.412.1441L,Cappellaro2015A&A...584A..62C,Perley2020ApJ...904...35P,Ma2025A&A...698A.306M,Ma2025A&A...698A.305M}, while  \citet{Eldridge2013MNRAS.436..774E} even estimated a much lower rate of 2.4\%. SNe IIn display  tremendous heterogeneity in their light curves and spectra \citep[see, e.g.,][]{Kiewe2012ApJ...744...10K,Pastorello2019A&A...628A..93P,Nyholm2020A&A...637A..73N,Reguitti2024A&A...686A.231R,Salmaso2025A&A...695A..29S}, with  absolute magnitudes at peak ranging from $M_{r} \approx -16$~mag for the faintest events to $M_{r} \approx -22$~mag for very luminous objects \citep[e.g.,][]{Nyholm2020A&A...637A..73N,Hiramatsu2024arXiv241107287H}.
The light-curve shapes are divided into three main subgroups on the basis of their morphology \citep[][]{Taddia2013A&A...555A..10T,Taddia2015A&A...580A.131T}: the plateau-like SNe IIn-P \citep[e.g.,][]{Kankare2012MNRAS.424..855K,Mauerhan2013MNRAS.431.2599M,Elias-Rosa2024A&A...686A..13E}, the fast and linearly (in magnitudes per day) declining SNe IIn-L \citep[e.g.,][]{Fassia2000MNRAS.318.1093F,DiCarlo2002ApJ...573..144D,Pastorello2002MNRAS.333...27P}, and the long-lasting SNe IIn \citep[e.g.,][]{Turatto1993MNRAS.262..128T,Zhang2012AJ....144..131Z,SN2005ip_Stritzinger2012ApJ...756..173S,SN2015da_Tartaglia2020A&A...635A..39T,SN2017hcc_Moran2023A&A...669A..51M}. 
Members of this last category are generally more luminous than SNe IIn-P, although without reaching the brightness of superluminous (SL) SNe. 

In addition, several studies suggest that SNe~IIn may be more prone to containing warm dust compared to other types of CC SNe, as they typically exhibit late-time ($\gtrsim100$ days) infrared (IR) emission associated with warm dust \citep[e.g.,][]{Fox2009ApJ...691..650F,Fox2010ApJ...725.1768F,Fox2011ApJ...741....7F,Miller2010MNRAS.404..305M,Miller2010AJ....139.2218M, Maeda2013ApJ...776....5M, Gall2014Natur.511..326G, Bevan2019MNRAS.485.5192B, Bevan2020ApJ...894..111B, Dwek2021ApJ...917...84D, Raphael2025ApJ...983..101B,
Shahbandeh2025ApJ...985..262S}. 
This photometric diversity, along with spectral information such as CSM and ejecta velocities, reflects a wide range of CSM geometric configurations, masses, and density profiles --- hence, in general, different properties of the progenitors and their local environments. 

The variety of SN IIn observables suggests the existence of multiple progenitor channels \citep[see, e.g.,][]{Dwarkadas2011MNRAS.412.1639D,Moriya2013MNRAS.435.1520M,Leloudas2015A&A...574A..61L,Smith2017hsn..book..403S,Ransome2025ApJ...987...13R}. In particular, luminous blue variables (LBVs) have been proposed as plausible progenitors by \citet{Kotak2006A&A...460L...5K}, \citet{Gal-Yam2007ApJ...656..372G}, and \citet{Gal-Yam2009Natur.458..865G}, as they expel significant amounts of mass during their eruptive phases at velocities comparable to those observed in the spectra of SNe~IIn \citep[][]{Kiewe2012ApJ...744...10K,Taddia2013A&A...555A..10T}. Specifically, this scenario is strongly supported by the direct identification of the progenitor of the Type IIn SN\,2005gl in pre-explosion {\it Hubble Space Telescope (HST)} archival images, and its subsequent disappearance in late-time observations of the SN site \citep{Gal-Yam2007ApJ...656..372G,Gal-Yam2009Natur.458..865G}. On the other hand,  the class of SN 2008S-like objects\footnote{They are often referred to as intermediate-luminosity red transients \citep[ILRTs; e.g.,][]{Botticella2009MNRAS.398.1041B,Cai2022Univ....8..493C}.} exhibits relatively narrow hydrogen lines similar to those of SNe~IIn, and their progenitors are typically enshrouded in dusty cocoons and have modest estimated initial masses of 8--15 \msun  \citep[e.g.,][]{Cai2018MNRAS.480.3424C,Cai2021A&A...654A.157C,Valerin2025A&A...695A..43V,Valerin2025A&A...695A..42V}. While \citet{Smith2009ApJ...697L..49S} interpreted these events as SN impostors analogous to the eruptions of LBVs, alternative explanations --- such as a super-asymptotic giant branch (AGB) stars \citep[e.g.,][]{Moriya2014A&A...569A..57M} or blue supergiants \citep[BSGs;][]{Moriya2024PASJ...76L..27M} as progenitors --- remain plausible.

On the other hand, the most extreme exemplars, such as overluminous and long-lasting events, significantly broaden our understanding of SNe IIn. For instance, superluminous SNe IIn require both a high kinetic energy in the inner shell and a large mass reservoir in the outer shell \citep[see, e.g.,][]{Moriya2013MNRAS.428.1020M,Dessart2015MNRAS.449.4304D}. Some good examples of such SNe IIn are
SN\,2008am \citep[][]{SN2008am_Chatzopoulos2011ApJ...729..143C}, SN 2010jl \citep[e.g.,][]{Andrews2011AJ....142...45A,Stoll2011ApJ...730...34S,Smith2011ApJ...732...63S,Smith2012AJ....143...17S,Zhang2012AJ....144..131Z,Maeda2013ApJ...776....5M,SN2010jl_Ofek2014ApJ...781...42O,Ofek2019PASP..131e4204O,Fransson2014ApJ...797..118F,Jencson2016MNRAS.456.2622J,Chugai2018MNRAS.481.3643C}, SN 2015da \citep[][]{SN2015da_Tartaglia2020A&A...635A..39T,SN2015da_Smith2024MNRAS.530..405S}, and more recently SN 2017hcc \citep[e.g.,][]{Prieto2017RNAAS...1...28P,Smith2020MNRAS.499.3544S,SN2017hcc_Moran2023A&A...669A..51M,Sethulakshmi2026ApJ..1001..169S}. In this study, we present observations of the luminous, slowly evolving Type IIn SN\,2019vxm, monitored for more than two years. Relevant datasets for this SN have been  published by \citet{Vallely2021MNRAS.500.5639V}, \citet{Tsvetkov2023AstBu..78..514T,Tsvetkov2024AN....34530166T}, \citet{Lane2026ApJ..1003...19L}, and \citet{Lelkes2026arXiv260523637L}, and will be discussed later.

This paper is structured as follows. Sect. \ref{section:Basic_information} describes the basic information on SN\,2019vxm, including discussions of distance and reddening. The photometric and spectroscopic analysis are presented in Sects. \ref{section:photometry}~and \ref{section:spectroscopy}, respectively. Sect. \ref{section:discussion} discusses our results.
We report all the photometric and spectroscopic observations, as well as the data-reduction techniques, in Appendix~\ref{appendix:data}.

\section{Basic information for SN\,2019vxm}
\label{section:Basic_information}

\begin{figure}[htbp]
\begin{center}
\includegraphics[width=\linewidth]{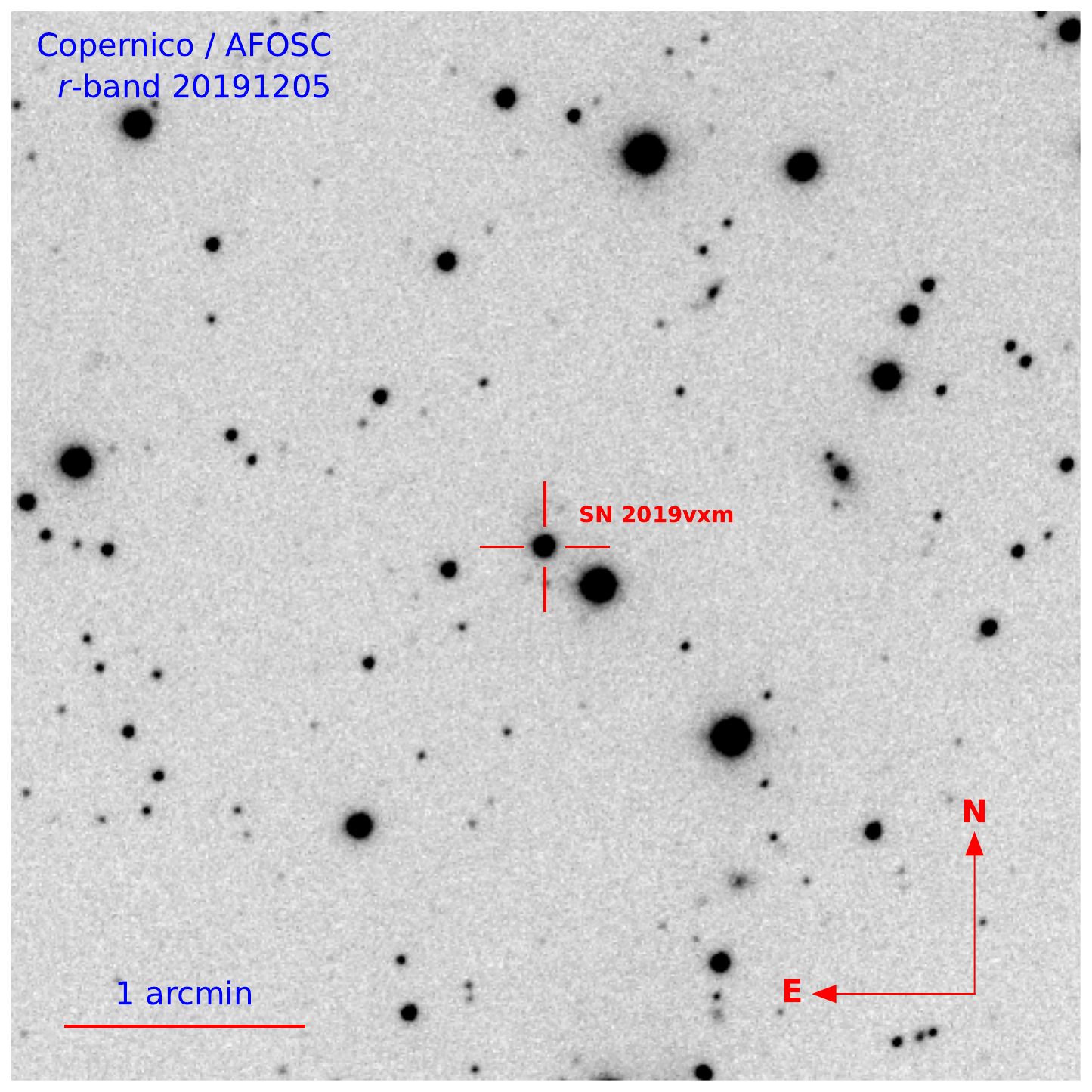}
\caption{SN~2019vxm in a Copernico $+$ AFOSC image taken with Sloan $r$ filter on 2019 December 5. The SN is marked at the crosshair, near the centre of the image.}
\label{fig:finderchart}
\end{center}
\end{figure}

\begin{figure}[htbp]
\begin{center}
\includegraphics[width=\linewidth]{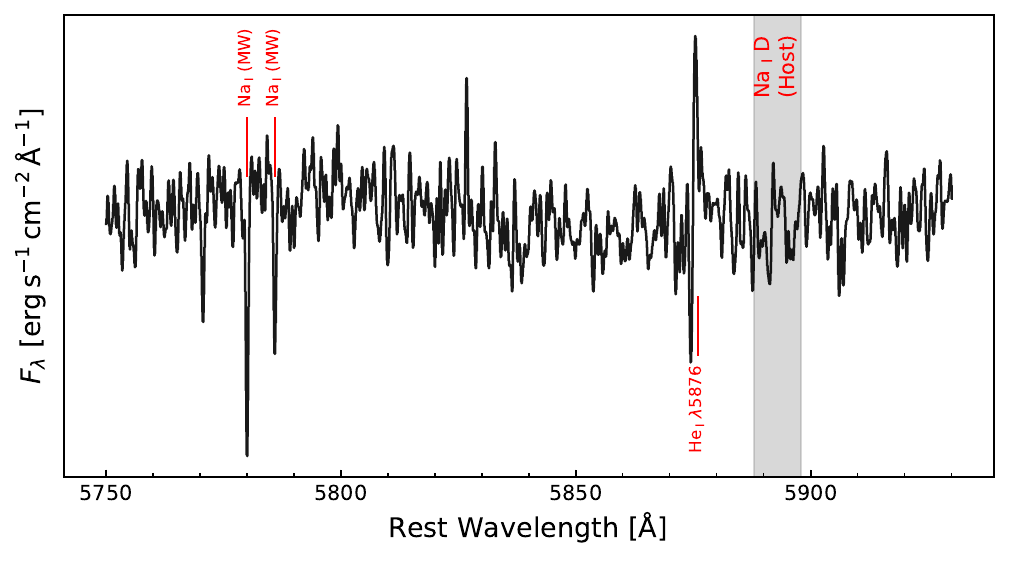}
\caption{Zoomed-in view of the Na\,{\sc i}\,D spectral region in the Copernico/Echelle spectrum taken on 2019 December 06 ($+$19.7~d). }
\label{fig:spec:HeNa}
\end{center}
\end{figure}

SN\,2019vxm (also known as ASASSN-19acc, ATLAS19bchy, Gaia19fje, and PS20nyb) was first announced by the All-Sky Automated Survey for Supernovae \citep[ASAS-SN;][]{Shappee2014AAS...22323603S, Kochanek2017PASP..129j4502K} on 2019 December 1 UTC (JD = 2,458,818.54; UTC dates are used throughout the paper), at $g\approx15.4$ mag \citep[AB mag;][]{Oke1983ApJ...266..713O} \citep{Stanek2019TNSTR2492....1S}. The source was later classified by \citet{Leadbeater2019TNSCR2506....1L} as a young Type IIn SN based on the similarity of its spectrum to that of SN~1998S about 20 days before maximum light.

SN\,2019vxm (RA~=~$19^{\rm hr}58^{\rm m}28\fs54$, Dec.~=~$+62\degr08\arcmin15\farcs83$; J2000) has a projected offset of 3.19 kpc from the core of its possible host galaxy SDSS J195828.83+620824.3, and is 8\farcs53 south and 2\farcs07 west of the galaxy centre (see Fig.~\ref{fig:finderchart}). 
Given the absence of a published redshift for the putative host galaxy, we adopt $z=0.019\pm0.001$ based on the fit to the early H$\alpha$ emission line of SN~2019vxm,  consistent with the value adopted by others \citep{Leadbeater2019TNSCR2506....1L,Tsvetkov2024AN....34530166T,Lane2026ApJ..1003...19L}. This redshift corresponds to a luminosity distance of $d_{L} = 79.2 \pm 4.2$~Mpc and a distance modulus of $\mu_{L} = 34.49\pm0.12$~mag. These values are calculated under the assumption of a standard $\Lambda$CDM cosmology with H$_{0} = 73~\mathrm{km\,s^{-1}\,Mpc^{-1}}$, $\Omega_{\mathrm{M}} = 0.27$, and $\Omega_{\Lambda} = 0.73$ \citep[][]{Spergel2007ApJS..170..377S}.

The Milky Way (MW) extinction toward SN\,2019vxm is $E(B - V)_{\mathrm{MW}} = 0.087\pm0.002$~mag  \citep{Schlafly2011ApJ...737..103S}, in agreement with the value inferred from the measurement of the equivalent width (EW) of the Galactic Na\,{\sc i}\,D $\lambda\lambda$5890, 5896 absorption lines in our early, high-resolution ($R\approx20,000$) Copernico/Echelle spectrum taken on 2019 December 06 ($+$19.7~d after explosion). To estimate the additional host-galaxy extinction, we checked the EW of the interstellar Na\,{\sc i}\,D absorption at the redshift of the host galaxy in the same spectrum but did not detect any Na\,{\sc i} lines. This is consistent with the findings of \citet{Tsvetkov2023AstBu..78..514T, Tsvetkov2024AN....34530166T} in their better-quality spectra. The detailed identification of Na\,{\sc i}\,D lines is shown in Fig. \ref{fig:spec:HeNa}.
Thus, we assume that the total line-of-sight extinction of SN\,2019vxm is equal to the Galactic value, $E(B - V)_{\mathrm{Total}} =0.087\pm0.002$~mag.

\section{Photometry}
\label{section:photometry}

We monitored the photometric evolution of SN~2019vxm for $\sim 700$ days from discovery in the optical band, and for $\sim 4$\,yr in the mid-infrared (MIR) domain.
{\it Swift}-UVOT observations were obtained in the ultraviolet (UV) and the blue optical filters, spanning a period from $\sim 16$ to  50 days after maximum light.
Basic information on the photometric observations of SN\,2019vxm and the data-reduction techniques  are provided in Sect. \ref{appendix:photodata}, while the multiband light curves are shown in Fig.~\ref{fig:phot:applc}.

\subsection{Apparent magnitude light curves} 
\label{section:apparentphotometry}

\begin{figure*}[htbp]
\begin{center}
\includegraphics[width=0.9\textwidth]{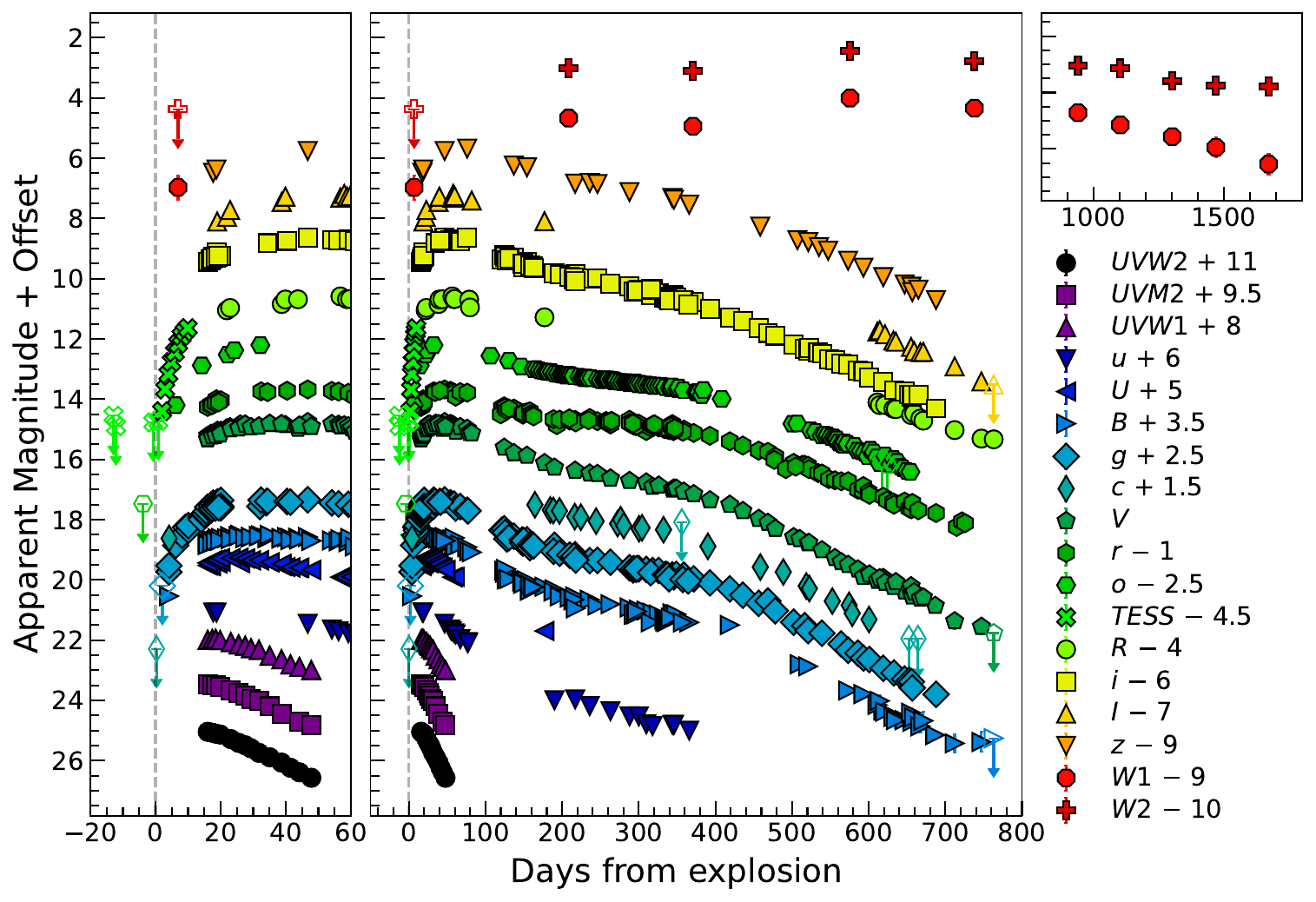}
\caption{Multiband (UV, optical, and MIR) apparent light curves of SN~2019vxm. Error bars are included for all photometric points, but the uncertainty is usually smaller than the marker size. A dotted vertical line indicates the epoch of explosion epoch.}
\label{fig:phot:applc}
\end{center}
\end{figure*}

We adopt the explosion time of MJD = 58804.03$\pm$0.30 precisely constrained by \citet{Lane2026ApJ..1003...19L}. From the {\it Swift} $V$-band rise time of $35.0^{+1.8}_{-1.7}$\,d estimated by \citet{Lane2026ApJ..1003...19L}, the maximum-light time is MJD  58839.0. 
For each UV band, the epoch of the first \textit{Swift}-UVOT  data point was adopted to estimate the decline slopes.
Notably, the $R$- and $I$-band photometry of SN 2019vxm is  concentrated around the peak luminosity and at phases later than 600 d (see Fig.~\ref{fig:phot:applc}); consequently, the decline slopes were computed only for these two bands at phases beyond 600 d.

The decline rates for different bands, summarised in Appendix Table~\ref{tab:decline_rate}, exhibit a somewhat unexpected heterogeneity across the filters.
After maximum light, SN~2019vxm displays a slow brightness decline with several changes in the decline rate. The UV light curves exhibit an approximately linear (in mag/day) decline, with rates significantly steeper than those in the optical bands (e.g., $\gamma_{0-40}$($UVW2$) $\approx 4.96$ mag/100 d). 
Over the period 0--100\,d after peak light, the blue optical filters fade more rapidly (e.g., $\gamma_{0-100}$($B$) $\approx 1.32$ mag/100 d), whereas the redder ones evolve more slowly (e.g., $\gamma_{0-100}$($r$) $\approx 0.61$ mag/100 d). 
Between 100 and 400\,d, the decline rates in the optical bands decrease more slowly (e.g., $\gamma_{100-400}$($r$) $\approx 0.25$ mag/100 d), indicating a plateau-like evolution. 
After 400\,d, the decline rates increase again, approaching the $^{56}$Co--$^{56}$Fe radioactive decay rate of 0.98 mag/100\,d (e.g., $\gamma_{400-700}$($r$) $\approx 0.83$ mag/100 d). As noted by \citet{Lelkes2026arXiv260523637L}, this apparent alignment of the decline rate with the $^{56}$Co--$^{56}$Fe slope is likely coincidental. 

In the MIR domain, the light curves of SN\,2019vxm exhibit a different evolutionary behaviour compared to those at shorter wavelengths (Fig.~\ref{fig:phot:applc}). The first detection is dated MJD~58810.94 (15.97$\pm$0.40 mag) in the $W1$ band, while the source was not detected at that epoch in the $W2$ band.
However, the timing of the MIR peak is poorly constrained with  the available MIR data. 
From the epoch of maximum in the $V$ band, the MIR light curves may exhibit a peak within the first 0--200~d; however, the sparse sampling prevents a robust determination. Between 200 and 300~d, the luminosity remains approximately constant, and is followed by a second phase of brightening extending to $\sim$500~d. Meanwhile, the optical light curves show a pronounced decline, which may indicate the formation of dust (see Sect.~\ref{mir} for a detailed discussion).
Given the limited number of data points and the nonmonotonic evolution, we refrain from reporting quantitative rise or decline rates. 
Between 500 and 750~d, both the $W1$ and $W2$ bands show a declining trend, with decline rates of $\gamma_{500-750}(W1) \approx 0.21$~mag/100\,d and $\gamma_{500-750}(W2) \approx 0.20$~mag/100\,d. At later epochs, between 1000 and 1700~d, the MIR luminosity continues to decline. The decline rate in the $W1$ band increases slightly to $\gamma_{1000-1700}(W1) \approx 0.24$~mag/100\,d, whereas that in the $W2$ band decreases to $\gamma_{1000-1700}(W2) \approx 0.12$~mag/100\,d. Owing to the sparse data coverage at these late epochs, these estimates are subject to large uncertainties.

\subsection{Absolute magnitude light curves}
\begin{figure}[htbp]
\begin{center}
\includegraphics[width=\linewidth]{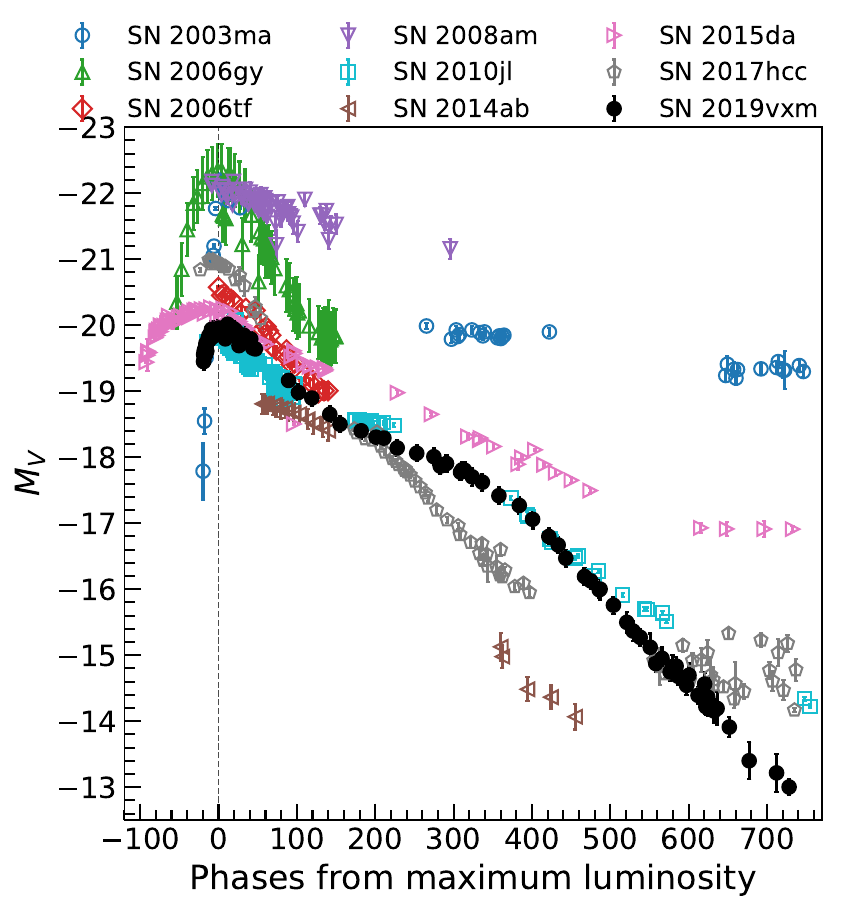}
\caption{Absolute $V$-band light curves of SN~2019vxm are compared with those of bright Type~IIn SNe and SLSN-II. For SNe 2006gy and 2008am, owing to the lack of sufficient $V$-band data, we use $R$- or $i$-band measurements instead.}
\label{fig:Absolute_magnitude}
\end{center}
\end{figure}

In this study, we compile a comparison sample of Type~IIn SNe to serve as a reference for the photometric and spectroscopic analysis of SN~2019vxm. The comparison objects either have luminosities similar to the ``optical'' luminosity of SN~2019vxm, or with a long evolution time.  All comparison SNe, listed in Appendix~Table~\ref{apptab:supm:cmpsinfo}, are shown with their absolute $V$-band light curves in Fig.~\ref{fig:Absolute_magnitude}.

Using the distance and extinction values from Sect.~\ref{section:Basic_information}, a second-order polynomial fit to the $V$-band light curve of SN~2019vxm yields a peak absolute magnitude of $M_{V} = -20.01 \pm 0.13$~mag at $\mathrm{MJD} = 58839.0 \pm 1.7$. While this luminosity is high in an absolute sense, it places SN~2019vxm at the fainter end of the comparison sample when SLSNe-IIn are included. Owing to their diverse progenitor channels and complex CSM interactions,  Type~IIn SNe exhibit a wide range of absolute magnitudes, spanning more than 6 mag \citep{Smith2017hsn..book..403S}. 
SN~2019vxm reaches a higher peak luminosity, approaching that of SNe~2010jl ($\sim -19.43$ mag) and 2015da ($\sim -20.2$ mag). Although it is not as luminous as SLSNe-IIn, SN~2019vxm exhibits a long-duration evolution similar to that of some extremely luminous SNe~IIn (e.g., SNe~2003ma and 2017hcc).
This is also consistent with the findings of \citet{Nyholm2020A&A...637A..73N} that the most luminous SNe IIn are generally found to be longer lasting.

\subsection{Colour evolution}
\label{subsection:Colour}

\begin{figure}[htbp]
\begin{center}
\includegraphics[width=\linewidth]{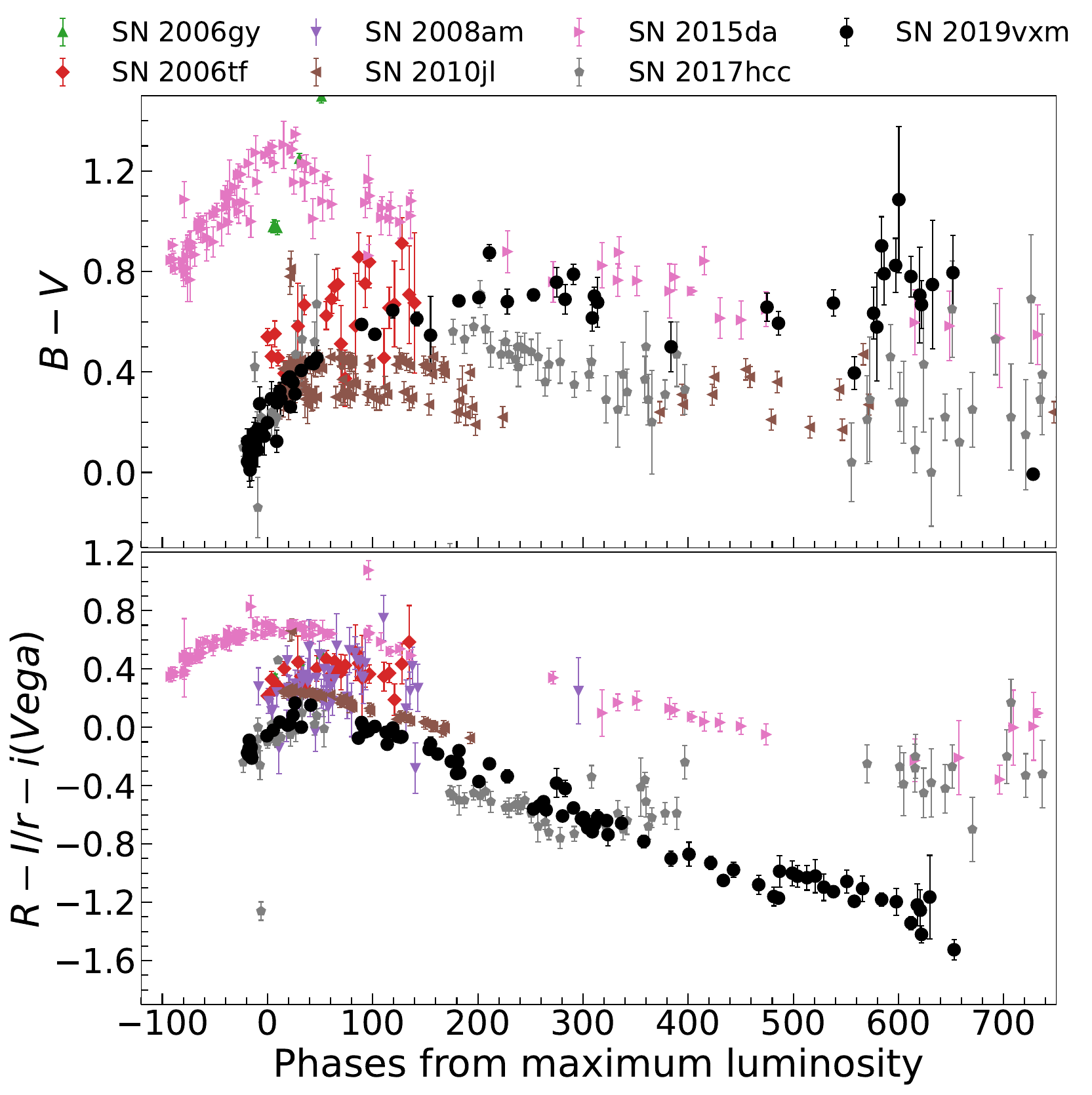}
\caption{Colour evolution of SN\,2019vxm, compared to a sample of SNe~IIn.
Upper panel: ($B-V$) colour evolution. Lower panel: ($R-I$) or ($r-i$) colour evolution.
The colour curves are corrected for both Galactic and host-galaxy extinction.}
\label{fig:colour_evolution}
\end{center}
\end{figure}

The intrinsic colour evolution of SN~2019vxm, compared with those of the SN~IIn comparison sample, is shown in Fig.~\ref{fig:colour_evolution}. 
From $\sim 20$~d before maximum light to $\sim 200$~d after, the $B-V$ colour increases from $\sim0.04$~mag to a peak value of $\sim0.87$~mag, indicating an overall trend for the object to become redder with time. 
Such colour evolution alone does not uniquely imply a decrease in temperature, as dust extinction can preferentially attenuate blue photons and thus produce a redder observed colour. Moreover, line blanketing, which becomes more effective as the temperature decreases and the ionisation state of the ejecta evolves, can further enhance the reddening. At later phases, the $B-V$ colour remains relatively stable ($\sim0.67$--0.80~mag).
Overall, the colour evolution of SN~2019vxm closely resembles that of SN~2017hcc. 
By contrast, the $B-V$ colour of SNe~2005ip and 2010jl  remains relatively stable ($\sim0.30$--0.50~mag) throughout the observed phases. The $B-V$ colour of SNe~2006tf and 2015da exhibits an increasing trend at early times; however, the $B-V$ colour of SN~2015da decreases after the peak before finally reaching a stable value.

In terms of the $r-i$ colour evolution, SN~2019vxm shows minor differences with respect to the other objects in the sample, particularly at phases later than 300~d (see the bottom panel in Fig.~\ref{fig:colour_evolution}). From $\sim 20$~d before maximum light to about 30~d after peak, the $r-i$ colour of SN 2019vxm increases from $-0.18$~mag to a peak value of $\sim0.17$~mag. Subsequently, the $r-i$ colour exhibits an apparent blueward trend, declining to $\sim-1.53$~mag by +700~d. This behaviour is likely driven by the increasing contribution of the H$\alpha$ emission to the $r$ band relative to the $i$ band, rather than reflecting an intrinsic continuum evolution. In agreement with this interpretation, the $g-r$ colour shows a reddening trend over the same phases. The early evolution of SN~2019vxm within the first $\sim400$~d is broadly consistent with that of SN~2017hcc, although at later phases the $r-i$ colour of SN~2019vxm becomes noticeably bluer. For the other comparison Type~IIn SNe, the $r-i$ colour evolution of SN~2015da and SN~2017hcc is broadly consistent, showing an initial reddening followed by a blueward trend and stabilising in the late phases. By contrast, the $r-i$ colour of SNe~2006tf and 2008am \citep{SN2006tf_Smith2008ApJ...686..467S, SN2008am_Chatzopoulos2011ApJ...729..143C} remains nearly constant from the earliest epochs, whereas SN~2010jl exhibits blueward evolution after maximum light, although the late-time data are unavailable.

At late phases ($>400$~d), SN~2019vxm exhibits strong H$\alpha$ emission that enhances the flux in the $r$ band relative to the other bands, resulting in a progressively bluer $r-i$ colour.
This strong H$\alpha$ emission is usually associated with  circumstellar interaction, suggesting that it may also be at play in SN~2019vxm during this stage.
None of the comparison Type~IIn SNe exhibit such a sustained blueward trend, although a similar behaviour was previously reported for other Type~II SNe \citep[e.g.,\ SNe~2016iog and 2018hfm;][]{Peng2026A&A...705A.104P, Zhang2022MNRAS.509.2013Z}, and attributed to the effects of CSM interaction.

\subsection{Pseudobolometric light curves} 

\begin{figure}[htbp]
\begin{center}
\includegraphics[width=\linewidth]{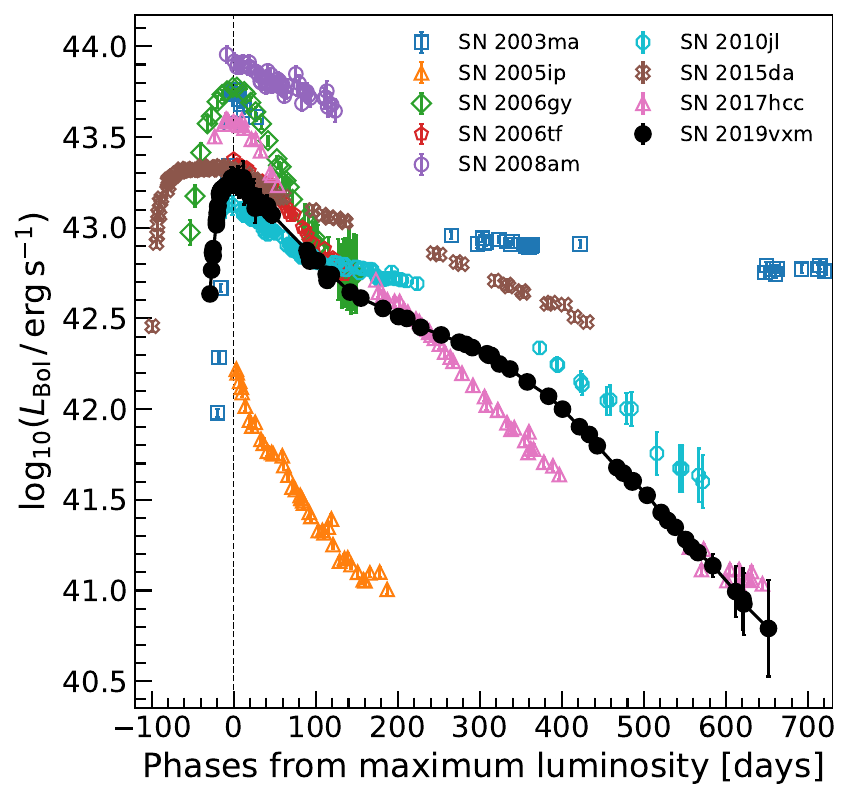}
\caption{Pseudobolometric light curves of SN~2019vxm and comparison bright SNe IIn and SLSN-II. The comparison objects have either similar luminosities to that of the ``optical'' luminosity of SN~2019vxm or with long evolution times. The luminosity is calculated by integrating from the $B$ to the $I/i$ bands.}
\label{fig:pbollc}
\end{center}
\end{figure}

To enable a robust comparison between SN\,2019vxm and other SNe~IIn, we constructed pseudobolometric light curves using the available photometry in the bands from $B$ to $z$, and derived the evolution of the luminosity through black-body fits to the spectral energy distribution (SED), as shown in Fig.~\ref{fig:pbollc}. The properties adopted for the comparison sample are summarised in Table~\ref{apptab:supm:cmpsinfo}.
Within the selected sample, most SNe~IIn reach peak luminosities above \(10^{43}\,\mathrm{erg\,s^{-1}}\). 
The peak luminosity of SN\,2019vxm is \(L \approx 1.59 \times 10^{43}\,\mathrm{erg\,s^{-1}}\), one order of magnitude higher than that of SN~2005ip (\(L \approx 1.66 \times 10^{42}\,\mathrm{erg\,s^{-1}}\)) and marginally exceeding that of SN~2010jl (\(L \approx 1.39 \times 10^{43}\,\mathrm{erg\,s^{-1}}\)). However, its luminosity remains systematically below those of the other SNe~IIn in the comparison sample.
SN\,2019vxm is more luminous than only SN~2010jl within the comparison sample, while most objects exhibit higher peak luminosities. 
Nevertheless, compared with the overall population of SNe~IIn, it remains toward the luminous end of the distribution \citep[$\sim 10^{42}$--$10^{44}\,\mathrm{erg\,s^{-1}}$;][]{Kiewe2012ApJ...744...10K, Taddia2013A&A...555A..10T, Smith2017hsn..book..403S}.

\subsection{Mid-infrared excess and dust formation}
\label{mir}
As mentioned in Sect. \ref{section:photometry},  \textit{WISE} observations revealed a significant late-time excess in the IR domain. In interacting SNe, large IR emission is frequently interpreted as a signature for the presence of dust.
Assuming that the source of the MIR emission is heated dust, we can estimate its mass and temperature by fitting the cool SED using simple models for graphite, silicates, and amorphous carbon composition \citep{Fox2010ApJ...725.1768F}. For simplicity, we assume an optically thin dust shell having mass $M_d$, composed by particles with radius $a$, and thermally
emitting at a single equilibrium temperature $T_d$, with both $M_d$ and $T_d$ as free parameters. The observed flux is given by
\begin{equation}
F_{\lambda} = \frac{M_d B_{\lambda}(T_d) \kappa_{\lambda}(a)}{D^2}\, , 
\end{equation}
where $B_{\lambda}$ is the Planck function for the temperature of the dust $T_d$, $\kappa_{\lambda}(a)$ is the dust opacity coefficient as a function of the wavelength for a given grain radius, and $D$ is the distance from the observer \citep{Hildebrand1983QJRAS..24..267H}. This approach has been  applied to other SNe IIn \citep{Fox2011ApJ...741....7F, SN2005ip_Stritzinger2012ApJ...756..173S, Fox2013AJ....146....2F, Maeda2013ApJ...776....5M, Szalai2019ApJS..241...38S, SN2015da_Tartaglia2020A&A...635A..39T}.
We perform our calculations for grains with size $a = 0.1\,\mu$m and 1.0\,$\mu$m for both graphite and silicates chemical compositions, searching for their associated $\kappa_{\lambda}(a)$ values from the Mie scattering derivations presented by \cite{Fox2010ApJ...725.1768F}\footnote{Values derived from \cite{Draine1984ApJ...285...89D} and \cite{Laor1993ApJ...402..441L}, recently updated by \cite{Valerin2025A&A...695A..42V}. For amorphous carbon, the coefficients are taken from \cite{Colangeli1995A&AS..113..561C} (see the ACAR curve in their Fig.~4).} (see their Fig. 4).
With these assumptions, we calculated the best fits of the grey body on the MIR SED of SN 2019vxm estimated from \textit{WISE} data for all available epochs. 
These estimates are largely uncertain owing to the SEDs being constructed from observations in only two filters in the MIR domain. 

The resulting dust masses and temperatures are summarised in Table \ref{tab:dust_properties} in Appendix \ref{sect:SuppMaterial}. 
We note a general trend of increasing dust mass (a confirmation that the process of dust formation is in progress) and decreasing dust temperature between phases $+6$ months and $+4.5$ yr.
At earlier phases (up to $+2.5$ yr), the dust masses range between 1 to $2 \times 10^{-3}$ \msol. The masses are similar for different dust species, except the case of graphite with $a = 1.0\,\mu$m, whose values are one order of magnitude smaller.
Given the assumption of an optically thin dust shell, the dust masses we derive must be considered as lower limits.
The dust temperatures are in the range 850--1300\,K (although mostly clustered on the hotter side), except for graphite with $a = 0.1\,\mu$m, which provides significantly lower temperatures, in the range 750--800\,K.
At later phases (between $+2.5$ and $+3.5$ yr), the dust masses are higher, (1.4--3.3) $\times 10^{-3}$ \msol, while the dust temperatures are lower (750--950\,K). As a reference, \cite{Bevan2019MNRAS.485.5192B} and \cite{Fox2010ApJ...725.1768F} estimated a dust mass of (2--5) $\times10^{-3}$ \msol for SN 2005ip at $\sim900$ d.
Finally, at the last epoch ($+4.5$ yr), we determine a dramatic increase in dust masses ($8\times10^{-3}$ to 0.01~\msol) and a further decrease in the dust temperature (450--600\,K). The last value is close to the 0.005--0.01~\msol found by \cite{Bevan2020ApJ...894..111B} for SN~2010jl at a similar phase (+1400 days).
Assuming a standard gas-to-dust ratio of 100:1 (but see \citealt{Sarangi2025ApJ...993...94S}), the dust masses we infer imply a total CSM mass of a few tenths of \msol and up to 1 \msol, but again these values should be considered as lower limits. We remark that our independent dust analysis is aligned with the results from \citet{Lelkes2026arXiv260523637L}.

\begin{figure}[htbp]
\begin{center}
\includegraphics[width=\linewidth]{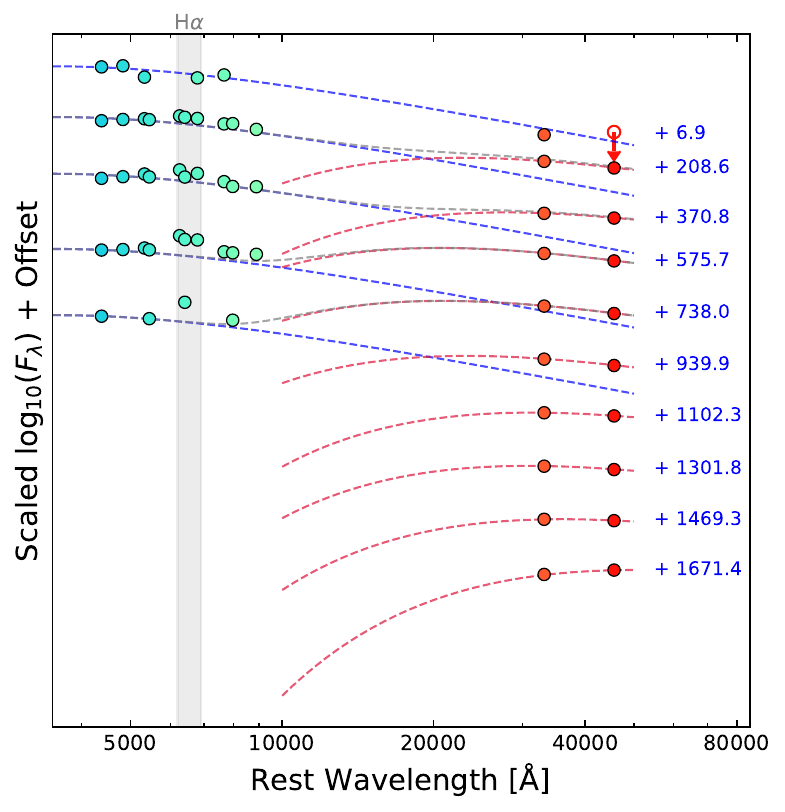}
\caption{SED evolution of SN~2019vxm. The dashed blue lines represent the best black-body fits to the optical bands, while the dashed red lines represent the best fits of the grey-body dust emission model for silicates of $a=0.1\,\mathrm{\mu m}$. Each model is overplotted with the SED, and SEDs are shifted vertically by an arbitrary constant for clarity. }
\label{fig:dust}
\end{center}
\end{figure}

\subsection{Light-curve modelling} 
\label{subsection:LC_modeling}

\begin{figure}[htbp]
\begin{center}
\includegraphics[width=\linewidth]{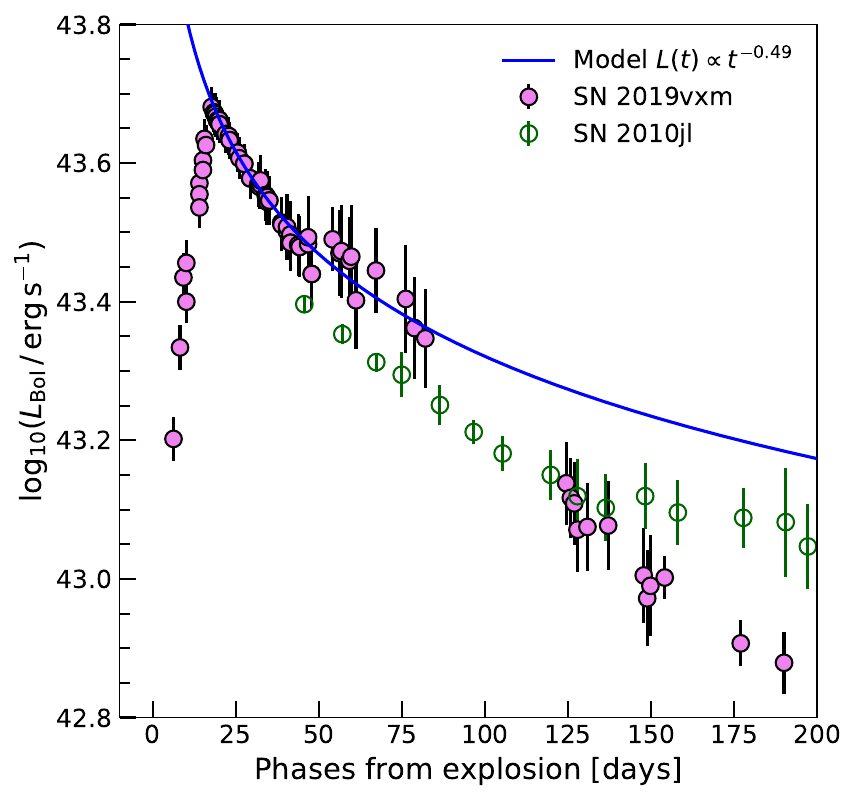}
\caption{Light-curve modelling of SN~2019vxm, overplotted with the comparison SN~2010jl. The light curve is fitted with a single power-law function, shown as a blue line.}
\label{fig:lcmdl}
\end{center}
\end{figure}

We estimate the CSM properties of SN~2019vxm through the bolometric light curve using the simple analytic light-curve model presented by \citet{Moriya2013MNRAS.435.1520M} to reproduce the bolometric light curves of SNe powered by interaction
between ejecta and dense CSM. The model fits well the bolometric light curve until 100~days after peak brightness, while the model's luminosity remains systematically higher than the observed light curve at later phases.
While our observations prove that an IR excess affects the bolometric light curve at phases later than 208~days, observations in the IR domain are not available at earlier stages. Consequently, an underestimated IR contribution between about 100 and 208~days can be responsible of the discrepancy between the observed bolometric light curve and the model's predictions. A similar approach was adopted for the analysis of SN~2010jl \citep[e.g.,][]{Fransson2014ApJ...797..118F,Moriya2013MNRAS.435.1520M,SN2010jl_Ofek2014ApJ...781...42O}. Our fit resulted in $L(t)=2\times 10^{44}(t/\mathrm{day})^{-0.49}~\mathrm{erg~s^{-1}}$. The fitted power-law luminosity evolution is similar to those obtained for SN~2010jl ($2.04\times 10^{44}(t/\mathrm{day})^{-0.486}~\mathrm{erg~s^{-1}}$; \citealt{Moriya2013MNRAS.435.1520M}).

If we assume that the CSM density structure follows $\rho_\mathrm{CSM}\propto r^{-2}$ from steady mass loss and the SN ejecta have a broken power-law density structure, $L(t)\propto t^{-0.49}$ indicates the outer SN ejecta density follows $\rho_\mathrm{ej} \propto r^{-8.1}$, which is consistent with the radiative envelope of LBVs \citep{1999ApJ...510..379M}. We assume that the inner SN ejecta density follows $\rho_\mathrm{ej} \propto r^{-1}$.
Then, adopting an approach similar to that of \citet{Moriya2013MNRAS.435.1520M}, and  assuming an SN ejecta mass of $5~\mathrm{M_\odot}$ and an explosion energy of $5\times 10^{51}~\mathrm{erg}$, we estimate a mass-loss rate of $0.05~\mathrm{M_\odot~yr^{-1}}$ with the assumed CSM velocity of $150~\mathrm{km~s^{-1}}$. If we set the ejecta mass to be $10~\mathrm{M_\odot}$, the mass-loss rate becomes $0.1~\mathrm{M_\odot~yr^{-1}}$, which is in agreement with the independent analysis from \citet{Lelkes2026arXiv260523637L} with their optically thick CSM model ($\dot{M}\sim$~0.2--1 \msun\,yr$^{-1}$). The estimated mass-loss rate for SN\,2019vxm is consistent with the value inferred from its early-phase spectrum modelling (see Sect. \ref{subsection:SpecComp}). This rate is also consistent with those of SN\,2010jl  \citep[$\sim 0.1$ \msun\,yr$^{-1}$;][]{Fransson2014ApJ...797..118F} and SN\,2015da  \citep[$\sim 0.6$\,\msun\,yr$^{-1}$;][]{SN2015da_Tartaglia2020A&A...635A..39T}.

\section{Spectroscopy}
\label{section:spectroscopy}

SN\,2019vxm was classified as a Type IIn SN on 2019 December~1 \citep{Leadbeater2019TNSCR2506....1L}, 20.3 days prior to $V$-band maximum light, and just 0.7 day after the official discovery. Three days later, we started a spectroscopic follow-up campaign of SN\,2019vxm, during which we collected 38 optical spectra spanning about 22 months of evolution. The data-reduction procedures applied to these spectra are outlined in Sect. \ref{appendix:specdata}. 
The log of the spectroscopic observations is reported in Table \ref{tab:speclog_2019vxm} (Appendix~\ref{appendix:SpecInfo}), along with instrumental configuration details.

\subsection{Low-resolution spectral sequence}
\label{subsection:spectrasequence}

\begin{figure*}[htbp]
\begin{center}
\includegraphics[width=0.9\linewidth]{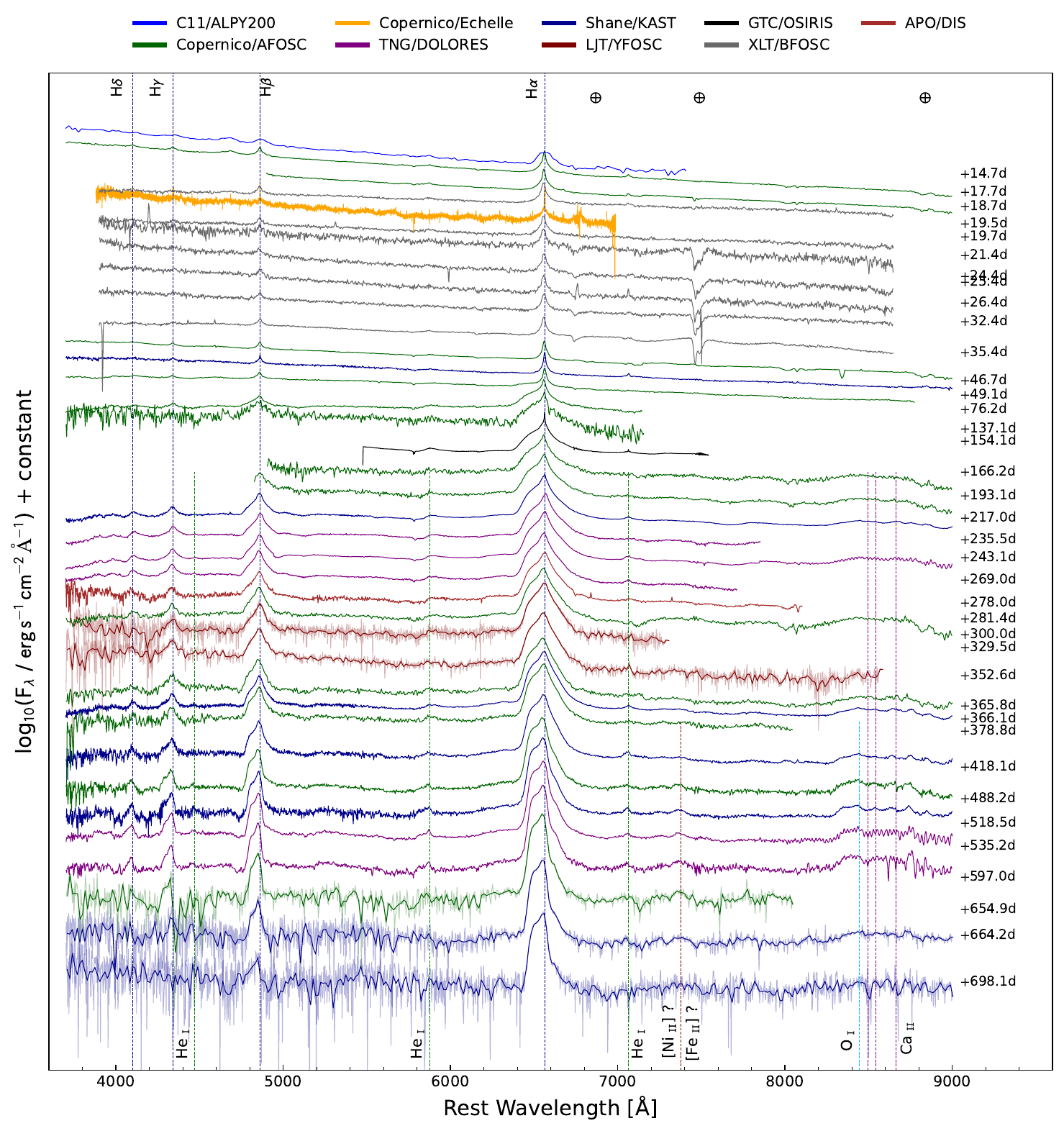}
\caption{The spectral evolution of SN~2019vxm. Some significant spectral features are marked with vertical-coloured lines. The marked phases are with respect to the time of the $V$-band maximum light. The spectra are redshift-corrected and calibrated to match the photometric data; vertical shifts have been applied for clarity. Telluric lines are indicated with encircled ``plus'' symbols. Data from different telescopes and/or instruments are presented in different colours. Spectra with a low S/N were re-binned with 15~\AA~each bin; the original (unbinned) spectra are displayed in lighter colours behind.}
\label{fig:spec:spevo}
\end{center}
\end{figure*}

The time series of 38 optical spectra of SN\,2019vxm is shown in Fig.~\ref{fig:spec:spevo}. 
In the following, we analyse in detail the spectral evolution during the three temporal windows identified in Sect. \ref{section:apparentphotometry} (see also Table \ref{tab:decline_rate}).
In the same three time windows, we also note major changes in the spectra of SN\,2019vxm, especially in the continuum temperature and the  \Ha profile.

{\bf Early phases: $< +100$~days:} The early-time spectra of SN\,2019vxm are characterised by a strong blue continuum with prominent, relatively narrow Balmer emission lines (from \Ha~to \hd), which are typically seen in Type IIn SNe. The obvious bump detected at $\sim 4600$--4700\,\AA\ in the first two spectra (at $+$14.73 and $+$17.68 days) is the classical flash-ionisation feature visible in a number of CC SNe, due to a blend of emission lines of \Ciii~$\lambda$4648, \Niii~$\lambda$4640, and \Heii~$\lambda$4686 (see the top-left panel of Fig. \ref{fig:SpecComp} in Sect. \ref{subsection:SpecComp}). In particular, these flash-ionisation signatures are frequently observed in SNe II within the first few days after the explosion \citep[e.g., SNe\,1998S, 2013fs, 2013cu, 2024cld, and PTF\,11iqb;][]{Leonard2000ApJ...536..239L, Shivvers2015ApJ...806..213S, Gal-Yam2014Natur.509..471G, Dessart2017A&A...605A..83D, Killestein2026MNRAS.548f2261K, Smith2015MNRAS.449.1876S}, indicating that enhanced mass loss just before the explosion is common among Type II SN progenitors \citep{Bruch2023ApJ...952..119B,Jacobson2024ApJ...970..189J}.
However, the flash-ionisation features remained detectable in SN~2019vxm until $\sim$17 days after the explosion. This prolonged duration suggests that the CSM environment of SN~2019vxm is highly dense, massive, and/or extended, indicating stronger mass-loss activity compared to typical SNe~II. Further details are shown in the top-left panel of Fig. \ref{fig:SpecComp} and discussed in Sect. \ref{subsection:SpecComp}. In addition, we detect weak emission features of \Hei~$\lambda$5876 and $\lambda$7065.

We estimate the temperature by fitting the continuum with a black-body function, following the methodology described by \citet{Cai2026A&A...707A.157C}. The best-fit results give a photospheric temperature of $T_{\rm BB} = 20,100_{-1,350}^{+1,450}~{\rm K}$ for the first spectrum at $+$14.7~d, which then falls rapidly to $T_{\rm BB} = 14,250_{-350}^{+400}~{\rm K}$ in the second spectrum ($+$17.7 d), and continues cooling to $T_{\rm BB} = 7,400_{-100}^{+100}~{\rm K}$ at $+$76.2 d.  Unfortunately, the narrowest emission components of the Balmer lines are unresolved in these early-time spectra, except for a high-resolution Copernico/Echelle spectrum at $+$19.7~d. 
However, using this spectrum, we note that both the observed  \Ha~and \Hb~profiles exhibit a narrow resolved component superimposed on a broader feature, both of which are centered at the rest wavelength. The \Heii~$\lambda$4686 emission line is still visible in this spectrum.
Detailed analysis of this Copernico/Echelle spectrum is presented in Sect. \ref{subsection:HighResolution}.

The other pre- and near-maximum spectra were mostly obtained with the 2.16\,m Xinglong telescope (XLT) (see Table \ref{tab:speclog_2019vxm} in Appendix \ref{appendix:specdata}), and do not show significant evolution. They are just marginally redder than earlier spectra, and flash-ionisation features have completely disappeared. In all cases, the low spectral resolution does not allow for a reliable measurement of the full width at half-maximum intensity (FWHM) velocity for the narrow component.
The highest-quality spectrum, in terms of both signal-to-noise ratio (S/N) and resolution, was obtained with the 3\,m Shane telescope at $+49.1$~days. The $\Ha$ profile in this spectrum is well-fitted by a two-component Gaussian function, comprising a relatively narrow emission component and a broader one. These two components were chosen purely based on their ability to accurately reproduce the observed line profile. The narrower component is unresolved, whereas the broad emission component exhibits a FWHM velocity of $2500\pm50\,\kms$ which is likely attributed to the electron-scattering wings \citep{Huang2018MNRAS.475.1261H}. 
After an observational gap of about one month, a new Copernico/AFOSC spectrum was obtained at $+76.2$~days. 
This spectrum is still quite blue and dominated by the unresolved Balmer components. However, a broader base in the \Ha~profile with a $v_{\mathrm{FWHM}}=9500\pm180~\kms$ is now more evident, and broad undulations are marginally visible in the blue region.  The broad component likely originates from the high-velocity component of the ejecta, as the outer part of the CSM gradually becomes optically thin. 
Furthermore, the emergence of metal lines in the blue region of this spectrum suggests that some of the inner parts of the environment are gradually becoming visible. These features indicate this spectrum is transitioning toward the intermediate phases.

{\bf Intermediate phases, $+$100 to $+$400 days:} The spectral monitoring of this phase starts after an observational gap of $\sim 2$ months.
In this period, we note a dramatic change in the \Ha\ and \Hb\ profiles: these are no longer dominated by the narrow component, as the broader component becomes progressively more prominent and exhibits an increasingly  asymmetric profile. 
In particular, the full line profile of \Ha~is now reproduced with two components. The broad component has a FWHM velocity range of 6100--8400 \kms\ and its centre is blueshifted by a maximum value of $\sim 1370$\,\kms\ with respect to the rest-frame position.  
Furthermore, the narrower emission is now resolved in most spectra, with a FWHM velocity range of 1200--1800\,\kms, suggesting the presence of an intermediate-width component or a broadening of the previously unresolved narrow emission. The nature of this component can be clarified by analysing a higher-resolution spectrum.
Specifically, we use a moderate-resolution GTC/R2500R spectrum at $+$166.2~d to pose tighter constraints on the wind velocity during this intermediate phase. 
This moderate-resolution spectrum shows a composite H$\alpha$ profile, suggesting three different kinematic components: a very narrow component attributed to the unshocked CSM, an intermediate-width component due to shocked material, and a broad component produced by the SN ejecta (see Sect. \ref{subsection:HighResolution}). The narrow and intermediate-width components cannot be distinguised in the low-resolution spectra.
In this spectrum, a narrow P~Cygni feature attributed to \ion{He}{I}~$\lambda5876$ is superposed on the broad P~Cygni profile of \ion{Na}{I}~D, while the \ion{He}{I}~$\lambda7065$ line is observed on the red side of the spectrum (see further details in Sect. \ref{subsection:HighResolution}).

Later spectra exhibit minor changes. Their spectral continuum is relatively red, maintaining an almost constant temperature, fluctuating between $T_{\rm BB} \approx~5650~{\rm K}$ and $T_{\rm BB} \approx~5150~{\rm K}$. Such  spectral evolution is consistent with the observed photometric evolution characterised by a sort of plateau. This behaviour is largely driven by hydrogen recombination occurring within the expanding cold dense shell (CDS). As the opacity outside the recombination front becomes negligible \citep{Arnett1989apj}, the recombination process dominates this zone and produces the broad, boxy emission lines observed.  Finally,
these spectra confirm the presence of \ion{He}{I}~$\lambda5876$ and \ion{He}{I}~$\lambda7065$ lines; in particular, \ion{He}{I}~$\lambda7065$ is clearly detected in the high-S/N spectra and exhibits minor evolution in its profile width. Additionally, the blend between \ion{He}{I}~$\lambda5876$ and \ion{Na}{I}~D becomes more prominent with time.
The blue region of the spectra exhibits a pseudocontinuum. This feature is likely caused by the line-blanketing effect within the expanding CDS, which primarily originates from metal lines such as \ion{Fe}{II}. Finally, the systematic blueshift of the \Ha~lines, the asymmetry in the hydrogen-line profiles, and the IR excess observed in the light curves after $+$210 d, reveal the signature of new dust formation. 

{\bf Late phases, $> +400$ days:}
At very late phases, the spectral continuum becomes redder and the \Caii~$\lambda$$\lambda$$\lambda$8498, 8542, 8662 near-infrared (NIR) triplet becomes visible, along with \ion{He}{I}~$\lambda5876$ and \ion{He}{I}~$\lambda7065$. 
We noticed that the profile of \ion{He}{I}~$\lambda7065$ broadens at very late times. This suggests that the increased prominence of the blended \Hei~$\lambda5876$ and \Nai\,D region is also primarily driven by the broadening of the \Hei~$\lambda5876$ component.
At very late phases, we can fit the \Ha~profiles with two Gaussians. The narrow component  remains at a roughly constant FWHM velocity of  $\sim1300$~\kms, whereas the broad component velocity gradually drops from $6950_{-240}^{+270}$ \kms~($+$294.5 d) to $3600_{-260}^{+260}$ \kms~($+$663.1~d). A broad P Cygni profile is clearly detected in the \Hb~line in the $+$518.5 d Shane/Kast spectrum. The broad and low-contrast P~Cygni absorption likely originates from the SN ejecta, exhibiting a velocity of $-7100\pm940$ \kms. Additionally, we observed that the \Hb~P~Cygni absorption feature at $+$535.2~days and $+$597.0~days exhibits double troughs. This could be explained by the existence of a high-velocity hydrogen layer, potentially caused by an asymmetric distribution of the ejecta. However, since a corresponding feature is not observed in \Ha~and the blue end of the spectra suffers from a low S/N ratio, the possibility of instrumental artifacts cannot be completely ruled out. Several lines formed in the inner ejecta region, such as \ion{O}{I}~$\lambda8446$, are also detected. This indicates that the outer layers of the ejecta have gradually become optically thin, and the photosphere receded from the CDS to the inner part. Furthermore, we observe an emission peak around 7400~\AA\ in our late-phase spectra. This feature is likely caused by [\ion{Fe}{II}] and [\ion{Ni}{II}] emission within the inner part of the ejecta. The overall spectral evolution is similar to that of other SNe IIn \citep[e.g., SN~2017hcc;][]{SN2017hcc_Moran2023A&A...669A..51M}.

In addition, as the spectral evolution progresses beyond $+$400~days, we note that the broad component of H$\alpha$ becomes narrower, a distinct asymmetry between the blue and red wings emerges, and the peak of the broad component shows a prominent blueshift.  
We suggest that this morphological change might be due to dust formation. This interpretation is corroborated by the photometric data from the same period: an excess in the MIR begins to appear after $\sim$370~days (see Sect. \ref{mir}), and the optical light curve exhibits a faster decline rate, which aligns with the obscuration and cooling effects caused by dust formation. A detailed discussion regarding the effect of dust formation on the spectroscopic observations can be found in Sect.~\ref{subsection:EvolutionH}. 
Additionally, \cite{Anderson2014MNRAS.441..671A} suggest that lines can form inside the photospheric radius derived from continuum fitting, since the opacity of the ejecta is mainly provided by electron scattering. Furthermore, the optically thick ejecta suppress emission originating from the receding regions.
However, we consider this explanation less likely, as the ejecta are gradually becoming optically thin during this epoch, which is inconsistent with the assumptions made by \cite{Anderson2014MNRAS.441..671A}.

\subsection{High- and medium-resolution spectra} 
\label{subsection:HighResolution}

\begin{figure}[htbp]
\begin{center}
\includegraphics[width=\linewidth]{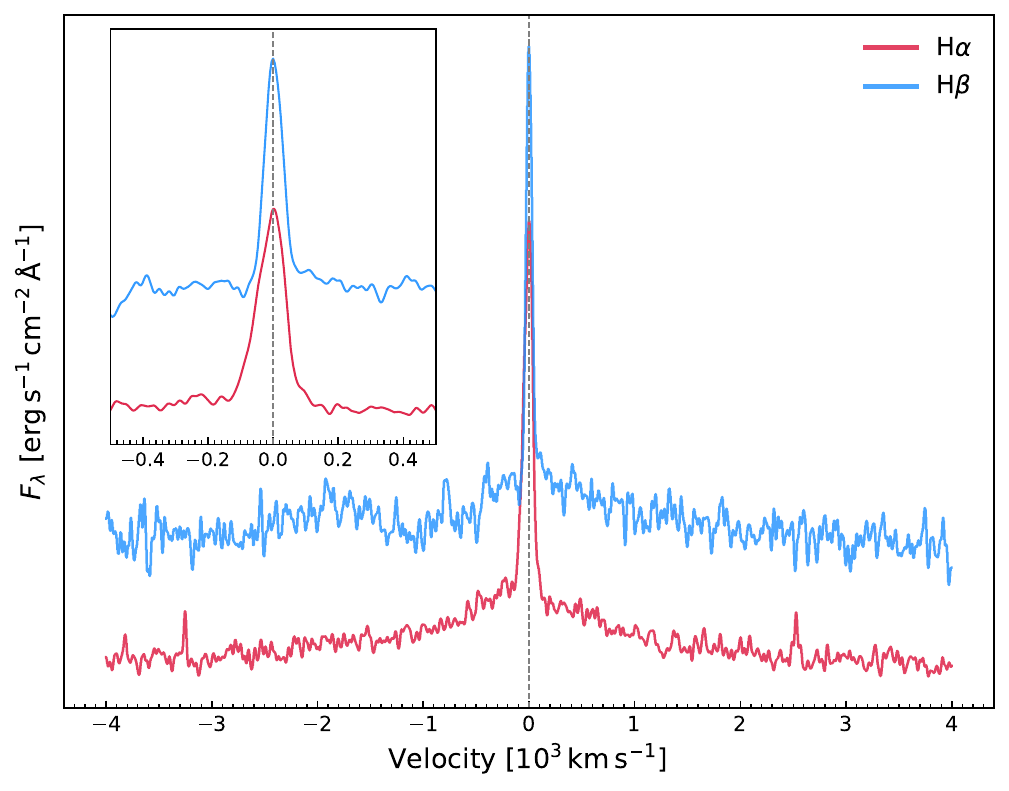}
\caption{H$\alpha$ and H$\beta$ profiles of SN\, 2019vxm (shown in blue and red colours, respectively) in velocity space, taken from the high-resolution Copernico/Echelle spectrum at $+$19.7~d. The inset in the upper-left corner provides a zoomed-in view of the P~Cygni profiles.}
\label{fig:spec:hrbalmerlines}
\end{center}
\end{figure}

\begin{figure}[htbp]
\begin{center}
\includegraphics[width=\linewidth]{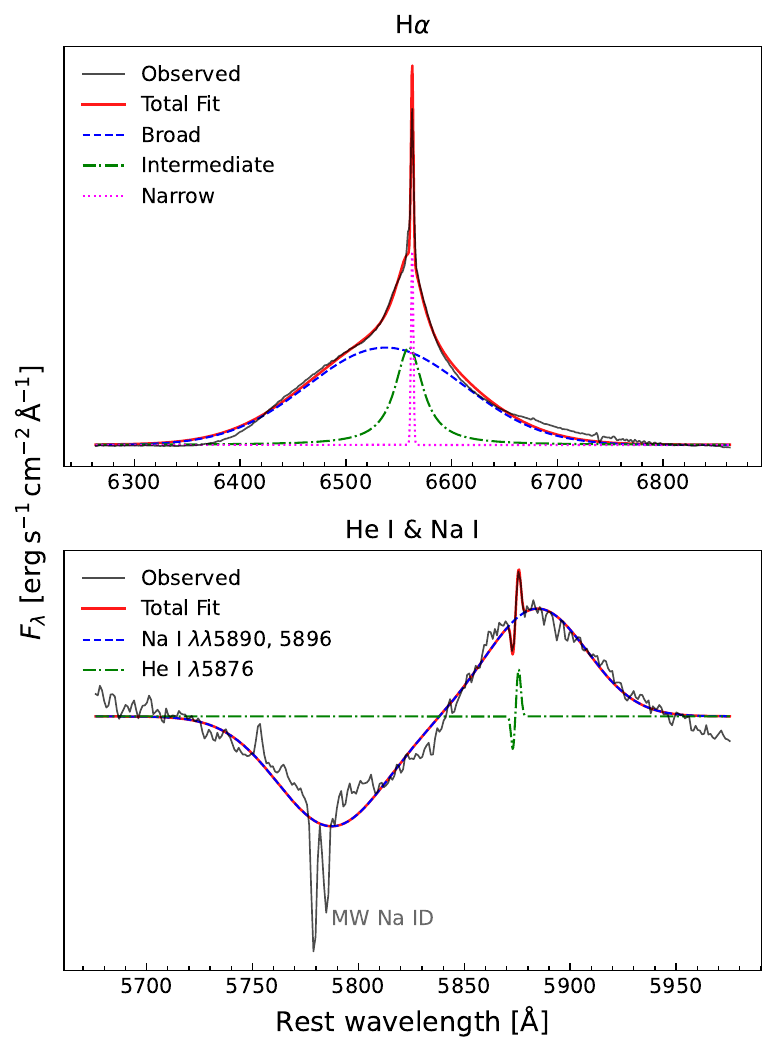}
\caption{Fits to the \Ha, \Hei, and \Nai~profiles of SN\, 2019vxm taken with the medium-resolution GTC/R2500R spectrum at $+$166.2~d. Upper panel: multicomponent fitting of H$\alpha$. Lower panel: multicomponent fitting of \ion{Na}{I}~D and \ion{He}{I}~$\lambda5876$. The \ion{Na}{I}~D lines from the interstellar medium (ISM) of the Milky Way (MW) are marked. Different components are marked with different colours.}
\label{fig:spec:mrbalmerlines}
\end{center}
\end{figure}

To constrain the velocity of the narrow component, a higher resolution spectrum is necessary.  With this motivation, we obtained a high-resolution ($R \approx 20,000$) Echelle spectrum with the Copernico 1.82\,m telescope on 2019 December 06 ($+$19.7~d; see the fifth spectrum from the top in Fig. \ref{fig:spec:spevo} and also in Fig. \ref{fig:spec:hrbalmerlines}) where we detected symmetric \Ha~and \Hb~profiles dominated by strong, narrow emission lines, on top of broader wings, most likely due to electron-scattering effects  \citep{Huang2018MNRAS.475.1261H}.  
As shown in Fig. \ref{fig:spec:hrbalmerlines}, a weak P~Cygni absorption is detected in \Hb. 
The velocity inferred from the position of the minimum of the narrow \Hb absorption is $-60\pm10$ \kms. Meanwhile, we clearly detect a narrow P~Cygni profile in the \Hei~$\lambda$5876 line (see Fig.~\ref{fig:spec:HeNa}), whose minimum is blueshifted by $-60\pm15$ \kms. Hence, this is  evidence for the presence of  slow, unshocked CSM arising from the pre-SN wind of the progenitor. We note that wind velocities of approximately 60 \kms\ lie at the bottom edge of wind velocities expected in LBVs, but also at the upper limit for RSG winds \citep[e.g., see Fig. 3 of][]{Smith2017hsn..book..403S}. 

The medium-resolution spectrum taken from GTC/R2500R at $+$166.2~days is shown in Fig.~\ref{fig:spec:mrbalmerlines}, which presents the region of the H$\alpha$ line (upper panel) and the \ion{He}{I}~$\lambda5876$ line, along with the \ion{Na}{I}~D lines (lower panel). We find that the H$\alpha$ profile can be described by two Gaussians and one Lorentzian (see detailed profile in the upper panel of Fig.~\ref{fig:spec:mrbalmerlines}). The narrowest Gaussian aligns well with the rest wavelength of  H$\alpha$ and shows a FWHM nearly coincident with the resolution limit ($\sim 3.4$~\AA; see Table \ref{tab:speclog_2019vxm}).
This can only give an upper limit to the velocity of the unshocked CSM ($<160$ \kms).
The intermediate component of H$\alpha$ is reproduced by a Lorentzian profile, which is primarily caused by natural  and collisional broadening in the dense CDS environment, showing a FWHM velocity of $\sim1450\,\mathrm{km\,s^{-1}}$. The broad component can be described by a Gaussian with a FWHM of $\sim8110\,\mathrm{km\,s^{-1}}$. This profile is suggested to form in the rapidly expanding ejecta, and the peak shows a blueshift of $\sim25.4$~\AA\ from the rest wavelength of H$\alpha$. The blueshifted peak of the high-velocity component of H$\alpha$ can be explained by the steep density profile of the ejecta and the possible formation of a CDS, which causes photons emitted from the receding side to be obscured by the optically thick material. Furthermore, since the opacity is primarily dominated by electron scattering, lines are able to form inside the photosphere \citep{Anderson2014MNRAS.441..671A}. As an alternative explanation, a strongly asymmetric ejection or a $^{56}$Ni clump in the direction of the observer can also be invoked. However, since the dust-forming event is detected around $+$208.6~days, and there were no observations in MIR bands (see Sect.~\ref{mir}), the possibility of asymmetric dust absorption in the red wings of H$\alpha$ cannot be entirely ruled out.

The profile of the blended \ion{He}{I}~$\lambda5876$ and \ion{Na}{I}~D lines can be described by two P~Cygni profiles (see details in the lower panel of Fig.~\ref{fig:spec:mrbalmerlines}). We fit the profile assuming that the continuum is constant within a narrow region. The narrow double absorption trough at $\sim5780$~\AA\ can be explained as the Galactic \ion{Na}{I}~D absorption, which shows a similar blueshift when corrected for the redshift of the SN. We suggest that the broad, low-contrast P~Cygni profile might originate from the \ion{Na}{I}~D lines in the SN ejecta, exhibiting a velocity of $\sim-5370\,\mathrm{km\,s^{-1}}$. We note that the velocity of this profile is lower than that of the broad H$\alpha$. This difference can be explained if the velocity of H$\alpha$ represents the outermost ejecta, whereas metal lines such as \ion{Na}{I} represent the bulk velocity of the ejecta. The narrow P~Cygni profile represents the \ion{He}{I}~$\lambda5876$ line. The velocity measured from the absorption component is $\sim-160\,\mathrm{km\,s^{-1}}$. We suggest that this profile forms in the outer unshocked CSM, since the velocity is comparable to the velocities measured from the high-resolution H$\beta$ and \Hei~$\lambda$5876 P~Cygni profiles in the aforementioned Echelle spectrum.

\subsection{Evolution of the H lines} 
\label{subsection:EvolutionH}

\begin{figure}[htbp]
\begin{center}
\includegraphics[width=\linewidth]{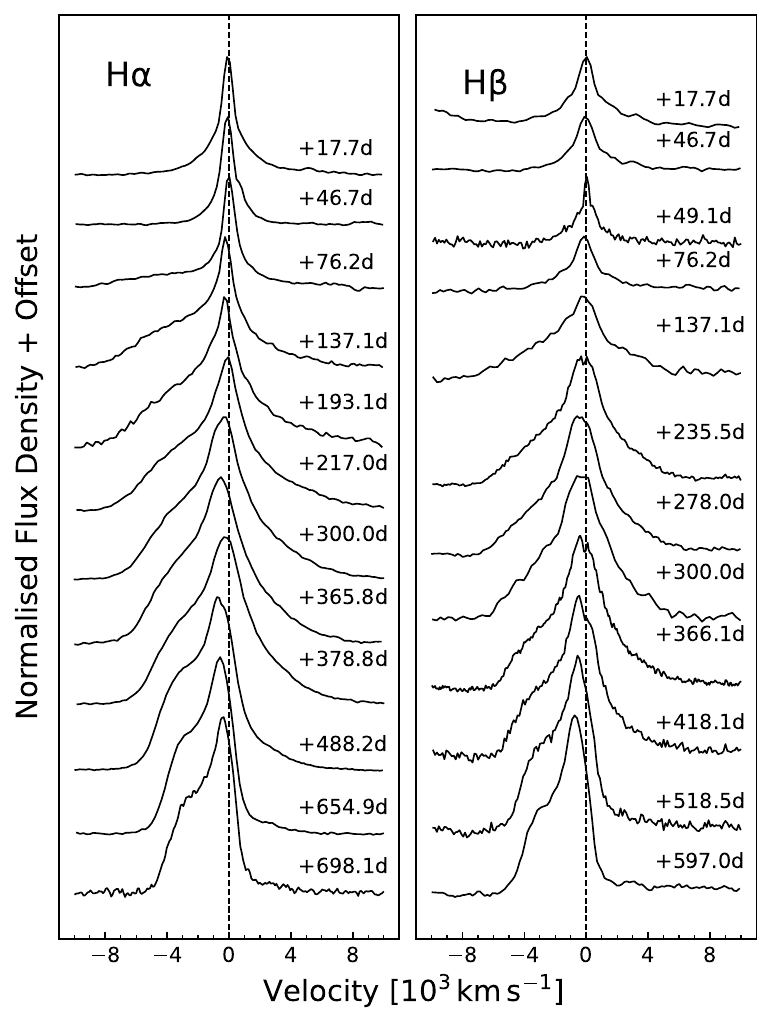}
\caption{Evolution of the \Ha and \Hb lines of SN~2019vxm. The selected spectra exhibit a relatively high S/N and are representative of the various evolutionary phases.}
\label{fig:spec:vphase}
\end{center}
\end{figure}

Figure~\ref{fig:spec:vphase} presents the temporal evolution of the $\ha$ and $\hb$ line profiles in the spectra of SN 2019vxm obtained from $+$17.7 d to $+$597.0 d relative to the explosion epoch. 
The line profiles are displayed in velocity space after the normalisation to the continuum and a vertical offset has been applied for clarity. The dashed vertical line marks zero velocity, corresponding to the rest frame. 

At early epochs ($\lesssim100$ d), the symmetric \Ha~and \Hb~profiles are well reproduced by a single Lorentzian function. The \Ha~and \Hb~profiles during this phase are symmetric, and no significant systematic blueshift is observed. The narrow component is resolved only in the  Copernico/Echelle spectrum at $+$19.7~d (see Sect. \ref{subsection:HighResolution}). We note that H emission lines undergoing electron scattering in a hot medium result in symmetric profiles, as the velocity broadening is driven by the microscopic, thermal velocities of the electrons \citep{Munch1948ApJ...108..116M}, as frequently observed in luminous SNe IIn \citep[e.g.,][]{Smith2012AJ....143...17S,SN2015da_Tartaglia2020A&A...635A..39T,SN2017hcc_Moran2023A&A...669A..51M}. 

Between +100 d and +400 d, the Balmer lines undergo substantial morphological evolution. The intermediate-width component of $\ha$ becomes more prominent and broader, and the line peak gradually shifts toward the blue. Simultaneously, the red wing is progressively suppressed, producing an increasingly asymmetric, blue-skewed profile. $\hb$ follows the same qualitative trend. 
The persistence of a strong intermediate component over several hundred days indicates sustained and powerful ejecta-CSM interaction and likely originates from the shocked material, by analogy with other long-lived Type IIn events such as SN~2010jl \citep{Fransson2014ApJ...797..118F}. The systematic blueshift and attenuation of the red wing are usually interpreted as effects of the increasing optical depth due to newly formed dust within the dense post-shock region, which preferentially obscures the receding emission \citep{Smith2008ApJ...680..568S}, or intrinsic asymmetry in the CSM distribution \citep{Fransson2014ApJ...797..118F, Andrews2016MNRAS.457.3241A}.

\begin{figure}[htbp]
\begin{center}
\includegraphics[width=\linewidth]{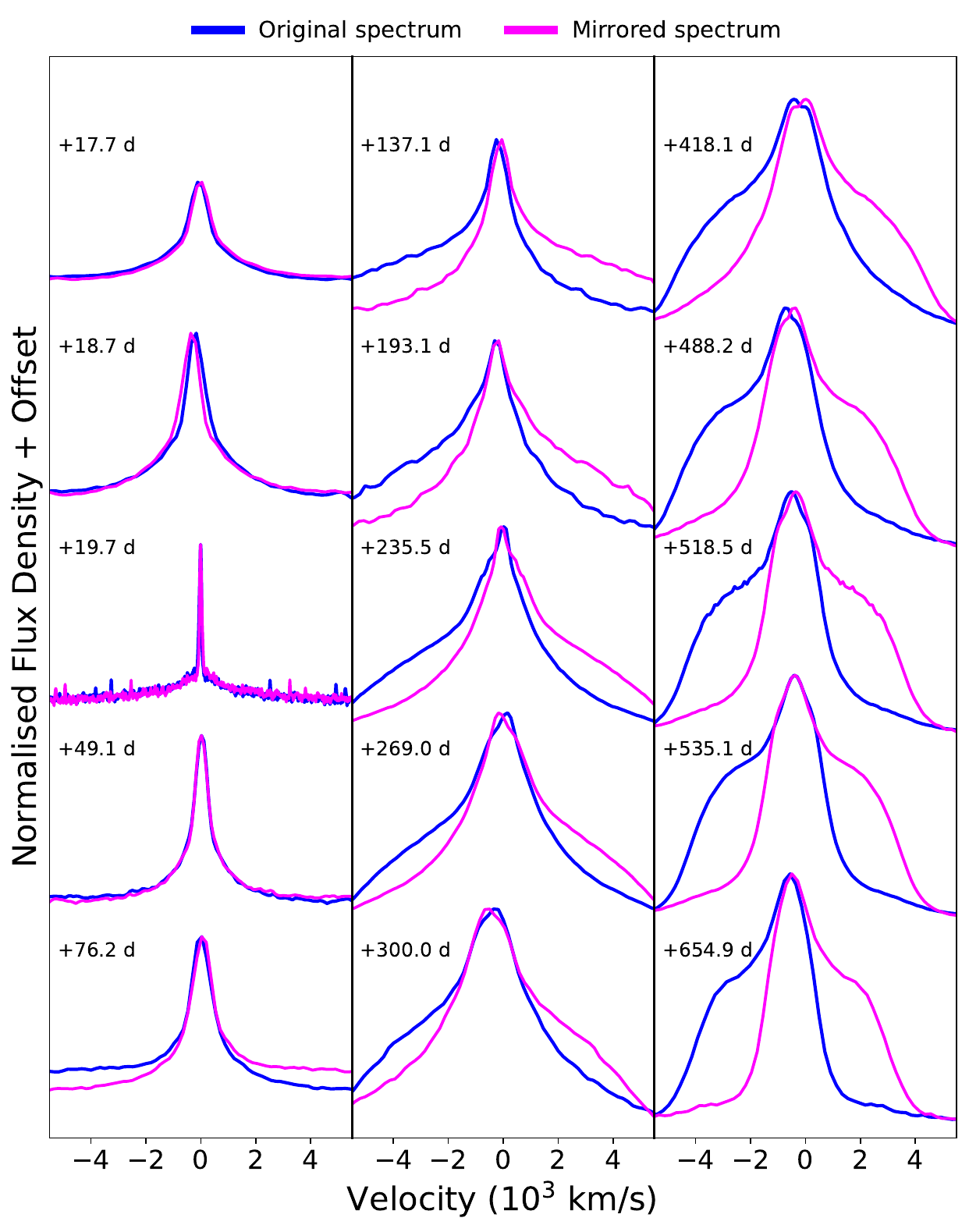}
\caption{\Ha~profiles of SN~2019vxm at selected epochs, redshifted to the line rest wavelength (blue) and mirrored with respect to the computed centroids (magenta). }
\label{fig:spec:mirror}
\end{center}
\end{figure}

At epochs later than +400 d, both $\ha$ and $\hb$ remain clearly detectable, confirming the long-lived nature of the interaction. Although the overall widths decrease slightly compared to the peak interaction phase, the profiles retain their asymmetric, blue-dominated morphology, with a continued suppression of the red wing. Notably, dust is expected to form within the inner, metal-rich ejecta in standard Type II SNe on a one-year timescale. However, dust formation can be more efficient and occur earlier in interacting SNe. In these cases \citep[e.g., SN\,2006jc, SN\,2016iog;][]{Mattila2008MNRAS.389..141M,Peng2026A&A...705A.104P}, dust forms within the compressed CDS at the ejecta-CSM interacting interface even at earlier phases, where the high density in the radiative post-shock layers facilitate the rapid post-shock gas cooling \citep[e.g.,][]{Smith2012AJ....143...17S,Sarangi2022ApJ...933...89S,Dessart2025A&A...698A.293D}. 
The concurrently observed signatures of systematic blueshift and wavelength-dependent asymmetries of the \Ha~and \Hb~emission lines (after $\sim$100 days), and IR brightening accompanied by optical fading (after $\sim$400 days), suggest new dust condensation in SN\,2019vxm. Similar features have also been observed in SN\,2010jl, where they were attributed to the formation of new dust \citep[see, e.g.,][]{Smith2012AJ....143...17S,Gall2014Natur.511..326G}. However, \citet{Fransson2014ApJ...797..118F} quantitatively interpret the blueshifted peak and symmetric \Ha~profiles in SN\,2010jl as resulting from a scattering medium moving coherently toward the observer. Following the methodology of \citet{Fransson2014ApJ...797..118F} 
 \citep[see also][]{SN2015da_Tartaglia2020A&A...635A..39T}, we mirrored the red wings of the \Ha~profiles with respect to the computed centroid at each epoch (Fig. \ref{fig:spec:mirror}). 
 The symmetric line profile with respect to the shifted centroid suggests that the observed blueshift of the \Ha~in SN\,2019vxm (see Fig. \ref{fig:spec:vphase}) is also a macroscopic velocity effect.   

\subsection{Comparison with Type IIn SN spectra}
\label{subsection:SpecComp}

\begin{figure*}[htbp]
\begin{center}
\includegraphics[width=0.9\linewidth]{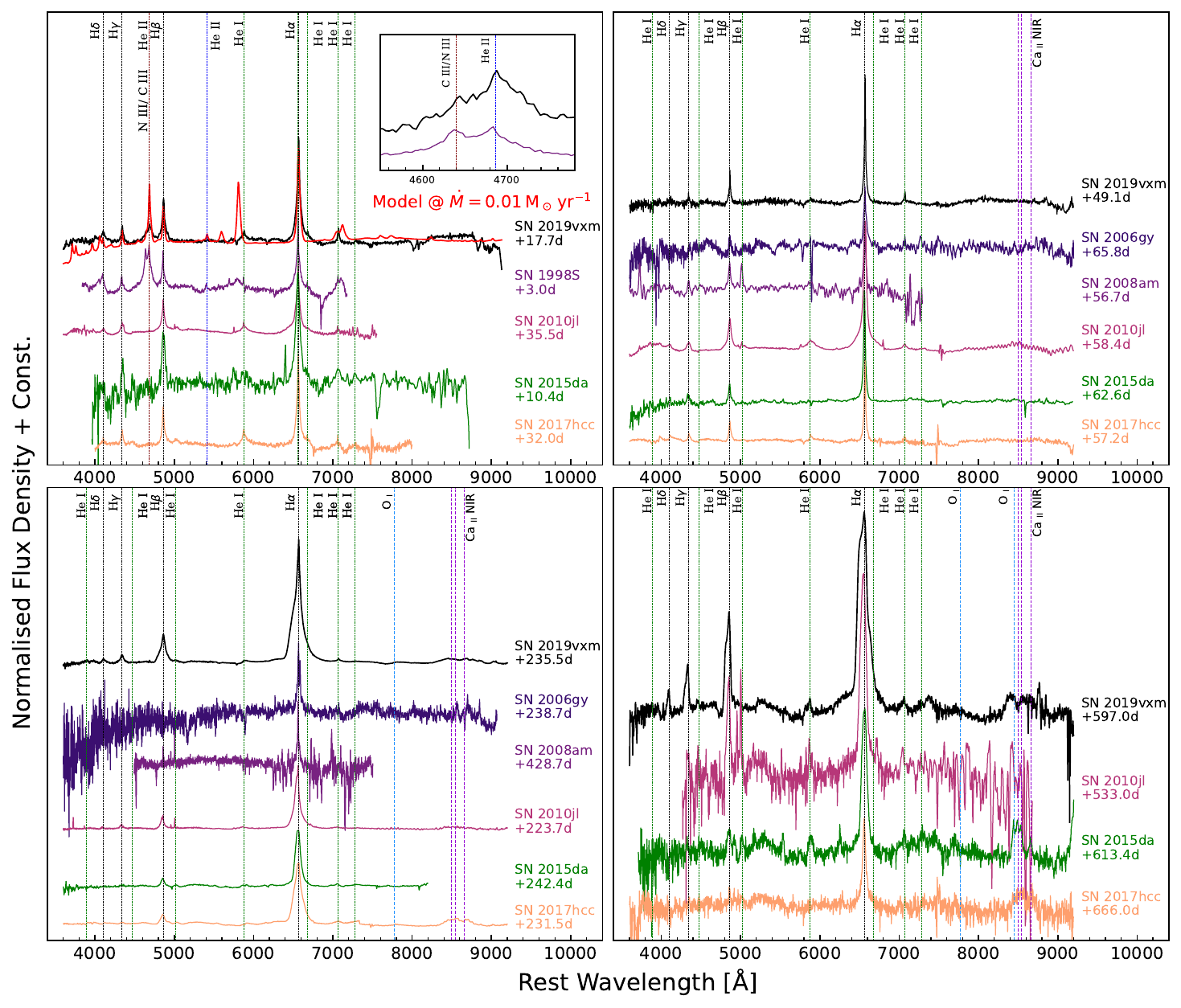}
\caption{Spectra of SN~2019vxm compared to similar Type IIn SNe and SLSNe-II at early ($0$--$40$~days), post-peak ($40$--$400$~days), and late phases (up to $\sim600$~days). Prominent spectral lines are indicated by colours. Phases are given relative to the estimated explosion epochs. All spectra are redshift-corrected, dereddened, normalised, and vertically offset for clarity. In the top-left panel, a model from \cite{Boian2020MNRAS.496.1325B} is overplotted (red line) on an early spectrum with prominent flash-ionisation features ($+$17.7~days after explosion). The inset shows a close-up view of the 4400--5000\,\AA\ region, highlighting the prominent flash-ionisation features. }
\label{fig:SpecComp}
\end{center}
\end{figure*}

We now compare the spectra of SN\,2019vxm with those of well-observed, slowly evolving SNe IIn in four distinct phases. The upper-left panel of Fig.~\ref{fig:SpecComp} presents a comparison of SN\,2019vxm with similar SNe~IIn and SLSNe-II at early phases. We selected the earliest spectrum of SN~2019vxm with adequate resolution, taken $+$17.7~days after the estimated explosion epoch. At these early phases, the spectra are dominated by strong and narrow Balmer lines.
High-ionisation lines, including \ion{N}{III} and \ion{C}{III}, are also present with moderate intensities in SN\,2019vxm, suggesting that the temperature is still high at $+$17.7~days. In the comparison sample, only the spectrum of SN~1998S taken $+$3.0~days after explosion exhibits similar \ion{N}{III} and \ion{C}{III} flash-ionisation features. In contrast, such features are not observed in the first spectra of SN\,2010jl and SN\,2017hcc, while only faint \ion{C}{III} and \ion{N}{III} lines are tentatively identified in the spectrum of SN~2015da. 
The narrow emission lines of \ion{He}{I} $\lambda5876$ and $\lambda7065$ are relatively faint compared to the Balmer lines, by analogy with other comparison objects. 

To constrain the mass-loss activity of the progenitor, we compared our early-phase spectrum with early interacting SN models from \cite{Boian2020MNRAS.496.1325B}. The theoretical models are convolved with a Gaussian kernel to match the resolution of our spectrum ($\sim 14$~\AA; see Table~\ref{tab:speclog_2019vxm}). The luminosity of the model was selected to match the observed bolometric luminosity at $+17.7$~days after explosion, corresponding to $L \approx 10^{10} {\rm L}_\odot$. The terminal velocity of the wind is assumed to be $v_{\infty} = 150\,\mathrm{km\,s^{-1}}$, which is consistent with the velocity measured from the P~Cygni absorption profiles of the \Hb~and \Hei~$\lambda$5876 lines in the high-resolution spectrum. The model grid provided by \cite{Boian2020MNRAS.496.1325B} terminates at $+3.7$~days after explosion, reaching an inner radius of $R_{\mathrm{in}} \approx 3.2 \times 10^{14}\,\mathrm{cm}$. 
Because this theoretical phase is earlier than our first spectroscopic observation, the intensity of the prominent Balmer lines in the model is expected to be overestimated. However, because the flash-ionisation lines remain clearly visible and the derived black-body temperature remains high ($14,260_{-330}^{+370}$~K), the physical conditions required to produce early-phase emission features are still present. Therefore, the temporal mismatch between the observation and the model does not fundamentally bias the results, and the order-of-magnitude estimation of the mass-loss rate remains physically justified.

Assuming solar abundances and accounting for the effects of the phase mismatch, we estimate the mass-loss rate of the progenitor of $\dot{M} \gtrsim 10^{-2}\,\mathrm{M_{\odot}\,yr^{-1}}$. We consider this result as a lower limit since the actual spectroscopic epoch is later than the model assumption. This mass-loss rate  is consistent with those derived for SN\,2010jl ($\sim 0.1$ \msun\,yr$^{-1}$) and SN\,2015da ($\sim 0.6$ \msun\,yr$^{-1}$), as reported by \citet[][]{Fransson2014ApJ...797..118F} and  \citet[][]{SN2015da_Tartaglia2020A&A...635A..39T}, respectively. The best-matching model (shown in Fig. \ref{fig:SpecComp} with a solid red line) exhibits H$\alpha$ and H$\beta$~line intensities comparable to those in the observed spectrum. Discrepancies primarily arise among the high-ionisation and \ion{He}{I} lines. These inconsistencies are likely caused by the phase mismatch, as the observed line-forming region is expected to be cooler than the region in the earlier-phase model. Furthermore, the specific helium abundance or the density difference of the line-forming regions might affect the intensity of \ion{He}{I}~$\lambda5876$, but this parameter is not a critical factor for estimating the overall mass-loss rate. These results indicate a high mass-loss rate, suggesting that the progenitor of SN~2019vxm was likely an LBV or a star in a binary system shedding its envelope during the binary Roche-lobe overflow (RLOF) phase \citep[see, e.g.,][]{Smith2014ARA&A..52..487S,Smith2017hsn..book..403S}.

The middle, late, and very late-phase spectral comparisons are presented in the remaining three panels of Fig.~\ref{fig:SpecComp}.
In the post-peak phase (40--100 days; upper-right panel), the spectra of SN\,2019vxm remain dominated by strong Balmer emission lines, indicating persistent ejecta-CSM interaction \citep{Schlegel1990MNRAS.244..269S, Kiewe2012ApJ...744...10K}. Compared with the early phase, the continuum progressively weakens while emission features become more prominent. The H$\alpha$ profile develops a broader base and shows mild asymmetry, resembling those observed in luminous SNe IIn such as SN~2010jl and SN~2015da at comparable phases \citep{Fransson2014ApJ...797..118F, SN2015da_Tartaglia2020A&A...635A..39T}.
The strength of \Ha~relative to \Hb~gradually increases; the observed Balmer decrement values are about 1.69 ($+$17.7 d) and 2.58 ($+$49.1 d), likely due to increasing optical-depth effects and possible dust attenuation in the dense post-shock region \citep{Smith2008ApJ...680..568S, Gall2011A&ARv..19...43G}. 
Helium lines, including \ion{He}{I} $\lambda5876$ and $\lambda7065$, remain weak compared with the Balmer lines, typical of the hydrogen-rich circumstellar environments characterising normal SNe~IIn \citep{Kiewe2012ApJ...744...10K, Smith2017hsn..book..403S}. The overall spectral morphology closely resembles other luminous and long-lived SNe~IIn, particularly SN~2010jl, suggesting sustained interaction within dense CSM.

During the late phase (200--400 days; lower-left panel), the continuum emission further fades and the spectra become increasingly emission-line dominated. The Balmer lines remain the most prominent features; in particular, H$\alpha$ exhibits pronounced intermediate and broad components similar to those seen in SNe~2005ip,  2010jl, and 2015da \citep{SN2005ip_Stritzinger2012ApJ...756..173S, Fransson2014ApJ...797..118F}. The line profiles exhibit increasing asymmetry, with the red wings being gradually suppressed relative to the blue wings. Such asymmetric profiles are usually interpreted as signatures of new  dust condensation in the post-shock region or anisotropic CSM geometry \citep{Smith2008ApJ...680..568S, Andrews2016MNRAS.457.3241A}. 
The \Caii~NIR triplet is observed in the spectrum of SN\,2019vxm, as well as in SN\,2006gy and SN\,2017hcc.  

At very late epochs ($\gtrsim$500 days; lower-right panel), SN\,2019vxm continues to exhibit strong Balmer emission, confirming the long-lived nature of the interaction. The continuum becomes very faint and the spectra are dominated by emission features. The H$\alpha$ line maintains a strong, intermediate-width component and a clearly blueshifted asymmetric profile. 
In contrast, SN~2015da and SN~2017hcc exhibit relatively weaker emission features at comparable epochs, suggesting less-sustained CSM interaction \citep{SN2015da_Tartaglia2020A&A...635A..39T,Chandra2022MNRAS.517.4151C}. The Ca~II NIR triplet becomes more visible at these epochs, indicating cooling ejecta and increasing importance of recombination emission. Overall, the spectral comparison indicates that SN\,2019vxm belongs to the class of luminous, strongly interacting SNe~IIn with long-lasting ejecta-CSM interaction, and evidence for a dense circumstellar environment shaped by intense pre-explosion mass loss and late-time dust condensation \citep{Smith2017hsn..book..403S, Fransson2014ApJ...797..118F}. 

\section{Discussion and conclusions}
\label{section:discussion}

In this paper, we have presented the photometric and spectroscopic properties of SN\,2019vxm, a luminous and long-lived Type IIn SN exhibiting slow spectroscopic and  photometric evolution, together with an evident MIR excess at late times. SN\,2019vxm shares obvious similarity with other luminous Type~IIn SNe, such as SN 2010jl and SN 2017hcc, especially in terms of its overall brightness and spectral evolution. 

The light curve of SN\,2019vxm exhibits a relatively slow rise to the peak brightness ($35.0\pm1.8$ days in the \textit{Swift} $V$ band), which is within the slow-rise SN~IIn population of \citet{Nyholm2020A&A...637A..73N}. This behaviour is consistent with interaction with massive and dense CSM due to the long diffusion timescale for photons travelling through optically thick CSM.
The SN is more luminous ($M_{\mathrm{V}} = -20.01 \pm 0.13 $~mag) than most SNe IIn \citep[e.g., the average peak brightness of $M_\mathrm{r}^{\mathrm{peak}} \approx -19.18 \pm 1.32~\mathrm{mag}$ in the sample of][]{Nyholm2020A&A...637A..73N}. The estimated ``optical'' peak luminosity of SN\,2019vxm is \(L \approx 1.59 \times 10^{43}\,\mathrm{erg\,s^{-1}}\). The rise is followed by a monotonic decline with slow evolution (e.g., $\overline{\gamma(r)}\approx 0.55$ mag/100 d) until the SN disappeared $\sim 660$ days after  maximum light. The light-curve evolution of SN\,2019vxm is consistent with the findings of \citet{Nyholm2020A&A...637A..73N}: higher-luminosity SNe IIn are generally found to be longer-lasting, while slowly rising SNe IIn generally decline more slowly in luminosity. 
In addition, we examined the $B-V$ and $r-i$ colour evolution of SN~2019vxm in comparison with several SNe~IIn. These objects show significant diversity in their colour behaviour over time. Such variation likely arises from differences in the physical processes that influence the thermal and radiative properties of the emitting regions.

To model the light curve of SN~2019vxm, we found a power-law decline ($L_{\mathrm{bol}}\propto t^{-0.49}$) until $\sim100$~days post-explosion, which is able to describe the early-time light curve, similar to SN~2010jl. The observational data deviate from the model after $\sim100$~days from explosion, a phenomenon potentially caused by dust formation as the ejecta cool, similar to SN~2010jl \citep{Moriya2013MNRAS.435.1520M}. Additionally, hydrogen recombination takes place during this phase as the temperature decreases to $\sim6000$~K, leading to a relatively flat evolution of the light curve, as the CDS forms and the broad H$\alpha$ emission line emerges.
After $\sim300$~days, the decline rate of the light curve becomes significantly steeper, caused by intense dust formation. A prominent MIR excess is also observed at this phase. Grey-body fitting to the SED in the MIR part suggests a lower mass limit of $\sim10^{-4}\,\mathrm{M_\odot}$ for graphite dust or $\sim10^{-3}\,\mathrm{M_\odot}$ for silicate dust, with temperatures of $\sim 1000$~K. Furthermore, because the CSM interaction weakens in the late phase as the expanding CDS becomes optically thin, radioactive heating likely dominates the tail of the light curve, resulting in a faster decline that tracks radioactive decay. From the model fitting, the estimated mass-loss rates are $\dot M\approx 5\times 10^{-2}\, \mathrm{M_\odot\,yr^{-1}}$ and $\dot M\approx 1\times 10^{-1}\, \mathrm{M_\odot\,yr^{-1}}$ with different parameter assumptions, which are considered to be high values, supporting that there was an intense mass-loss activity decades before the explosion. However, we note that
the lower bound of mass-loss rate is 
approximately one order of magnitude lower than the values inferred for SN\,2010jl ($\sim 0.1$ \msun\,yr$^{-1}$) and SN\,2015da ($\sim 0.6$ \msun\,yr$^{-1}$), while the upper bound is roughly consistent with these objects.

The early-time spectra of SN~2019vxm from $+$14.7\,d to $+$76.2\,d show very blue continua ($T \approx 7400$--20,100\,K) and prominent, narrow Balmer emission lines. The earliest two spectra are characterised by the prominent flash-ionisation features of \Ciii, \Niii, and \Heii, from which we estimate a lower limit for the mass-loss rate of the progenitor of $\dot{M} \gtrsim 10^{-2}\,\mathrm{M_{\odot}\,yr^{-1}}$. These features were also observed in the early spectra of SN~1998S \citep{Fassia2000MNRAS.318.1093F, Anupama2001A&A...367..506A}. However, the emergence of these features in SN~2019vxm occurs at a much later epoch compared to SN~1998S. Given that the diffusion timescale in the CSM interaction is expressed as $t_{\mathrm{CSM}}\propto \kappa M_{\rm CSM}/cR_{\rm CSM}$, this long-lasting flash ionisation can be explained by massive or dense CSM. Additionally, the presence of a highly extended CSM environment cannot be ruled out. This estimate is consistent with the value inferred from the light-curve modelling and is also similar to that of SN\,2010jl (see details in Sect. \ref{subsection:LC_modeling}).
The short duration ($\lesssim5$ days) of these flash-ionisation features indicates that the CSM is spatially confined and originates from enhanced mass loss occurring shortly (months to a few years) before the explosion. 

The Echelle spectrum may show a narrow P Cygni absorption imposed on \Hb~(clearly detected in \Hei~$\lambda$5876 line), indicative of absorption by slowly moving optically thick CSM along the line of sight with a velocity of $60\pm10$ \kms. This velocity falls marginally below the lower end of values typically inferred for LBVs \citep[$10^{2}-10^{3}$ \kms; ][]{Smith2011MNRAS.415..773S,Smith2017hsn..book..403S}, while it is slightly higher than velocities observed for red supergiants \citep[RSG, 10--50 \kms; ][]{Jura1990ApJS...73..769J,Smith2017hsn..book..403S}. 
However, the mass-loss rates inferred from both the light curve and early spectral modelling are approximately three orders of magnitude higher than typical values of RSG winds. This makes an RSG progenitor unlikely, but the rates are consistent with an LBV or binary RLOF scenario. 
Nevertheless, pre-SN superwinds might be occurring in RSGs  \citep[e.g., see the case of the Galactic RSG VY Canis Majoris;][]{Smith2009AJ....137.3558S}. Taken together, these lines of evidence (similar to that of SNe\,2010jl and 2015da) favour a very massive star as a progenitor for SN\,2019vxm, most likely an LBV. However, the possibility that the progenitor was an extreme RSG star with a superwind, or a post-RSG/LBV transitional star, cannot be entirely ruled out.
 
After $+41.2$~days, a broad emission component emerges in the Balmer lines, marking the formation of the CDS. At epochs later than $+200$~days post-explosion, the broad component narrows, with the FWHM velocity exhibiting a linear decline. This gradual narrowing of the profile suggests that the outer interaction region becomes increasingly optically thin, accompanied by the recession of the photosphere. Moreover, the peak of the broad component consistently exhibits a blueshift throughout the evolution. In the early phase, similar to other Type II SNe, this blueshift can be explained by line formation occurring within the photosphere radius by fitting the continuum; the probability of photon destruction is reduced in a medium where the opacity is dominated by electron scattering \citep{Anderson2014MNRAS.441..671A}. In addition, increasing optical depth in the receding direction will also produce an asymmetric profile, and suppress the red-wing emission. 
At late phases, this asymmetric profile is likely driven by dust formation, which attenuates the red wing of the emission lines \citep{Smith2008ApJ...680..568S, Peng2026A&A...705A.104P}. An asymmetric CSM distribution like in SN~2017hcc can also exhibit such features, as mentioned by \cite{Fransson2014ApJ...797..118F} and \citet{Andrews2016MNRAS.457.3241A}.

\citet{Lane2026ApJ..1003...19L} identified the host of SN \, 2019vxm as a dwarf galaxy with a total stellar mass of $\log_{10}(M_g/{\rm M_{\odot}}) = 7.72^{+0.37}_{-0.55}$, consistent with the results of the environmental trends reported by \citet[][]{Schulze2018MNRAS.473.1258S} --- SLSNe~IIn are typically hosted in dwarf galaxies ($M_{\odot}\lesssim10^{10}\, \msun$) with only a few exceptions, such as the case of SN \,2006gy. In addition, \citet{Lane2026ApJ..1003...19L} estimated the stellar metallicity of $\log_{10}(Z_*/Z_{\odot}) = -0.13^{+0.22}_{-0.26}$ for the SN host galaxy, consistent with the findings of \citet{Moriya2023A&A...677A..20M} that higher mass-loss rates and wind velocities tend to occur in more metal-poor environments. 
This is in conflict with the expectation from radiation-driven winds, indicating that the CSM is not a result of these winds. On the other hand, \citet{Lane2026ApJ..1003...19L} identified a spatial and temporal coincidence between SN 2019vxm and an X-ray burst (GRB191117A) at shock breakout, finding a 3.3$\sigma$ association confidence. The short duration (on the order of seconds) of the X-ray transient indicates that the interaction is most likely occurring with dense, asymmetric CSM.

Our MIR photometry of SN~2019vxm spans 7--1690 days from the explosion, and shows an IR excess starting from 210 days. 
At the first epoch, at a phase of $\sim$7 d, a single black body adequately fits the SED, but a second, cooler black body is required to reproduce the SEDs after $\sim$210 days (see Fig. \ref{fig:dust} in Sect. \ref{mir}). This reveals that there is a general trend of increasing dust mass and decreasing dust temperature with the phase. 
It thereby seems likely that the IR emission in SN~2019vxm originated from the newly formed dust similar to a number of other overluminous Type IIn SNe, such as SN 2010jl and SN 2017hcc. 
 
In summary, we propose that SN\,2019vxm originated from the explosion of an LBV embedded in  structured, asymmetric CSM. Nevertheless, the possibility of an extreme RSG progenitor with a superwind, or a post-RSG/LBV transitional star, cannot be entirely excluded. Although SN\,2019vxm was studied through multiwavelength observations, we expect further observations  (particularly in the IR domain) can improve our understanding of its properties. For instance, multiband IR photometry and spectra covering multiple epochs at very late phases, combined with detailed modelling of spectral line profiles, would refine estimates of the dust properties. New spectroscopic instrumentation such as SOXS \citep[Son Of X-Shooter; ][]{Radhakrishnan2024arXiv240717288R}, installed on the NTT at the La Silla Observatory in Chile, will be crucial for expanding the spectral monitoring of SN\,2019vxm-like events toward the NIR region. Even more, existing IR space facilities such as the {\it James Webb Space Telescope} can further extend their monitoring to longer wavelengths and to very late phases of their evolution.


\bibliographystyle{aa}
\bibliography{RefIIn} 

@ARTICLE{1999ApJ...510..379M,
       author = {{Matzner}, Christopher D. and {McKee}, Christopher F.},
        title = "{The Expulsion of Stellar Envelopes in Core-Collapse Supernovae}",
      journal = {\apj},
         year = 1999,
        month = jan,
       volume = {510},
       number = {1},
        pages = {379-403},
          doi = {10.1086/306571},
archivePrefix = {arXiv},
       eprint = {astro-ph/9807046},
 primaryClass = {astro-ph},
       adsurl = {https://ui.adsabs.harvard.edu/abs/1999ApJ...510..379M}
}

@ARTICLE{Miller1993,
       author = {{Miller}, J.~S. and {Stone}, R.~P.~S.},
        title = "{}",
      journal = {Lick Obs. Tech. Rep. 66},
         year = 1993,
       volume = {},
        pages = {}
}

@ARTICLE{Filippenko1982PASP...94..715F,
       author = {{Filippenko}, A.~V.},
        title = "{The importance of atmospheric differential refraction in spectrophotometry.}",
      journal = {\pasp},
         year = 1982,
        month = aug,
       volume = {94},
        pages = {715-721},
          doi = {10.1086/131052},
       adsurl = {https://ui.adsabs.harvard.edu/abs/1982PASP...94..715F}
}

@ARTICLE{Hildebrand1983QJRAS..24..267H,
       author = {{Hildebrand}, R.~H.},
        title = "{The determination of cloud masses and dust characteristics from submillimetre thermal emission.}",
      journal = {\qjras},
         year = 1983,
        month = sep,
       volume = {24},
        pages = {267-282},
       adsurl = {https://ui.adsabs.harvard.edu/abs/1983QJRAS..24..267H}
}

@ARTICLE{Filippenko1997ARA&A..35..309F,
   author = {{Filippenko}, A.~V.},
    title = "{Optical Spectra of Supernovae}",
  journal = {\araa},
     year = 1997,
   volume = 35,
    pages = {309-355},
      doi = {10.1146/annurev.astro.35.1.309},
   adsurl = {http://adsabs.harvard.edu/abs/1997ARA%26A..35..309F}
}

@ARTICLE{Draine1984ApJ...285...89D,
       author = {{Draine}, B.~T. and {Lee}, H.~M.},
        title = "{Optical Properties of Interstellar Graphite and Silicate Grains}",
      journal = {\apj},
         year = 1984,
        month = oct,
       volume = {285},
        pages = {89},
          doi = {10.1086/162480},
       adsurl = {https://ui.adsabs.harvard.edu/abs/1984ApJ...285...89D}
}

@ARTICLE{Laor1993ApJ...402..441L,
       author = {{Laor}, Ari and {Draine}, Bruce T.},
        title = "{Spectroscopic Constraints on the Properties of Dust in Active Galactic Nuclei}",
      journal = {\apj},
         year = 1993,
        month = jan,
       volume = {402},
        pages = {441},
          doi = {10.1086/172149},
       adsurl = {https://ui.adsabs.harvard.edu/abs/1993ApJ...402..441L}
}

@ARTICLE{Colangeli1995A&AS..113..561C,
       author = {{Colangeli}, L. and {Mennella}, V. and {Palumbo}, P. and {Rotundi}, A. and {Bussoletti}, E.},
        title = "{Mass extinction coefficients of various submicron amorphous carbon grains: Tabulated values from 40 NM to 2 mm.}",
      journal = {\aaps},
         year = 1995,
        month = nov,
       volume = {113},
        pages = {561},
       adsurl = {https://ui.adsabs.harvard.edu/abs/1995A&AS..113..561C}
}

@ARTICLE{Schlegel1990MNRAS.244..269S,
   author = {{Schlegel}, E.~M.},
    title = "{A new subclass of Type II supernovae?}",
  journal = {\mnras},
     year = 1990,
    month = may,
   volume = 244,
    pages = {269-271},
   adsurl = {http://adsabs.harvard.edu/abs/1990MNRAS.244..269S}
}

@ARTICLE{Chevalier1994ApJ...420..268C,
       author = {{Chevalier}, Roger A. and {Fransson}, Claes},
        title = "{Emission from Circumstellar Interaction in Normal Type II Supernovae}",
      journal = {\apj},
         year = 1994,
        month = jan,
       volume = {420},
        pages = {268},
          doi = {10.1086/173557},
       adsurl = {https://ui.adsabs.harvard.edu/abs/1994ApJ...420..268C}
}

@ARTICLE{Chugai1997Ap&SS.252..225C,
       author = {{Chugai}, N.~N.},
        title = "{Supernovae in dense winds}",
      journal = {\apss},
         year = 1997,
        month = mar,
       volume = {252},
       number = {1-2},
        pages = {225-236},
          doi = {10.1023/A:1000847125928},
       adsurl = {https://ui.adsabs.harvard.edu/abs/1997Ap&SS.252..225C}
}

@ARTICLE{Pastorello2002MNRAS.333...27P,
       author = {{Pastorello}, A. and {Turatto}, M. and {Benetti}, S. and {Cappellaro}, E. and {Danziger}, I.~J. and {Mazzali}, P.~A. and {Patat}, F. and {Filippenko}, A.~V. and {Schlegel}, D.~J. and {Matheson}, T.},
        title = "{The type IIn supernova 1995G: interaction with the circumstellar medium}",
      journal = {\mnras},
         year = 2002,
        month = jun,
       volume = {333},
       number = {1},
        pages = {27-38},
          doi = {10.1046/j.1365-8711.2002.05366.x},
archivePrefix = {arXiv},
       eprint = {astro-ph/0201483},
 primaryClass = {astro-ph},
       adsurl = {https://ui.adsabs.harvard.edu/abs/2002MNRAS.333...27P}
}

@ARTICLE{Pastorello2019A&A...628A..93P,
       author = {{Pastorello}, A. and {Reguitti}, A. and {Morales-Garoffolo}, A. and {Cano}, Z. and {Prentice}, S.~J. and {Hiramatsu}, D. and {Burke}, J. and {Kankare}, E. and {Kotak}, R. and {Reynolds}, T. and {Smartt}, S.~J. and {Bose}, S. and {Chen}, P. and {Congiu}, E. and {Dong}, S. and {Geier}, S. and {Gromadzki}, M. and {Hsiao}, E.~Y. and {Kumar}, S. and {Ochner}, P. and {Pignata}, G. and {Tomasella}, L. and {Wang}, L. and {Arcavi}, I. and {Ashall}, C. and {Callis}, E. and {de Ugarte Postigo}, A. and {Fraser}, M. and {Hosseinzadeh}, G. and {Howell}, D.~A. and {Inserra}, C. and {Kann}, D.~A. and {Mason}, E. and {Mazzali}, P.~A. and {McCully}, C. and {Rodr{\'\i}guez}, {\'O}. and {Phillips}, M.~M. and {Smith}, K.~W. and {Tartaglia}, L. and {Th{\"o}ne}, C.~C. and {Wevers}, T. and {Young}, D.~R. and {Pumo}, M.~L. and {Lowe}, T.~B. and {Magnier}, E.~A. and {Wainscoat}, R.~J. and {Waters}, C. and {Wright}, D.~E.},
        title = "{A luminous stellar outburst during a long-lasting eruptive phase first, and then SN IIn 2018cnf}",
      journal = {\aap},
         year = 2019,
        month = aug,
       volume = {628},
          eid = {A93},
        pages = {A93},
          doi = {10.1051/0004-6361/201935420},
archivePrefix = {arXiv},
       eprint = {1906.00814},
 primaryClass = {astro-ph.SR},
       adsurl = {https://ui.adsabs.harvard.edu/abs/2019A&A...628A..93P}
}

@ARTICLE{Silverman2012MNRAS.425.1789S,
       author = {{Silverman}, Jeffrey M. and {Foley}, Ryan J. and {Filippenko}, Alexei V. and {Ganeshalingam}, Mohan and {Barth}, Aaron J. and {Chornock}, Ryan and {Griffith}, Christopher V. and {Kong}, Jason J. and {Lee}, Nicholas and {Leonard}, Douglas C. and {Matheson}, Thomas and {Miller}, Emily G. and {Steele}, Thea N. and {Barris}, Brian J. and {Bloom}, Joshua S. and {Cobb}, Bethany E. and {Coil}, Alison L. and {Desroches}, Louis-Benoit and {Gates}, Elinor L. and {Ho}, Luis C. and {Jha}, Saurabh W. and {Kandrashoff}, Michael T. and {Li}, Weidong and {Mandel}, Kaisey S. and {Modjaz}, Maryam and {Moore}, Matthew R. and {Mostardi}, Robin E. and {Papenkova}, Marina S. and {Park}, Sung and {Perley}, Daniel A. and {Poznanski}, Dovi and {Reuter}, Cassie A. and {Scala}, James and {Serduke}, Franklin J.~D. and {Shields}, Joseph C. and {Swift}, Brandon J. and {Tonry}, John L. and {Van Dyk}, Schuyler D. and {Wang}, Xiaofeng and {Wong}, Diane S.},
        title = "{Berkeley Supernova Ia Program - I. Observations, data reduction and spectroscopic sample of 582 low-redshift Type Ia supernovae}",
      journal = {\mnras},
         year = 2012,
        month = sep,
       volume = {425},
       number = {3},
        pages = {1789-1818},
          doi = {10.1111/j.1365-2966.2012.21270.x},
archivePrefix = {arXiv},
       eprint = {1202.2128},
 primaryClass = {astro-ph.CO},
       adsurl = {https://ui.adsabs.harvard.edu/abs/2012MNRAS.425.1789S}
}

@ARTICLE{Leadbeater2019TNSCR2506....1L,
       author = {{Leadbeater}, R.},
        title = "{Transient Classification Report for 2019-12-02}",
      journal = {Transient Name Server Classification Report},
         year = 2019,
        month = dec,
       volume = {2019-2506},
       adsurl = {https://ui.adsabs.harvard.edu/abs/2019TNSCR2506....1L}
}

@ARTICLE{Stanek2019TNSTR2492....1S,
       author = {{Stanek}, K.~Z.},
        title = "{ASAS-SN Transient Discovery Report for 2019-12-01}",
      journal = {Transient Name Server Discovery Report},
         year = 2019,
        month = dec,
       volume = {2019-2492},
        pages = {1},
       adsurl = {https://ui.adsabs.harvard.edu/abs/2019TNSTR2492....1S}
}

@ARTICLE{Lane2026ApJ..1003...19L,
       author = {{Lane}, Zachary G. and {Ridden-Harper}, Ryan and {Rest}, Sofia and {Rest}, Armin and {Ransome}, Conor L. and {Wang}, Qinan and {Montilla}, Clarinda and {Steed}, Micaela and {Andreoni}, Igor and {Armstrong}, Patrick and {Brown}, Peter J. and {Cooke}, Jeffrey and {Coulter}, David A. and {Fox}, Ori and {Freeburn}, James and {Galoppo}, Marco and {Gal-Yam}, Avishay and {Goldberg}, Jared A. and {Harvey-Hawes}, Christopher and {Hiramatsu}, Daichi and {Hounsell}, Rebekah and {Howell}, D. Andrew and {Leicester}, Brayden and {Lelkes}, Kl{\'a}ra and {Linial}, Itai and {Luisi}, Jaime and {McCully}, Curtis and {Moln{\'a}r}, L{\'a}szl{\'o} and {Moore}, Thomas and {Mourier}, Pierre and {Nugent}, Anya E. and {O'Neill}, David and {Roxburgh}, Hugh and {Shukawa}, Koji and {Smartt}, Stephen J. and {Smith}, Nathan and {Smith}, Ken W. and {Subrayan}, Bhagya M. and {Carrasco}, Sebastian Vergara and {Villar}, V. Ashley and {Vink{\'o}}, J{\'o}zsef and {Wasserman}, Tal and {Zenati}, Yossef and {Zimmerman}, Erez A.},
        title = "{SN 2019vxm: A Shocking Coincidence between Fermi and TESS}",
      journal = {\apj},
         year = 2026,
        month = may,
       volume = {1003},
       number = {1},
          eid = {19},
        pages = {19},
          doi = {10.3847/1538-4357/ae6245},
archivePrefix = {arXiv},
       eprint = {2511.15975},
 primaryClass = {astro-ph.HE},
       adsurl = {https://ui.adsabs.harvard.edu/abs/2026ApJ..1003...19L}
}

@ARTICLE{Tsvetkov2024AN....34530166T,
       author = {{Tsvetkov}, Dmitry Yu and {Pavlyuk}, Nickolay N. and {Dodin}, Alexandr V. and {Shatsky}, Nickolay I. and {Potanin}, Sergey A. and {Ikonnikova}, Nataliya P. and {Burlak}, Marina A. and {Belinskii}, Aleksandr A. and {Volkov}, Igor M. and {Echeistov}, Vsevolod A.},
        title = "{SNe 2019vxm and 2020tlf: two luminous type II Supernovae with signatures of circumstellar interaction}",
      journal = {Astronomische Nachrichten},
         year = 2024,
        month = may,
       volume = {345},
       number = {4},
          eid = {e20230166},
        pages = {e20230166},
          doi = {10.1002/asna.20230166},
       adsurl = {https://ui.adsabs.harvard.edu/abs/2024AN....34530166T}
}

@ARTICLE{Tsvetkov2023AstBu..78..514T,
       author = {{Tsvetkov}, D. Yu. and {Pavlyuk}, N.~N. and {Echeistov}, V.~A. and {Baklanov}, P.~V.},
        title = "{Progress in Supernovae Studies with the 2.5-m Telescope at the Caucasus Mountain Observatory of SAI MSU}",
      journal = {Astrophysical Bulletin},
         year = 2023,
        month = dec,
       volume = {78},
       number = {4},
        pages = {514-534},
          doi = {10.1134/S1990341323700232},
       adsurl = {https://ui.adsabs.harvard.edu/abs/2023AstBu..78..514T}
}

@ARTICLE{Vallely2021MNRAS.500.5639V,
       author = {{Vallely}, P.~J. and {Kochanek}, C.~S. and {Stanek}, K.~Z. and {Fausnaugh}, M. and {Shappee}, B.~J.},
        title = "{High-cadence, early-time observations of core-collapse supernovae from the TESS prime mission}",
      journal = {\mnras},
         year = 2021,
        month = feb,
       volume = {500},
       number = {4},
        pages = {5639-5656},
          doi = {10.1093/mnras/staa3675},
archivePrefix = {arXiv},
       eprint = {2010.06596},
 primaryClass = {astro-ph.HE},
       adsurl = {https://ui.adsabs.harvard.edu/abs/2021MNRAS.500.5639V}
}

@INPROCEEDINGS{Shappee2014AAS...22323603S,
       author = {{Shappee}, Benjamin and {Prieto}, J. and {Stanek}, K.~Z. and {Kochanek}, C.~S. and {Holoien}, T. and {Jencson}, J. and {Basu}, U. and {Beacom}, J.~F. and {Szczygiel}, D. and {Pojmanski}, G. and {Brimacombe}, J. and {Dubberley}, M. and {Elphick}, M. and {Foale}, S. and {Hawkins}, E. and {Mullins}, D. and {Rosing}, W. and {Ross}, R. and {Walker}, Z.},
        title = "{All Sky Automated Survey for SuperNovae (ASAS-SN or ``Assassin'')}",
    booktitle = {American Astronomical Society Meeting Abstracts \#223},
         year = 2014,
       series = {},
       volume = {223},
        month = jan,
          eid = {236.03},
        pages = {236.03},
       adsurl = {https://ui.adsabs.harvard.edu/abs/2014AAS...22323603S}
}

@ARTICLE{Kochanek2017PASP..129j4502K,
       author = {{Kochanek}, C.~S. and {Shappee}, B.~J. and {Stanek}, K.~Z. and {Holoien}, T.~W. -S. and {Thompson}, Todd A. and {Prieto}, J.~L. and {Dong}, Subo and {Shields}, J.~V. and {Will}, D. and {Britt}, C. and {Perzanowski}, D. and {Pojma{\'n}ski}, G.},
        title = "{The All-Sky Automated Survey for Supernovae (ASAS-SN) Light Curve Server v1.0}",
      journal = {\pasp},
         year = 2017,
        month = oct,
       volume = {129},
       number = {980},
        pages = {104502},
          doi = {10.1088/1538-3873/aa80d9},
archivePrefix = {arXiv},
       eprint = {1706.07060},
 primaryClass = {astro-ph.SR},
       adsurl = {https://ui.adsabs.harvard.edu/abs/2017PASP..129j4502K}
}

@MISC{HEAsoft2014ascl.soft08004N,
       author = {{Nasa Heasarc}},
        title = "{HEAsoft: Unified Release of FTOOLS and XANADU}",
 howpublished = {Astrophysics Source Code Library, record ascl:1408.004},
         year = 2014,
        month = aug,
          eid = {ascl:1408.004},
        pages = {ascl:1408.004},
archivePrefix = {ascl},
       eprint = {1408.004},
       adsurl = {https://ui.adsabs.harvard.edu/abs/2014ascl.soft08004N}
}

@ARTICLE{Cai2018MNRAS.480.3424C,
   author = {{Cai}, Y. -Z. and {Pastorello}, A. and {Fraser}, M. and {Botticella}, M.~T. and 
	{Gall}, C. and {Arcavi}, I. and {Benetti}, S. and {Cappellaro}, E. and 
	{Elias-Rosa}, N. and {Harmanen}, J. and {Hosseinzadeh}, G. and 
	{Howell}, D.~A. and {Isern}, J. and {Kangas}, T. and {Kankare}, E. and 
	{Kuncarayakti}, H. and {Lundqvist}, P. and {Mattila}, S. and 
	{McCully}, C. and {Reynolds}, T.~M. and {Somero}, A. and {Stritzinger}, M.~D. and 
	{Terreran}, G.},
    title = "{AT 2017be - a new member of the class of intermediate-luminosity red transients}",
  journal = {\mnras},
archivePrefix = "arXiv",
   eprint = {1807.11676},
 primaryClass = "astro-ph.HE",
     year = 2018,
    month = nov,
   volume = 480,
    pages = {3424-3445},
      doi = {10.1093/mnras/sty2070},
   adsurl = {http://adsabs.harvard.edu/abs/2018MNRAS.480.3424C}
}

@ARTICLE{Chambers2016arXiv161205560C,
       author = {{Chambers}, K.~C. and {Magnier}, E.~A. and {Metcalfe}, N. and {Flewelling}, H.~A. and {Huber}, M.~E. and {Waters}, C.~Z. and {Denneau}, L. and {Draper}, P.~W. and {Farrow}, D. and {Finkbeiner}, D.~P. and {Holmberg}, C. and {Koppenhoefer}, J. and {Price}, P.~A. and {Rest}, A. and {Saglia}, R.~P. and {Schlafly}, E.~F. and {Smartt}, S.~J. and {Sweeney}, W. and {Wainscoat}, R.~J. and {Burgett}, W.~S. and {Chastel}, S. and {Grav}, T. and {Heasley}, J.~N. and {Hodapp}, K.~W. and {Jedicke}, R. and {Kaiser}, N. and {Kudritzki}, R. -P. and {Luppino}, G.~A. and {Lupton}, R.~H. and {Monet}, D.~G. and {Morgan}, J.~S. and {Onaka}, P.~M. and {Shiao}, B. and {Stubbs}, C.~W. and {Tonry}, J.~L. and {White}, R. and {Ba{\~n}ados}, E. and {Bell}, E.~F. and {Bender}, R. and {Bernard}, E.~J. and {Boegner}, M. and {Boffi}, F. and {Botticella}, M.~T. and {Calamida}, A. and {Casertano}, S. and {Chen}, W. -P. and {Chen}, X. and {Cole}, S. and {Deacon}, N. and {Frenk}, C. and {Fitzsimmons}, A. and {Gezari}, S. and {Gibbs}, V. and {Goessl}, C. and {Goggia}, T. and {Gourgue}, R. and {Goldman}, B. and {Grant}, P. and {Grebel}, E.~K. and {Hambly}, N.~C. and {Hasinger}, G. and {Heavens}, A.~F. and {Heckman}, T.~M. and {Henderson}, R. and {Henning}, T. and {Holman}, M. and {Hopp}, U. and {Ip}, W. -H. and {Isani}, S. and {Jackson}, M. and {Keyes}, C.~D. and {Koekemoer}, A.~M. and {Kotak}, R. and {Le}, D. and {Liska}, D. and {Long}, K.~S. and {Lucey}, J.~R. and {Liu}, M. and {Martin}, N.~F. and {Masci}, G. and {McLean}, B. and {Mindel}, E. and {Misra}, P. and {Morganson}, E. and {Murphy}, D.~N.~A. and {Obaika}, A. and {Narayan}, G. and {Nieto-Santisteban}, M.~A. and {Norberg}, P. and {Peacock}, J.~A. and {Pier}, E.~A. and {Postman}, M. and {Primak}, N. and {Rae}, C. and {Rai}, A. and {Riess}, A. and {Riffeser}, A. and {Rix}, H.~W. and {R{\"o}ser}, S. and {Russel}, R. and {Rutz}, L. and {Schilbach}, E. and {Schultz}, A.~S.~B. and {Scolnic}, D. and {Strolger}, L. and {Szalay}, A. and {Seitz}, S. and {Small}, E. and {Smith}, K.~W. and {Soderblom}, D.~R. and {Taylor}, P. and {Thomson}, R. and {Taylor}, A.~N. and {Thakar}, A.~R. and {Thiel}, J. and {Thilker}, D. and {Unger}, D. and {Urata}, Y. and {Valenti}, J. and {Wagner}, J. and {Walder}, T. and {Walter}, F. and {Watters}, S.~P. and {Werner}, S. and {Wood-Vasey}, W.~M. and {Wyse}, R.},
        title = "{The Pan-STARRS1 Surveys}",
      journal = {arXiv e-prints},
         year = 2016,
        month = dec,
          eid = {arXiv:1612.05560},
        pages = {arXiv:1612.05560},
archivePrefix = {arXiv},
       eprint = {1612.05560},
 primaryClass = {astro-ph.IM},
       adsurl = {https://ui.adsabs.harvard.edu/abs/2016arXiv161205560C}
}

@ARTICLE{Flewelling2020ApJS..251....7F,
       author = {{Flewelling}, H.~A. and {Magnier}, E.~A. and {Chambers}, K.~C. and {Heasley}, J.~N. and {Holmberg}, C. and {Huber}, M.~E. and {Sweeney}, W. and {Waters}, C.~Z. and {Calamida}, A. and {Casertano}, S. and {Chen}, X. and {Farrow}, D. and {Hasinger}, G. and {Henderson}, R. and {Long}, K.~S. and {Metcalfe}, N. and {Narayan}, G. and {Nieto-Santisteban}, M.~A. and {Norberg}, P. and {Rest}, A. and {Saglia}, R.~P. and {Szalay}, A. and {Thakar}, A.~R. and {Tonry}, J.~L. and {Valenti}, J. and {Werner}, S. and {White}, R. and {Denneau}, L. and {Draper}, P.~W. and {Hodapp}, K.~W. and {Jedicke}, R. and {Kaiser}, N. and {Kudritzki}, R.~P. and {Price}, P.~A. and {Wainscoat}, R.~J. and {Chastel}, S. and {McLean}, B. and {Postman}, M. and {Shiao}, B.},
        title = "{The Pan-STARRS1 Database and Data Products}",
      journal = {\apjs},
         year = 2020,
        month = nov,
       volume = {251},
       number = {1},
          eid = {7},
        pages = {7},
          doi = {10.3847/1538-4365/abb82d},
archivePrefix = {arXiv},
       eprint = {1612.05243},
 primaryClass = {astro-ph.IM},
       adsurl = {https://ui.adsabs.harvard.edu/abs/2020ApJS..251....7F}
}

@ARTICLE{Magnier2020ApJS..251....3M,
       author = {{Magnier}, Eugene A. and {Chambers}, K.~C. and {Flewelling}, H.~A. and {Hoblitt}, J.~C. and {Huber}, M.~E. and {Price}, P.~A. and {Sweeney}, W.~E. and {Waters}, C.~Z. and {Denneau}, L. and {Draper}, P.~W. and {Hodapp}, K.~W. and {Jedicke}, R. and {Kaiser}, N. and {Kudritzki}, R. -P. and {Metcalfe}, N. and {Stubbs}, C.~W. and {Wainscoat}, R.~J.},
        title = "{The Pan-STARRS Data-processing System}",
      journal = {\apjs},
         year = 2020,
        month = nov,
       volume = {251},
       number = {1},
          eid = {3},
        pages = {3},
          doi = {10.3847/1538-4365/abb829},
archivePrefix = {arXiv},
       eprint = {1612.05240},
 primaryClass = {astro-ph.IM},
       adsurl = {https://ui.adsabs.harvard.edu/abs/2020ApJS..251....3M}
}

@ARTICLE{Tonry2018PASP..130f4505T,
       author = {{Tonry}, J.~L. and {Denneau}, L. and {Heinze}, A.~N. and {Stalder}, B. and {Smith}, K.~W. and {Smartt}, S.~J. and {Stubbs}, C.~W. and {Weiland}, H.~J. and {Rest}, A.},
        title = "{ATLAS: A High-cadence All-sky Survey System}",
      journal = {\pasp},
         year = 2018,
        month = jun,
       volume = {130},
       number = {988},
        pages = {064505},
          doi = {10.1088/1538-3873/aabadf},
archivePrefix = {arXiv},
       eprint = {1802.00879},
 primaryClass = {astro-ph.IM},
       adsurl = {https://ui.adsabs.harvard.edu/abs/2018PASP..130f4505T}
}

@ARTICLE{Smith2020PASP..132h5002S,
       author = {{Smith}, K.~W. and {Smartt}, S.~J. and {Young}, D.~R. and {Tonry}, J.~L. and {Denneau}, L. and {Flewelling}, H. and {Heinze}, A.~N. and {Weiland}, H.~J. and {Stalder}, B. and {Rest}, A. and {Stubbs}, C.~W. and {Anderson}, J.~P. and {Chen}, T. -W. and {Clark}, P. and {Do}, A. and {F{\"o}rster}, F. and {Fulton}, M. and {Gillanders}, J. and {McBrien}, O.~R. and {O'Neill}, D. and {Srivastav}, S. and {Wright}, D.~E.},
        title = "{Design and Operation of the ATLAS Transient Science Server}",
      journal = {\pasp},
         year = 2020,
        month = aug,
       volume = {132},
       number = {1014},
          eid = {085002},
        pages = {085002},
          doi = {10.1088/1538-3873/ab936e},
archivePrefix = {arXiv},
       eprint = {2003.09052},
 primaryClass = {astro-ph.IM},
       adsurl = {https://ui.adsabs.harvard.edu/abs/2020PASP..132h5002S}
}

@ARTICLE{Shingles2021TNSAN...7....1S,
       author = {{Shingles}, L. and {Smith}, K.~W. and {Young}, D.~R. and {Smartt}, S.~J. and {Tonry}, J. and {Denneau}, L. and {Heinze}, A. and {Weiland}, H. and {Flewelling}, H. and {Stalder}, B. and {Clocchiatti}, A. and {F{\"o}rster}, F. and {Pignata}, G. and {Rest}, A. and {Anderson}, J. and {Stubbs}, C. and {Erasmus}, N.},
        title = "{Release of the ATLAS Forced Photometry server for public use}",
      journal = {Transient Name Server AstroNote},
         year = 2021,
        month = jan,
       volume = {7},
        pages = {1-7},
       adsurl = {https://ui.adsabs.harvard.edu/abs/2021TNSAN...7....1S}
}

@ARTICLE{Spergel2007ApJS..170..377S,
	author = {{Spergel}, D.~N. and {Bean}, R. and {Dor{\'e}}, O. and {Nolta}, M.~R. and {Bennett}, C.~L. and {Dunkley}, J. and {Hinshaw}, G. and {Jarosik}, N. and {Komatsu}, E. and {Page}, L. and {Peiris}, H.~V. and {Verde}, L. and {Halpern}, M. and {Hill}, R.~S. and {Kogut}, A. and {Limon}, M. and {Meyer}, S.~S. and {Odegard}, N. and {Tucker}, G.~S. and {Weiland}, J.~L. and {Wollack}, E. and {Wright}, E.~L.},
	title = "{Three-Year Wilkinson Microwave Anisotropy Probe (WMAP) Observations: Implications for Cosmology}",
	journal = {\apjs},
	year = 2007,
	month = jun,
	volume = {170},
	number = {2},
	pages = {377-408},
	doi = {10.1086/513700},
	archivePrefix = {arXiv},
	eprint = {astro-ph/0603449},
	primaryClass = {astro-ph},
	adsurl = {https://ui.adsabs.harvard.edu/abs/2007ApJS..170..377S}
}

@ARTICLE{Schlafly2011ApJ...737..103S,
       author = {{Schlafly}, Edward F. and {Finkbeiner}, Douglas P.},
        title = "{Measuring Reddening with Sloan Digital Sky Survey Stellar Spectra and Recalibrating SFD}",
      journal = {\apj},
         year = 2011,
        month = aug,
       volume = {737},
       number = {2},
          eid = {103},
        pages = {103},
          doi = {10.1088/0004-637X/737/2/103},
archivePrefix = {arXiv},
       eprint = {1012.4804},
 primaryClass = {astro-ph.GA},
       adsurl = {https://ui.adsabs.harvard.edu/abs/2011ApJ...737..103S}
}

@ARTICLE{SN2003ma_Rest2011ApJ...729...88R,
       author = {{Rest}, A. and {Foley}, R.~J. and {Gezari}, S. and {Narayan}, G. and {Draine}, B. and {Olsen}, K. and {Huber}, M.~E. and {Matheson}, T. and {Garg}, A. and {Welch}, D.~L. and et al.},
        title = "{Pushing the Boundaries of Conventional Core-collapse Supernovae: The Extremely Energetic Supernova SN 2003ma}",
      journal = {\apj},
         year = 2011,
        month = mar,
       volume = {729},
       number = {2},
          eid = {88},
        pages = {88},
          doi = {10.1088/0004-637X/729/2/88},
archivePrefix = {arXiv},
       eprint = {0911.2002},
 primaryClass = {astro-ph.CO},
       adsurl = {https://ui.adsabs.harvard.edu/abs/2011ApJ...729...88R}
}

@ARTICLE{SN2005ip_Stritzinger2012ApJ...756..173S,
       author = {{Stritzinger}, Maximilian and {Taddia}, Francesco and {Fransson}, Claes and {Fox}, Ori D. and {Morrell}, Nidia and {Phillips}, M.~M. and {Sollerman}, Jesper and {Anderson}, J.~P. and {Boldt}, Luis and {Brown}, Peter J. and et al.},
        title = "{Multi-wavelength Observations of the Enduring Type IIn Supernovae 2005ip and 2006jd}",
      journal = {\apj},
         year = 2012,
        month = sep,
       volume = {756},
       number = {2},
          eid = {173},
        pages = {173},
          doi = {10.1088/0004-637X/756/2/173},
archivePrefix = {arXiv},
       eprint = {1206.5575},
 primaryClass = {astro-ph.CO},
       adsurl = {https://ui.adsabs.harvard.edu/abs/2012ApJ...756..173S}
}

@ARTICLE{SN2006gy_Smith2007ApJ...666.1116S,
       author = {{Smith}, Nathan and {Li}, Weidong and {Foley}, Ryan J. and {Wheeler}, J. Craig and {Pooley}, David and {Chornock}, Ryan and {Filippenko}, Alexei V. and {Silverman}, Jeffrey M. and {Quimby}, Robert and {Bloom}, Joshua S. and et al.},
        title = "{SN 2006gy: Discovery of the Most Luminous Supernova Ever Recorded, Powered by the Death of an Extremely Massive Star like {\ensuremath{\eta}} Carinae}",
      journal = {\apj},
         year = 2007,
        month = sep,
       volume = {666},
       number = {2},
        pages = {1116-1128},
          doi = {10.1086/519949},
archivePrefix = {arXiv},
       eprint = {astro-ph/0612617},
 primaryClass = {astro-ph},
       adsurl = {https://ui.adsabs.harvard.edu/abs/2007ApJ...666.1116S}
}

@ARTICLE{SN2006gy_Ofek2007ApJ...659L..13O,
       author = {{Ofek}, E.~O. and {Cameron}, P.~B. and {Kasliwal}, M.~M. and {Gal-Yam}, A. and {Rau}, A. and {Kulkarni}, S.~R. and {Frail}, D.~A. and {Chandra}, P. and {Cenko}, S.~B. and {Soderberg}, A.~M. and et al.},
        title = "{SN 2006gy: An Extremely Luminous Supernova in the Galaxy NGC 1260}",
      journal = {\apjl},
         year = 2007,
        month = apr,
       volume = {659},
       number = {1},
        pages = {L13-L16},
          doi = {10.1086/516749},
archivePrefix = {arXiv},
       eprint = {astro-ph/0612408},
 primaryClass = {astro-ph},
       adsurl = {https://ui.adsabs.harvard.edu/abs/2007ApJ...659L..13O}
}

@ARTICLE{SN2006tf_Smith2008ApJ...686..467S,
       author = {{Smith}, Nathan and {Chornock}, Ryan and {Li}, Weidong and {Ganeshalingam}, Mohan and {Silverman}, Jeffrey M. and {Foley}, Ryan J. and {Filippenko}, Alexei V. and {Barth}, Aaron J.},
        title = "{SN 2006tf: Precursor Eruptions and the Optically Thick Regime of Extremely Luminous Type IIn Supernovae}",
      journal = {\apj},
         year = 2008,
        month = oct,
       volume = {686},
       number = {1},
        pages = {467-484},
          doi = {10.1086/591021},
archivePrefix = {arXiv},
       eprint = {0804.0042},
 primaryClass = {astro-ph},
       adsurl = {https://ui.adsabs.harvard.edu/abs/2008ApJ...686..467S}
}

@ARTICLE{SN2010jl_Ofek2014ApJ...781...42O,
       author = {{Ofek}, Eran O. and {Zoglauer}, Andreas and {Boggs}, Steven E. and {Barri{\'e}re}, Nicolas M. and {Reynolds}, Stephen P. and {Fryer}, Chris L. and {Harrison}, Fiona A. and {Cenko}, S. Bradley and {Kulkarni}, Shrinivas R. and {Gal-Yam}, Avishay and et al.},
        title = "{SN 2010jl: Optical to Hard X-Ray Observations Reveal an Explosion Embedded in a Ten Solar Mass Cocoon}",
      journal = {\apj},
         year = 2014,
        month = jan,
       volume = {781},
       number = {1},
          eid = {42},
        pages = {42},
          doi = {10.1088/0004-637X/781/1/42},
archivePrefix = {arXiv},
       eprint = {1307.2247},
 primaryClass = {astro-ph.HE},
       adsurl = {https://ui.adsabs.harvard.edu/abs/2014ApJ...781...42O}
}

@ARTICLE{SN2008am_Chatzopoulos2011ApJ...729..143C,
       author = {{Chatzopoulos}, E. and {Wheeler}, J. Craig and {Vinko}, J. and {Quimby}, R. and {Robinson}, E.~L. and {Miller}, A.~A. and {Foley}, R.~J. and {Perley}, D.~A. and {Yuan}, F. and {Akerlof}, C. and et al.},
        title = "{SN 2008am: A Super-luminous Type IIn Supernova}",
      journal = {\apj},
         year = 2011,
        month = mar,
       volume = {729},
       number = {2},
          eid = {143},
        pages = {143},
          doi = {10.1088/0004-637X/729/2/143},
archivePrefix = {arXiv},
       eprint = {1101.3581},
 primaryClass = {astro-ph.HE},
       adsurl = {https://ui.adsabs.harvard.edu/abs/2011ApJ...729..143C}
}

@ARTICLE{SN2015da_Tartaglia2020A&A...635A..39T,
       author = {{Tartaglia}, L. and {Pastorello}, A. and {Sollerman}, J. and {Fransson}, C. and {Mattila}, S. and {Fraser}, M. and {Taddia}, F. and {Tomasella}, L. and {Turatto}, M. and {Morales-Garoffolo}, A. and et al.},
        title = "{The long-lived Type IIn SN 2015da: Infrared echoes and strong interaction within an extended massive shell}",
      journal = {\aap},
         year = 2020,
        month = mar,
       volume = {635},
          eid = {A39},
        pages = {A39},
          doi = {10.1051/0004-6361/201936553},
archivePrefix = {arXiv},
       eprint = {1908.08580},
 primaryClass = {astro-ph.HE},
       adsurl = {https://ui.adsabs.harvard.edu/abs/2020A&A...635A..39T}
}

@ARTICLE{SN2015da_Smith2024MNRAS.530..405S,
       author = {{Smith}, Nathan and {Andrews}, Jennifer E. and {Milne}, Peter and {Filippenko}, Alexei V. and {Brink}, Thomas G. and {Kelly}, Patrick L. and {Yuk}, Heechan and {Jencson}, Jacob E.},
        title = "{SN 2015da: late-time observations of a persistent superluminous Type IIn supernova with post-shock dust formation}",
      journal = {\mnras},
         year = 2024,
        month = may,
       volume = {530},
       number = {1},
        pages = {405-423},
          doi = {10.1093/mnras/stae726},
archivePrefix = {arXiv},
       eprint = {2312.00253},
 primaryClass = {astro-ph.HE},
       adsurl = {https://ui.adsabs.harvard.edu/abs/2024MNRAS.530..405S}
}

@ARTICLE{SN2017hcc_Moran2023A&A...669A..51M,
       author = {{Moran}, S. and {Fraser}, M. and {Kotak}, R. and {Pastorello}, A. and {Benetti}, S. and {Brennan}, S.~J. and {Guti{\'e}rrez}, C.~P. and {Kankare}, E. and {Kuncarayakti}, H. and {Mattila}, S. and et al.},
        title = "{A long life of excess: The interacting transient SN 2017hcc}",
      journal = {\aap},
         year = 2023,
        month = jan,
       volume = {669},
          eid = {A51},
        pages = {A51},
          doi = {10.1051/0004-6361/202244565},
archivePrefix = {arXiv},
       eprint = {2210.14076},
 primaryClass = {astro-ph.HE},
       adsurl = {https://ui.adsabs.harvard.edu/abs/2023A&A...669A..51M}
}

@INCOLLECTION{Smith2017hsn..book..403S,
       author = {{Smith}, Nathan},
        title = "{Interacting Supernovae: Types IIn and Ibn}",
    booktitle = {Handbook of Supernovae},
         year = 2017,
       editor = {{Alsabti}, Athem W. and {Murdin}, Paul},
    publisher = {Springer Nature Link},
        pages = {403},
          doi = {10.1007/978-3-319-21846-5_38},
       adsurl = {https://ui.adsabs.harvard.edu/abs/2017hsn..book..403S}
}

@ARTICLE{Peng2026A&A...705A.104P,
       author = {{Peng}, Z.-H. and {Benetti}, S. and {Cai}, Y.-Z. and {Pastorello}, A. and {Valerin}, G. and {Reguitti}, A. and {Fiore}, A. and {Fang}, Q.-L. and {Wang}, Z.-Y. and {Berton}, M. and {Borsato}, L. and {Cappellaro}, E. and {Congiu}, E. and {Elias-Rosa}, N. and {Granata}, V. and {Isern}, J. and {La Mura}, G. and {Ochner}, P. and {Raddi}, R. and {Terreran}, G. and {Tomasella}, L. and {Turatto}, M. and {Yan}, S.-Y. and {Pei}, S.-P. and {Wu}, C.-Y. and {Zha}, S. and {Wang}, X.-F. and {Wang}, B. and {Pan}, Y.},
        title = "{SN 2016iog: A fast-declining Type II-L supernova with an ultra-faint tail persistently interacting with circumstellar material}",
      journal = {\aap},
         year = 2026,
        month = jan,
       volume = {705},
          eid = {A104},
        pages = {A104},
          doi = {10.1051/0004-6361/202556330},
archivePrefix = {arXiv},
       eprint = {2511.12929},
 primaryClass = {astro-ph.SR},
       adsurl = {https://ui.adsabs.harvard.edu/abs/2026A&A...705A.104P}
}

@ARTICLE{Zhang2022MNRAS.509.2013Z,
       author = {{Zhang}, Xinghan and {Wang}, Xiaofeng and {Sai}, Hanna and {Niculescu-Duvaz}, Maria and {Filippenko}, Alexei V. and {Zheng}, WeiKang and {Brink}, T.~G. and {Lin}, Han and {Zhang}, Jicheng and {Cai}, Yongzhi and {Mo}, Jun and {Zhang}, Jujia and {Baron}, E. and {DerKacy}, J.~M. and {Huang}, F. and {Zhang}, T.-M.},
        title = "{SN 2018hfm: a low-energy Type II supernova with prominent signatures of circumstellar interaction and dust formation}",
      journal = {\mnras},
         year = 2022,
        month = jan,
       volume = {509},
       number = {2},
        pages = {2013-2032},
          doi = {10.1093/mnras/stab3007},
archivePrefix = {arXiv},
       eprint = {2110.10440},
 primaryClass = {astro-ph.HE},
       adsurl = {https://ui.adsabs.harvard.edu/abs/2022MNRAS.509.2013Z}
}

@dataset{Cutri2012yCat.2311....0C,
       author = {{Cutri}, R.~M. and {et al.}},
        title = "{VizieR Online Data Catalog: WISE All-Sky Data Release (Cutri+ 2012)}",
 howpublished = {VizieR On-line Data Catalog: II/311.  Originally published in: 2012wise.rept....1C},
         year = 2012,
        month = apr,
          eid = {II/311},
       adsurl = {https://ui.adsabs.harvard.edu/abs/2012yCat.2311....0C}
}

@ARTICLE{Leonard2000ApJ...536..239L,
       author = {{Leonard}, Douglas C. and {Filippenko}, Alexei V. and {Barth}, Aaron J. and {Matheson}, Thomas},
        title = "{Evidence for Asphericity in the Type IIN Supernova SN 1998S}",
      journal = {\apj},
         year = 2000,
        month = jun,
       volume = {536},
       number = {1},
        pages = {239-254},
          doi = {10.1086/308910},
archivePrefix = {arXiv},
       eprint = {astro-ph/9908040},
 primaryClass = {astro-ph},
       adsurl = {https://ui.adsabs.harvard.edu/abs/2000ApJ...536..239L}
}

@ARTICLE{Gal-Yam2014Natur.509..471G,
       author = {{Gal-Yam}, Avishay and {Arcavi}, I. and {Ofek}, E.~O. and {Ben-Ami}, S. and {Cenko}, S.~B. and {Kasliwal}, M.~M. and {Cao}, Y. and {Yaron}, O. and {Tal}, D. and {Silverman}, J.~M. and {Horesh}, A. and {De Cia}, A. and {Taddia}, F. and {Sollerman}, J. and {Perley}, D. and {Vreeswijk}, P.~M. and {Kulkarni}, S.~R. and {Nugent}, P.~E. and {Filippenko}, A.~V. and {Wheeler}, J.~C.},
        title = "{A Wolf-Rayet-like progenitor of SN 2013cu from spectral observations of a stellar wind}",
      journal = {\nat},
         year = 2014,
        month = may,
       volume = {509},
       number = {7501},
        pages = {471-474},
          doi = {10.1038/nature13304},
archivePrefix = {arXiv},
       eprint = {1406.7640},
 primaryClass = {astro-ph.HE},
       adsurl = {https://ui.adsabs.harvard.edu/abs/2014Natur.509..471G}
}

@ARTICLE{Dessart2017A&A...605A..83D,
       author = {{Dessart}, Luc and {Hillier}, D. John and {Audit}, Edouard},
        title = "{Explosion of red-supergiant stars: Influence of the atmospheric structure on shock breakout and early-time supernova radiation}",
      journal = {\aap},
         year = 2017,
        month = sep,
       volume = {605},
          eid = {A83},
        pages = {A83},
          doi = {10.1051/0004-6361/201730942},
archivePrefix = {arXiv},
       eprint = {1704.01697},
 primaryClass = {astro-ph.SR},
       adsurl = {https://ui.adsabs.harvard.edu/abs/2017A&A...605A..83D}
}

@ARTICLE{Smith2015MNRAS.449.1876S,
       author = {{Smith}, Nathan and {Mauerhan}, Jon C. and {Cenko}, S. Bradley and {Kasliwal}, Mansi M. and {Silverman}, Jeffrey M. and {Filippenko}, Alexei V. and {Gal-Yam}, Avishay and {Clubb}, Kelsey I. and {Graham}, Melissa L. and {Leonard}, Douglas C. and {Horst}, J. Chuck and {Williams}, G. Grant and {Andrews}, Jennifer E. and {Kulkarni}, Shrinivas R. and {Nugent}, Peter and {Sullivan}, Mark and {Maguire}, Kate and {Xu}, Dong and {Ben-Ami}, Sagi},
        title = "{PTF11iqb: cool supergiant mass-loss that bridges the gap between Type IIn and normal supernovae}",
      journal = {\mnras},
         year = 2015,
        month = may,
       volume = {449},
       number = {2},
        pages = {1876-1896},
          doi = {10.1093/mnras/stv354},
archivePrefix = {arXiv},
       eprint = {1501.02820},
 primaryClass = {astro-ph.HE},
       adsurl = {https://ui.adsabs.harvard.edu/abs/2015MNRAS.449.1876S}
}

@ARTICLE{Shivvers2015ApJ...806..213S,
       author = {{Shivvers}, Isaac and {Groh}, Jose H. and {Mauerhan}, Jon C. and {Fox}, Ori D. and {Leonard}, Douglas C. and {Filippenko}, Alexei V.},
        title = "{Early Emission from the Type IIn Supernova 1998S at High Resolution}",
      journal = {\apj},
         year = 2015,
        month = jun,
       volume = {806},
       number = {2},
          eid = {213},
        pages = {213},
          doi = {10.1088/0004-637X/806/2/213},
archivePrefix = {arXiv},
       eprint = {1408.1404},
 primaryClass = {astro-ph.HE},
       adsurl = {https://ui.adsabs.harvard.edu/abs/2015ApJ...806..213S}
}

@ARTICLE{Killestein2026MNRAS.548f2261K,
       author = {{Killestein}, T.~L. and {Pursiainen}, M. and {Kotak}, R. and {Charalampopoulos}, P. and {Lyman}, J. and {Ackley}, K. and {Belkin}, S. and {Coppejans}, D.~L. and {Davies}, B. and {Dyer}, M.~J. and {Galbany}, L. and {Godson}, B. and {Jarvis}, D. and {Koivisto}, N. and {Kumar}, A. and {Magee}, M. and {Mitchell}, M. and {O'Neill}, D. and {Sahu}, A. and {Warwick}, B. and {Breton}, R.~P. and {Butterley}, T. and {Cai}, Y.-Z. and {Casares}, J. and {Dhillon}, V.~S. and {Elias-Rosa}, N. and {Fraser}, M. and {Galloway}, D.~K. and {Gompertz}, B. and {Gonz{\'a}lez-Ba{\~n}uelos}, M. and {Guti{\'e}rrez}, C.~P. and {Kangas}, T. and {Kankare}, E. and {Kelsey}, L. and {Kravtsov}, T. and {Leloudas}, G. and {Littlefair}, S.~P. and {Matilainen}, K. and {Mattila}, S. and {Nagao}, T. and {Noysena}, K. and {Nuttall}, L.~K. and {O'Brien}, P. and {Pollacco}, D. and {Ramsay}, G. and {Reguitti}, A. and {Reynolds}, T.~M. and {Salmaso}, I. and {Starling}, R.~L.~C. and {Steeghs}, D. and {Stritzinger}, M. and {Ulaczyk}, K. and {Valerin}, G. and {Wang}, Z.-Y. and {Wilson}, R.},
        title = "{SN 2024cld: unveiling the complex mass-loss histories of evolved supergiant progenitors to core collapse supernovae}",
      journal = {\mnras},
         year = 2026,
        month = may,
       volume = {548},
       number = {2},
          eid = {staf2261},
        pages = {staf2261},
          doi = {10.1093/mnras/staf2261},
archivePrefix = {arXiv},
       eprint = {2510.27631},
 primaryClass = {astro-ph.HE},
       adsurl = {https://ui.adsabs.harvard.edu/abs/2026MNRAS.548f2261K}
}

@ARTICLE{Bruch2023ApJ...952..119B,
       author = {{Bruch}, Rachel J. and {Gal-Yam}, Avishay and {Yaron}, Ofer and {Chen}, Ping and {Strotjohann}, Nora L. and {Irani}, Ido and {Zimmerman}, Erez and {Schulze}, Steve and {Yang}, Yi and {Kim}, Young-Lo and {Bulla}, Mattia and {Sollerman}, Jesper and {Rigault}, Mickael and {Ofek}, Eran and {Soumagnac}, Maayane and {Masci}, Frank J. and {Fremling}, Christoffer and {Perley}, Daniel and {Nordin}, Jakob and {Cenko}, S. Bradley and {Ho}, Anna Y.~Q. and {Adams}, S. and {Adreoni}, Igor and {Bellm}, Eric C. and {Blagorodnova}, Nadia and {Burdge}, Kevin and {De}, Kishalay and {Dekany}, Richard G. and {Dhawan}, Suhail and {Drake}, Andrew J. and {Duev}, Dmitry A. and {Graham}, Matthew and {Graham}, Melissa L. and {Jencson}, Jacob and {Karamehmetoglu}, Emir and {Kasliwal}, Mansi M. and {Kulkarni}, Shrinivas and {Miller}, A.~A. and {Neill}, James D. and {Prince}, Thomas A. and {Riddle}, Reed and {Rusholme}, Benjamin and {Sharma}, Y. and {Smith}, Roger and {Sravan}, Niharika and {Taggart}, Kirsty and {Walters}, Richard and {Yan}, Lin},
        title = "{The Prevalence and Influence of Circumstellar Material around Hydrogen-rich Supernova Progenitors}",
      journal = {\apj},
         year = 2023,
        month = aug,
       volume = {952},
       number = {2},
          eid = {119},
        pages = {119},
          doi = {10.3847/1538-4357/acd8be},
archivePrefix = {arXiv},
       eprint = {2212.03313},
 primaryClass = {astro-ph.HE},
       adsurl = {https://ui.adsabs.harvard.edu/abs/2023ApJ...952..119B}
}

@ARTICLE{Cai2026A&A...707A.157C,
       author = {{Cai}, Y.-Z. and {Pastorello}, A. and {Maeda}, K. and {Zhao}, J.-W. and {Wang}, Z.-Y. and {Peng}, Z.-H. and {Reguitti}, A. and {Tartaglia}, L. and {Filippenko}, A.~V. and {Pan}, Y. and {Valerin}, G. and {Kumar}, B. and {Wang}, Z. and {Fraser}, M. and {Anderson}, J.~P. and {Benetti}, S. and {Bose}, S. and {Brink}, T.~G. and {Cappellaro}, E. and {Chen}, T.-W. and {Chen}, X.-L. and {Elias-Rosa}, N. and {Esamdin}, A. and {Gal-Yam}, A. and {Gonz{\'a}lez-Ba{\~n}uelos}, M. and {Gromadzki}, M. and {Guti{\'e}rrez}, C.~P. and {Inserra}, C. and {Iskandar}, A. and {Kangas}, T. and {Kankare}, E. and {Kravtsov}, T. and {Kuncarayakti}, H. and {Li}, L.-P. and {Liu}, C.-X. and {Liu}, X.-K. and {Lundqvist}, P. and {Matilainen}, K. and {Mattila}, S. and {Moran}, S. and {M{\"u}ller-Bravo}, T.~E. and {Nagao}, T. and {Petrushevska}, T. and {Pignata}, G. and {Salmaso}, I. and {Smartt}, S.~J. and {Sollerman}, J. and {Srivastav}, S. and {Stritzinger}, M.~D. and {Wang}, L.-T. and {Yan}, S.-Y. and {Yang}, Y. and {Yang}, Y.-P. and {Zheng}, W. and {Zou}, X.-Z. and {Chen}, L.-Y. and {Du}, X.-L. and {Fang}, Q.-L. and {Fiore}, A. and {Ragosta}, F. and {Zha}, S. and {Zhang}, J.-J. and {Liu}, X.-W. and {Bai}, J.-M. and {Wang}, B. and {Wang}, X.-F.},
        title = "{Massive stars exploding in a He-rich circumstellar medium: XII. SN 2024acyl: A fast, linearly declining Type Ibn supernova with early flash-ionisation features}",
      journal = {\aap},
         year = 2026,
        month = mar,
       volume = {707},
          eid = {A157},
        pages = {A157},
          doi = {10.1051/0004-6361/202558014},
archivePrefix = {arXiv},
       eprint = {2511.04337},
 primaryClass = {astro-ph.SR},
       adsurl = {https://ui.adsabs.harvard.edu/abs/2026A&A...707A.157C}
}

@ARTICLE{Kiewe2012ApJ...744...10K,
       author = {{Kiewe}, Michael and {Gal-Yam}, Avishay and {Arcavi}, Iair and {Leonard}, Douglas C. and {Emilio Enriquez}, J. and {Cenko}, S. Bradley and {Fox}, Derek B. and {Moon}, Dae-Sik and {Sand}, David J. and {Soderberg}, Alicia M. and {CCCP}, The},
        title = "{Caltech Core-Collapse Project (CCCP) Observations of Type IIn Supernovae: Typical Properties and Implications for Their Progenitor Stars}",
      journal = {\apj},
         year = 2012,
        month = jan,
       volume = {744},
       number = {1},
          eid = {10},
        pages = {10},
          doi = {10.1088/0004-637X/744/1/10},
archivePrefix = {arXiv},
       eprint = {1010.2689},
 primaryClass = {astro-ph.CO},
       adsurl = {https://ui.adsabs.harvard.edu/abs/2012ApJ...744...10K}
}

@ARTICLE{Fransson2014ApJ...797..118F,
       author = {{Fransson}, Claes and {Ergon}, Mattias and {Challis}, Peter J. and {Chevalier}, Roger A. and {France}, Kevin and {Kirshner}, Robert P. and {Marion}, G.~H. and {Milisavljevic}, Dan and {Smith}, Nathan and {Bufano}, Filomena and {Friedman}, Andrew S. and {Kangas}, Tuomas and {Larsson}, Josefin and {Mattila}, Seppo and {Benetti}, Stefano and {Chornock}, Ryan and {Czekala}, Ian and {Soderberg}, Alicia and {Sollerman}, Jesper},
        title = "{High-density Circumstellar Interaction in the Luminous Type IIn SN 2010jl: The First 1100 Days}",
      journal = {\apj},
         year = 2014,
        month = dec,
       volume = {797},
       number = {2},
          eid = {118},
        pages = {118},
          doi = {10.1088/0004-637X/797/2/118},
archivePrefix = {arXiv},
       eprint = {1312.6617},
 primaryClass = {astro-ph.HE},
       adsurl = {https://ui.adsabs.harvard.edu/abs/2014ApJ...797..118F}
}

@ARTICLE{Smith2008ApJ...680..568S,
       author = {{Smith}, Nathan and {Foley}, Ryan J. and {Filippenko}, Alexei V.},
        title = "{Dust Formation and He II {\ensuremath{\lambda}}4686 Emission in the Dense Shell of the Peculiar Type Ib Supernova 2006jc}",
      journal = {\apj},
         year = 2008,
        month = jun,
       volume = {680},
       number = {1},
        pages = {568-579},
          doi = {10.1086/587860},
archivePrefix = {arXiv},
       eprint = {0704.2249},
 primaryClass = {astro-ph},
       adsurl = {https://ui.adsabs.harvard.edu/abs/2008ApJ...680..568S}
}

@ARTICLE{Andrews2016MNRAS.457.3241A,
       author = {{Andrews}, J.~E. and {Krafton}, Kelsie M. and {Clayton}, Geoffrey C. and {Montiel}, E. and {Wesson}, R. and {Sugerman}, Ben E.~K. and {Barlow}, M.~J. and {Matsuura}, M. and {Drass}, H.},
        title = "{Early dust formation and a massive progenitor for SN 2011ja?}",
      journal = {\mnras},
         year = 2016,
        month = apr,
       volume = {457},
       number = {3},
        pages = {3241-3253},
          doi = {10.1093/mnras/stw164},
archivePrefix = {arXiv},
       eprint = {1509.06379},
 primaryClass = {astro-ph.SR},
       adsurl = {https://ui.adsabs.harvard.edu/abs/2016MNRAS.457.3241A}
}

@ARTICLE{Boian2020MNRAS.496.1325B,
       author = {{Boian}, Ioana and {Groh}, Jose H.},
        title = "{Progenitors of early-time interacting supernovae}",
      journal = {\mnras},
         year = 2020,
        month = aug,
       volume = {496},
       number = {2},
        pages = {1325-1342},
          doi = {10.1093/mnras/staa1540},
archivePrefix = {arXiv},
       eprint = {2001.07651},
 primaryClass = {astro-ph.SR},
       adsurl = {https://ui.adsabs.harvard.edu/abs/2020MNRAS.496.1325B}
}

@ARTICLE{Li2011MNRAS.412.1441L,
       author = {{Li}, Weidong and {Leaman}, Jesse and {Chornock}, Ryan and {Filippenko}, Alexei V. and {Poznanski}, Dovi and {Ganeshalingam}, Mohan and {Wang}, Xiaofeng and {Modjaz}, Maryam and {Jha}, Saurabh and {Foley}, Ryan J. and {Smith}, Nathan},
        title = "{Nearby supernova rates from the Lick Observatory Supernova Search - II. The observed luminosity functions and fractions of supernovae in a complete sample}",
      journal = {\mnras},
         year = 2011,
        month = apr,
       volume = {412},
       number = {3},
        pages = {1441-1472},
          doi = {10.1111/j.1365-2966.2011.18160.x},
archivePrefix = {arXiv},
       eprint = {1006.4612},
 primaryClass = {astro-ph.SR},
       adsurl = {https://ui.adsabs.harvard.edu/abs/2011MNRAS.412.1441L}
}

@ARTICLE{Ma2025A&A...698A.305M,
       author = {{Ma}, Xiaoran and {Wang}, Xiaofeng and {Mo}, Jun and {Howell}, D. Andrew and {Pellegrino}, Craig and {Zhang}, Jujia and {Yan}, Shengyu and {Arcavi}, Iair and {Chen}, Zhihao and {Farah}, Joseph and {Padilla Gonzalez}, Estefania and {Guo}, Fangzhou and {Hiramatsu}, Daichi and {Li}, Gaici and {Lin}, Han and {Liu}, Jialian and {McCully}, Curtis and {Newsome}, Megan and {Sai}, Hanna and {Terreran}, Giacomo and {Xiang}, Danfeng and {Zhang}, Xinhan and {Zhang}, Tianmeng},
        title = "{Supernovae at distances <40 Mpc: I. Catalogues and fractions of supernovae in a complete sample}",
      journal = {\aap},
         year = 2025,
        month = jun,
       volume = {698},
          eid = {A305},
        pages = {A305},
          doi = {10.1051/0004-6361/202452684},
archivePrefix = {arXiv},
       eprint = {2504.04393},
 primaryClass = {astro-ph.HE},
       adsurl = {https://ui.adsabs.harvard.edu/abs/2025A&A...698A.305M}
}

@ARTICLE{Ma2025A&A...698A.306M,
       author = {{Ma}, Xiaoran and {Wang}, Xiaofeng and {Mo}, Jun and {Andrew Howell}, D. and {Pellegrino}, Craig and {Zhang}, Jujia and {Wu}, Chengyuan and {Yan}, Shengyu and {Liu}, Dongdong and {Arcavi}, Iair and {Chen}, Zhihao and {Farah}, Joseph and {Padilla Gonzalez}, Estefania and {Guo}, Fangzhou and {Hiramatsu}, Daichi and {Li}, Gaici and {Lin}, Han and {Liu}, Jialian and {McCully}, Curtis and {Newsome}, Megan and {Sai}, Hanna and {Terreran}, Giacomo and {Xiang}, Danfeng and {Zhang}, Xinhan},
        title = "{Supernovae at distances <40 Mpc: II. Supernova rate in the local Universe}",
      journal = {\aap},
         year = 2025,
        month = jun,
       volume = {698},
          eid = {A306},
        pages = {A306},
          doi = {10.1051/0004-6361/202452685},
archivePrefix = {arXiv},
       eprint = {2504.04507},
 primaryClass = {astro-ph.HE},
       adsurl = {https://ui.adsabs.harvard.edu/abs/2025A&A...698A.306M}
}

@ARTICLE{Cappellaro2015A&A...584A..62C,
       author = {{Cappellaro}, E. and {Botticella}, M.~T. and {Pignata}, G. and {Grado}, A. and {Greggio}, L. and {Limatola}, L. and {Vaccari}, M. and {Baruffolo}, A. and {Benetti}, S. and {Bufano}, F. and {Capaccioli}, M. and {Cascone}, E. and {Covone}, G. and {De Cicco}, D. and {Falocco}, S. and {Della Valle}, M. and {Jarvis}, M. and {Marchetti}, L. and {Napolitano}, N.~R. and {Paolillo}, M. and {Pastorello}, A. and {Radovich}, M. and {Schipani}, P. and {Spiro}, S. and {Tomasella}, L. and {Turatto}, M.},
        title = "{Supernova rates from the SUDARE VST-OmegaCAM search. I. Rates per unit volume}",
      journal = {\aap},
         year = 2015,
        month = dec,
       volume = {584},
          eid = {A62},
        pages = {A62},
          doi = {10.1051/0004-6361/201526712},
archivePrefix = {arXiv},
       eprint = {1509.04496},
 primaryClass = {astro-ph.CO},
       adsurl = {https://ui.adsabs.harvard.edu/abs/2015A&A...584A..62C}
}

@ARTICLE{Eldridge2013MNRAS.436..774E,
       author = {{Eldridge}, John J. and {Fraser}, Morgan and {Smartt}, Stephen J. and {Maund}, Justyn R. and {Crockett}, R. Mark},
        title = "{The death of massive stars - II. Observational constraints on the progenitors of Type Ibc supernovae}",
      journal = {\mnras},
         year = 2013,
        month = nov,
       volume = {436},
       number = {1},
        pages = {774-795},
          doi = {10.1093/mnras/stt1612},
archivePrefix = {arXiv},
       eprint = {1301.1975},
 primaryClass = {astro-ph.SR},
       adsurl = {https://ui.adsabs.harvard.edu/abs/2013MNRAS.436..774E}
}

@ARTICLE{Smith2011MNRAS.412.1522S,
       author = {{Smith}, Nathan and {Li}, Weidong and {Filippenko}, Alexei V. and {Chornock}, Ryan},
        title = "{Observed fractions of core-collapse supernova types and initial masses of their single and binary progenitor stars}",
      journal = {\mnras},
         year = 2011,
        month = apr,
       volume = {412},
       number = {3},
        pages = {1522-1538},
          doi = {10.1111/j.1365-2966.2011.17229.x},
archivePrefix = {arXiv},
       eprint = {1006.3899},
 primaryClass = {astro-ph.HE},
       adsurl = {https://ui.adsabs.harvard.edu/abs/2011MNRAS.412.1522S}
}

@ARTICLE{Smartt2009ARA&A..47...63S,
       author = {{Smartt}, Stephen J.},
        title = "{Progenitors of Core-Collapse Supernovae}",
      journal = {\araa},
         year = 2009,
        month = sep,
       volume = {47},
       number = {1},
        pages = {63-106},
          doi = {10.1146/annurev-astro-082708-101737},
archivePrefix = {arXiv},
       eprint = {0908.0700},
 primaryClass = {astro-ph.SR},
       adsurl = {https://ui.adsabs.harvard.edu/abs/2009ARA&A..47...63S}
}

@ARTICLE{Nyholm2020A&A...637A..73N,
       author = {{Nyholm}, A. and {Sollerman}, J. and {Tartaglia}, L. and {Taddia}, F. and {Fremling}, C. and {Blagorodnova}, N. and {Filippenko}, A.~V. and {Gal-Yam}, A. and {Howell}, D.~A. and {Karamehmetoglu}, E. and {Kulkarni}, S.~R. and {Laher}, R. and {Leloudas}, G. and {Masci}, F. and {Kasliwal}, M.~M. and {Mor{\r{a}}}, K. and {Moriya}, T.~J. and {Ofek}, E.~O. and {Papadogiannakis}, S. and {Quimby}, R. and {Rebbapragada}, U. and {Schulze}, S.},
        title = "{Type IIn supernova light-curve properties measured from an untargeted survey sample}",
      journal = {\aap},
         year = 2020,
        month = may,
       volume = {637},
          eid = {A73},
        pages = {A73},
          doi = {10.1051/0004-6361/201936097},
archivePrefix = {arXiv},
       eprint = {1906.05812},
 primaryClass = {astro-ph.SR},
       adsurl = {https://ui.adsabs.harvard.edu/abs/2020A&A...637A..73N}
}

@ARTICLE{Fraser2020RSOS....700467F,
       author = {{Fraser}, Morgan},
        title = "{Supernovae and transients with circumstellar interaction}",
      journal = {Royal Society Open Science},
         year = 2020,
        month = jul,
       volume = {7},
       number = {7},
          eid = {200467},
        pages = {200467},
          doi = {10.1098/rsos.200467},
       adsurl = {https://ui.adsabs.harvard.edu/abs/2020RSOS....700467F}
}

@ARTICLE{Taddia2013A&A...555A..10T,
       author = {{Taddia}, F. and {Stritzinger}, M.~D. and {Sollerman}, J. and {Phillips}, M.~M. and {Anderson}, J.~P. and {Boldt}, L. and {Campillay}, A. and {Castell{\'o}n}, S. and {Contreras}, C. and {Folatelli}, G. and {Hamuy}, M. and {Heinrich-Josties}, E. and {Krzeminski}, W. and {Morrell}, N. and {Burns}, C.~R. and {Freedman}, W.~L. and {Madore}, B.~F. and {Persson}, S.~E. and {Suntzeff}, N.~B.},
        title = "{Carnegie Supernova Project: Observations of Type IIn supernovae}",
      journal = {\aap},
         year = 2013,
        month = jul,
       volume = {555},
          eid = {A10},
        pages = {A10},
          doi = {10.1051/0004-6361/201321180},
archivePrefix = {arXiv},
       eprint = {1304.3038},
 primaryClass = {astro-ph.CO},
       adsurl = {https://ui.adsabs.harvard.edu/abs/2013A&A...555A..10T}
}

@ARTICLE{Taddia2015A&A...580A.131T,
       author = {{Taddia}, F. and {Sollerman}, J. and {Fremling}, C. and {Pastorello}, A. and {Leloudas}, G. and {Fransson}, C. and {Nyholm}, A. and {Stritzinger}, M.~D. and {Ergon}, M. and {Roy}, R. and {Migotto}, K.},
        title = "{Metallicity at the explosion sites of interacting transients}",
      journal = {\aap},
         year = 2015,
        month = aug,
       volume = {580},
          eid = {A131},
        pages = {A131},
          doi = {10.1051/0004-6361/201525989},
archivePrefix = {arXiv},
       eprint = {1505.04719},
 primaryClass = {astro-ph.HE},
       adsurl = {https://ui.adsabs.harvard.edu/abs/2015A&A...580A.131T}
}

@ARTICLE{Kankare2012MNRAS.424..855K,
       author = {{Kankare}, E. and {Ergon}, M. and {Bufano}, F. and {Spyromilio}, J. and {Mattila}, S. and {Chugai}, N.~N. and {Lundqvist}, P. and {Pastorello}, A. and {Kotak}, R. and {Benetti}, S. and {Botticella}, M.-T. and {Cumming}, R.~J. and {Fransson}, C. and {Fraser}, M. and {Leloudas}, G. and {Miluzio}, M. and {Sollerman}, J. and {Stritzinger}, M. and {Turatto}, M. and {Valenti}, S.},
        title = "{SN 2009kn - the twin of the Type IIn supernova 1994W}",
      journal = {\mnras},
         year = 2012,
        month = aug,
       volume = {424},
       number = {2},
        pages = {855-873},
          doi = {10.1111/j.1365-2966.2012.21224.x},
archivePrefix = {arXiv},
       eprint = {1205.0353},
 primaryClass = {astro-ph.SR},
       adsurl = {https://ui.adsabs.harvard.edu/abs/2012MNRAS.424..855K}
}

@ARTICLE{Elias-Rosa2024A&A...686A..13E,
       author = {{Elias-Rosa}, N. and {Brennan}, S.~J. and {Benetti}, S. and {Cappellaro}, E. and {Pastorello}, A. and {Kozyreva}, A. and {Lundqvist}, P. and {Fraser}, M. and {Anderson}, J.~P. and {Cai}, Y.-Z. and {Chen}, T.-W. and {Dennefeld}, M. and {Gromadzki}, M. and {Guti{\'e}rrez}, C.~P. and {Ihanec}, N. and {Inserra}, C. and {Kankare}, E. and {Kotak}, R. and {Mattila}, S. and {Moran}, S. and {M{\"u}ller-Bravo}, T.~E. and {Pessi}, P.~J. and {Pignata}, G. and {Reguitti}, A. and {Reynolds}, T.~M. and {Smartt}, S.~J. and {Smith}, K. and {Tartaglia}, L. and {Valerin}, G. and {de Boer}, T. and {Chambers}, K. and {Gal-Yam}, A. and {Gao}, H. and {Geier}, S. and {Mazzali}, P.~A. and {Nicholl}, M. and {Ragosta}, F. and {Rest}, A. and {Yaron}, O. and {Young}, D.~R.},
        title = "{SN 2020pvb: A Type IIn-P supernova with a precursor outburst}",
      journal = {\aap},
         year = 2024,
        month = jun,
       volume = {686},
          eid = {A13},
        pages = {A13},
          doi = {10.1051/0004-6361/202348790},
archivePrefix = {arXiv},
       eprint = {2402.02924},
 primaryClass = {astro-ph.SR},
       adsurl = {https://ui.adsabs.harvard.edu/abs/2024A&A...686A..13E}
}

@ARTICLE{Mauerhan2013MNRAS.431.2599M,
       author = {{Mauerhan}, Jon C. and {Smith}, Nathan and {Silverman}, Jeffrey M. and {Filippenko}, Alexei V. and {Morgan}, Adam N. and {Cenko}, S. Bradley and {Ganeshalingam}, Mohan and {Clubb}, Kelsey I. and {Bloom}, Joshua S. and {Matheson}, Thomas and {Milne}, Peter},
        title = "{SN 2011ht: confirming a class of interacting supernovae with plateau light curves (Type IIn-P)}",
      journal = {\mnras},
         year = 2013,
        month = may,
       volume = {431},
       number = {3},
        pages = {2599-2611},
          doi = {10.1093/mnras/stt360},
archivePrefix = {arXiv},
       eprint = {1209.0821},
 primaryClass = {astro-ph.SR},
       adsurl = {https://ui.adsabs.harvard.edu/abs/2013MNRAS.431.2599M}
}

@ARTICLE{DiCarlo2002ApJ...573..144D,
       author = {{Di Carlo}, E. and {Massi}, F. and {Valentini}, G. and {Di Paola}, A. and {D'Alessio}, F. and {Brocato}, E. and {Guidubaldi}, D. and {Dolci}, M. and {Pedichini}, F. and {Speziali}, R. and {Li Causi}, G. and {Caratti o Garatti}, A. and {Cappellaro}, E. and {Turatto}, M. and {Arkharov}, A.~A. and {Gnedin}, Y. and {Larionov}, V.~M. and {Benetti}, S. and {Pastorello}, A. and {Aretxaga}, I. and {Chavushyan}, V. and {Vega}, O. and {Danziger}, I.~J. and {Tornamb{\'e}}, A.},
        title = "{Optical and Infrared Observations of the Supernova SN 1999el}",
      journal = {\apj},
         year = 2002,
        month = jul,
       volume = {573},
       number = {1},
        pages = {144-156},
          doi = {10.1086/340496},
archivePrefix = {arXiv},
       eprint = {astro-ph/0203041},
 primaryClass = {astro-ph},
       adsurl = {https://ui.adsabs.harvard.edu/abs/2002ApJ...573..144D}
}

@ARTICLE{Fassia2000MNRAS.318.1093F,
       author = {{Fassia}, A. and {Meikle}, W.~P.~S. and {Vacca}, W.~D. and {Kemp}, S.~N. and {Walton}, N.~A. and {Pollacco}, D.~L. and {Smartt}, S. and {Oscoz}, A. and {Arag{\'o}n-Salamanca}, A. and {Bennett}, S. and {Hawarden}, T.~G. and {Alonso}, A. and {Alcalde}, D. and {Pedrosa}, A. and {Telting}, J. and {Arevalo}, M.~J. and {Deeg}, H.~J. and {Garz{\'o}n}, F. and {G{\'o}mez-Rold{\'a}n}, A. and {G{\'o}mez}, G. and {Guti{\'e}rrez}, C. and {L{\'o}pez}, S. and {Rozas}, M. and {Serra-Ricart}, M. and {Zapatero-Osorio}, M.~R.},
        title = "{Optical and infrared photometry of the Type IIn SN 1998S: days 11-146}",
      journal = {\mnras},
         year = 2000,
        month = nov,
       volume = {318},
       number = {4},
        pages = {1093-1104},
          doi = {10.1046/j.1365-8711.2000.03797.x},
archivePrefix = {arXiv},
       eprint = {astro-ph/0006080},
 primaryClass = {astro-ph},
       adsurl = {https://ui.adsabs.harvard.edu/abs/2000MNRAS.318.1093F}
}

@ARTICLE{Turatto1993MNRAS.262..128T,
       author = {{Turatto}, M. and {Cappellaro}, E. and {Danziger}, I.~J. and {Benetti}, S. and {Gouiffes}, C. and {della Valle}, M.},
        title = "{The type II supernova 1988Z in MCG +03-28-022 : increasingevidence of interaction of supernova ejecta with a circumstellar wind.}",
      journal = {\mnras},
         year = 1993,
        month = may,
       volume = {262},
        pages = {128-140},
          doi = {10.1093/mnras/262.1.128},
       adsurl = {https://ui.adsabs.harvard.edu/abs/1993MNRAS.262..128T}
}

@ARTICLE{Zhang2012AJ....144..131Z,
       author = {{Zhang}, Tianmeng and {Wang}, Xiaofeng and {Wu}, Chao and {Chen}, Juncheng and {Chen}, Jia and {Liu}, Qin and {Huang}, Fang and {Liang}, Jide and {Zhao}, Xulin and {Lin}, Lin and {Wang}, Min and {Dennefeld}, Michel and {Zhang}, Jujia and {Zhai}, Meng and {Wu}, Hong and {Fan}, Zhou and {Zou}, Hu and {Zhou}, Xu and {Ma}, Jun},
        title = "{Type IIn Supernova SN 2010jl: Optical Observations for over 500 Days after Explosion}",
      journal = {\aj},
         year = 2012,
        month = nov,
       volume = {144},
       number = {5},
          eid = {131},
        pages = {131},
          doi = {10.1088/0004-6256/144/5/131},
archivePrefix = {arXiv},
       eprint = {1208.6078},
 primaryClass = {astro-ph.SR},
       adsurl = {https://ui.adsabs.harvard.edu/abs/2012AJ....144..131Z}
}

@ARTICLE{Stoll2011ApJ...730...34S,
       author = {{Stoll}, R. and {Prieto}, J.~L. and {Stanek}, K.~Z. and {Pogge}, R.~W. and {Szczygie{\l}}, D.~M. and {Pojma{\'n}ski}, G. and {Antognini}, J. and {Yan}, H.},
        title = "{SN 2010jl in UGC 5189: Yet Another Luminous Type IIn Supernova in a Metal-poor Galaxy}",
      journal = {\apj},
         year = 2011,
        month = mar,
       volume = {730},
       number = {1},
          eid = {34},
        pages = {34},
          doi = {10.1088/0004-637X/730/1/34},
archivePrefix = {arXiv},
       eprint = {1012.3461},
 primaryClass = {astro-ph.CO},
       adsurl = {https://ui.adsabs.harvard.edu/abs/2011ApJ...730...34S}
}

@ARTICLE{Fox2009ApJ...691..650F,
       author = {{Fox}, Ori and {Skrutskie}, Michael F. and {Chevalier}, Roger A. and {Kanneganti}, Srikrishna and {Park}, Chan and {Wilson}, John and {Nelson}, Matthew and {Amirhadji}, Jason and {Crump}, Danielle and {Hoeft}, Alexi and {Provence}, Sydney and {Sargeant}, Benjamin and {Sop}, Joel and {Tea}, Matthew and {Thomas}, Steven and {Woolard}, Kyle},
        title = "{Near-Infrared Photometry of the Type IIn SN 2005ip: The Case for Dust Condensation}",
      journal = {\apj},
         year = 2009,
        month = jan,
       volume = {691},
       number = {1},
        pages = {650-660},
          doi = {10.1088/0004-637X/691/1/650},
archivePrefix = {arXiv},
       eprint = {0807.3555},
 primaryClass = {astro-ph},
       adsurl = {https://ui.adsabs.harvard.edu/abs/2009ApJ...691..650F}
}

@ARTICLE{Fox2010ApJ...725.1768F,
       author = {{Fox}, Ori D. and {Chevalier}, Roger A. and {Dwek}, Eli and {Skrutskie}, Michael F. and {Sugerman}, Ben E.~K. and {Leisenring}, Jarron M.},
        title = "{Disentangling the Origin and Heating Mechanism of Supernova Dust: Late-time Spitzer Spectroscopy of the Type IIn SN 2005ip}",
      journal = {\apj},
         year = 2010,
        month = dec,
       volume = {725},
       number = {2},
        pages = {1768-1778},
          doi = {10.1088/0004-637X/725/2/1768},
archivePrefix = {arXiv},
       eprint = {1005.4682},
 primaryClass = {astro-ph.HE},
       adsurl = {https://ui.adsabs.harvard.edu/abs/2010ApJ...725.1768F}
}

@ARTICLE{Fox2011ApJ...741....7F,
       author = {{Fox}, Ori D. and {Chevalier}, Roger A. and {Skrutskie}, Michael F. and {Soderberg}, Alicia M. and {Filippenko}, Alexei V. and {Ganeshalingam}, Mohan and {Silverman}, Jeffrey M. and {Smith}, Nathan and {Steele}, Thea N.},
        title = "{A Spitzer Survey for Dust in Type IIn Supernovae}",
      journal = {\apj},
         year = 2011,
        month = nov,
       volume = {741},
       number = {1},
          eid = {7},
        pages = {7},
          doi = {10.1088/0004-637X/741/1/7},
archivePrefix = {arXiv},
       eprint = {1104.5012},
 primaryClass = {astro-ph.SR},
       adsurl = {https://ui.adsabs.harvard.edu/abs/2011ApJ...741....7F}
}

@ARTICLE{Miller2010AJ....139.2218M,
       author = {{Miller}, A.~A. and {Smith}, N. and {Li}, W. and {Bloom}, J.~S. and {Chornock}, R. and {Filippenko}, A.~V. and {Prochaska}, J.~X.},
        title = "{New Observations of the Very Luminous Supernova 2006gy: Evidence for Echoes}",
      journal = {\aj},
         year = 2010,
        month = jun,
       volume = {139},
       number = {6},
        pages = {2218-2229},
          doi = {10.1088/0004-6256/139/6/2218},
archivePrefix = {arXiv},
       eprint = {0906.2201},
 primaryClass = {astro-ph.CO},
       adsurl = {https://ui.adsabs.harvard.edu/abs/2010AJ....139.2218M}
}

@ARTICLE{Miller2010MNRAS.404..305M,
       author = {{Miller}, A.~A. and {Silverman}, J.~M. and {Butler}, N.~R. and {Bloom}, J.~S. and {Chornock}, R. and {Filippenko}, A.~V. and {Ganeshalingam}, M. and {Klein}, C.~R. and {Li}, W. and {Nugent}, P.~E. and {Smith}, N. and {Steele}, T.~N.},
        title = "{SN 2008iy: an unusual Type IIn Supernova with an enduring 400-d rise time}",
      journal = {\mnras},
         year = 2010,
        month = may,
       volume = {404},
       number = {1},
        pages = {305-317},
          doi = {10.1111/j.1365-2966.2010.16280.x},
archivePrefix = {arXiv},
       eprint = {0911.4719},
 primaryClass = {astro-ph.HE},
       adsurl = {https://ui.adsabs.harvard.edu/abs/2010MNRAS.404..305M}
}

@ARTICLE{Dwarkadas2011MNRAS.412.1639D,
       author = {{Dwarkadas}, V.~V.},
        title = "{On luminous blue variables as the progenitors of core-collapse supernovae, especially Type IIn supernovae}",
      journal = {\mnras},
         year = 2011,
        month = apr,
       volume = {412},
       number = {3},
        pages = {1639-1649},
          doi = {10.1111/j.1365-2966.2010.18001.x},
archivePrefix = {arXiv},
       eprint = {1011.3484},
 primaryClass = {astro-ph.SR},
       adsurl = {https://ui.adsabs.harvard.edu/abs/2011MNRAS.412.1639D}
}

@ARTICLE{Leloudas2015A&A...574A..61L,
       author = {{Leloudas}, G. and {Hsiao}, E.~Y. and {Johansson}, J. and {Maeda}, K. and {Moriya}, T.~J. and {Nordin}, J. and {Petrushevska}, T. and {Silverman}, J.~M. and {Sollerman}, J. and {Stritzinger}, M.~D. and {Taddia}, F. and {Xu}, D.},
        title = "{Supernova spectra below strong circumstellar interaction}",
      journal = {\aap},
         year = 2015,
        month = feb,
       volume = {574},
          eid = {A61},
        pages = {A61},
          doi = {10.1051/0004-6361/201322035},
archivePrefix = {arXiv},
       eprint = {1306.1549},
 primaryClass = {astro-ph.SR},
       adsurl = {https://ui.adsabs.harvard.edu/abs/2015A&A...574A..61L}
}

@ARTICLE{Moriya2013MNRAS.435.1520M,
       author = {{Moriya}, Takashi J. and {Maeda}, Keiichi and {Taddia}, Francesco and {Sollerman}, Jesper and {Blinnikov}, Sergei I. and {Sorokina}, Elena I.},
        title = "{An analytic bolometric light curve model of interaction-powered supernovae and its application to Type IIn supernovae}",
      journal = {\mnras},
         year = 2013,
        month = oct,
       volume = {435},
       number = {2},
        pages = {1520-1535},
          doi = {10.1093/mnras/stt1392},
archivePrefix = {arXiv},
       eprint = {1307.2644},
 primaryClass = {astro-ph.HE},
       adsurl = {https://ui.adsabs.harvard.edu/abs/2013MNRAS.435.1520M}
}

@ARTICLE{Kotak2006A&A...460L...5K,
       author = {{Kotak}, R. and {Vink}, J.~S.},
        title = "{Luminous blue variables as the progenitors of supernovae with quasi-periodic radio modulations}",
      journal = {\aap},
         year = 2006,
        month = dec,
       volume = {460},
       number = {2},
        pages = {L5-L8},
          doi = {10.1051/0004-6361:20065800},
archivePrefix = {arXiv},
       eprint = {astro-ph/0610095},
 primaryClass = {astro-ph},
       adsurl = {https://ui.adsabs.harvard.edu/abs/2006A&A...460L...5K}
}

@ARTICLE{Gal-Yam2009Natur.458..865G,
       author = {{Gal-Yam}, A. and {Leonard}, D.~C.},
        title = "{A massive hypergiant star as the progenitor of the supernova SN 2005gl}",
      journal = {\nat},
         year = 2009,
        month = apr,
       volume = {458},
       number = {7240},
        pages = {865-867},
          doi = {10.1038/nature07934},
       adsurl = {https://ui.adsabs.harvard.edu/abs/2009Natur.458..865G}
}

@ARTICLE{Salmaso2025A&A...695A..29S,
       author = {{Salmaso}, I. and {Cappellaro}, E. and {Tartaglia}, L. and {Anderson}, J.~P. and {Benetti}, S. and {Bronikowski}, M. and {Cai}, Y.-Z. and {Charalampopoulos}, P. and {Chen}, T.-W. and {Concepcion}, E. and {Elias-Rosa}, N. and {Galbany}, L. and {Gromadzki}, M. and {Guti{\'e}rrez}, C.~P. and {Kankare}, E. and {Lundqvist}, P. and {Matilainen}, K. and {Mazzali}, P.~A. and {Moran}, S. and {M{\"u}ller-Bravo}, T.~E. and {Nicholl}, M. and {Pastorello}, A. and {Pessi}, P.~J. and {Pessi}, T. and {Petrushevska}, T. and {Pignata}, G. and {Reguitti}, A. and {Sollerman}, J. and {Srivastav}, S. and {Stritzinger}, M. and {Tomasella}, L. and {Valerin}, G.},
        title = "{The diversity of strongly interacting Type IIn supernovae}",
      journal = {\aap},
         year = 2025,
        month = mar,
       volume = {695},
          eid = {A29},
        pages = {A29},
          doi = {10.1051/0004-6361/202451764},
archivePrefix = {arXiv},
       eprint = {2410.06111},
 primaryClass = {astro-ph.HE},
       adsurl = {https://ui.adsabs.harvard.edu/abs/2025A&A...695A..29S}
}

@ARTICLE{Reguitti2024A&A...686A.231R,
       author = {{Reguitti}, A. and {Pignata}, G. and {Pastorello}, A. and {Dastidar}, R. and {Reichart}, D.~E. and {Haislip}, J.~B. and {Kouprianov}, V.~V.},
        title = "{Searching for precursor activity of Type IIn supernovae}",
      journal = {\aap},
         year = 2024,
        month = jun,
       volume = {686},
          eid = {A231},
        pages = {A231},
          doi = {10.1051/0004-6361/202348679},
archivePrefix = {arXiv},
       eprint = {2403.10398},
 primaryClass = {astro-ph.HE},
       adsurl = {https://ui.adsabs.harvard.edu/abs/2024A&A...686A.231R}
}

@ARTICLE{Gall2011A&ARv..19...43G,
       author = {{Gall}, C. and {Hjorth}, J. and {Andersen}, A.~C.},
        title = "{Production of dust by massive stars at high redshift}",
      journal = {\aapr},
         year = 2011,
        month = sep,
       volume = {19},
          eid = {43},
        pages = {43},
          doi = {10.1007/s00159-011-0043-7},
archivePrefix = {arXiv},
       eprint = {1108.0403},
 primaryClass = {astro-ph.CO},
       adsurl = {https://ui.adsabs.harvard.edu/abs/2011A&ARv..19...43G}
}

@ARTICLE{Chandra2022MNRAS.517.4151C,
       author = {{Chandra}, Poonam and {Chevalier}, Roger A. and {James}, Nicholas J.~H. and {Fox}, Ori D.},
        title = "{The luminous type IIn supernova SN 2017hcc: Infrared bright, X-ray, and radio faint}",
      journal = {\mnras},
         year = 2022,
        month = dec,
       volume = {517},
       number = {3},
        pages = {4151-4161},
          doi = {10.1093/mnras/stac2915},
archivePrefix = {arXiv},
       eprint = {2210.03212},
 primaryClass = {astro-ph.HE},
       adsurl = {https://ui.adsabs.harvard.edu/abs/2022MNRAS.517.4151C}
}

@ARTICLE{Moriya2023A&A...677A..20M,
       author = {{Moriya}, Takashi J. and {Galbany}, Llu{\'\i}s and {Jim{\'e}nez-Palau}, Cristina and {Anderson}, Joseph P. and {Kuncarayakti}, Hanindyo and {S{\'a}nchez}, Sebasti{\'a}n F. and {Lyman}, Joseph D. and {Pessi}, Thallis and {Prieto}, Jose L. and {Kochanek}, Christopher S. and {Dong}, Subo and {Chen}, Ping},
        title = "{Environmental dependence of Type IIn supernova properties}",
      journal = {\aap},
         year = 2023,
        month = sep,
       volume = {677},
          eid = {A20},
        pages = {A20},
          doi = {10.1051/0004-6361/202346703},
archivePrefix = {arXiv},
       eprint = {2306.09647},
 primaryClass = {astro-ph.HE},
       adsurl = {https://ui.adsabs.harvard.edu/abs/2023A&A...677A..20M}
}

@ARTICLE{Perley2020ApJ...904...35P,
       author = {{Perley}, Daniel A. and {Fremling}, Christoffer and {Sollerman}, Jesper and {Miller}, Adam A. and {Dahiwale}, Aishwarya S. and {Sharma}, Yashvi and {Bellm}, Eric C. and {Biswas}, Rahul and {Brink}, Thomas G. and {Bruch}, Rachel J. and {De}, Kishalay and {Dekany}, Richard and {Drake}, Andrew J. and {Duev}, Dmitry A. and {Filippenko}, Alexei V. and {Gal-Yam}, Avishay and {Goobar}, Ariel and {Graham}, Matthew J. and {Graham}, Melissa L. and {Ho}, Anna Y.~Q. and {Irani}, Ido and {Kasliwal}, Mansi M. and {Kim}, Young-Lo and {Kulkarni}, S.~R. and {Mahabal}, Ashish and {Masci}, Frank J. and {Modak}, Shaunak and {Neill}, James D. and {Nordin}, Jakob and {Riddle}, Reed L. and {Soumagnac}, Maayane T. and {Strotjohann}, Nora L. and {Schulze}, Steve and {Taggart}, Kirsty and {Tzanidakis}, Anastasios and {Walters}, Richard S. and {Yan}, Lin},
        title = "{The Zwicky Transient Facility Bright Transient Survey. II. A Public Statistical Sample for Exploring Supernova Demographics}",
      journal = {\apj},
         year = 2020,
        month = nov,
       volume = {904},
       number = {1},
          eid = {35},
        pages = {35},
          doi = {10.3847/1538-4357/abbd98},
archivePrefix = {arXiv},
       eprint = {2009.01242},
 primaryClass = {astro-ph.HE},
       adsurl = {https://ui.adsabs.harvard.edu/abs/2020ApJ...904...35P}
}

@ARTICLE{Bevan2019MNRAS.485.5192B,
       author = {{Bevan}, A. and {Wesson}, R. and {Barlow}, M.~J. and {De Looze}, I. and {Andrews}, J.~E. and {Clayton}, G.~C. and {Krafton}, K. and {Matsuura}, M. and {Milisavljevic}, D.},
        title = "{A decade of ejecta dust formation in the Type IIn SN 2005ip}",
      journal = {\mnras},
         year = 2019,
        month = jun,
       volume = {485},
       number = {4},
        pages = {5192-5206},
          doi = {10.1093/mnras/stz679},
archivePrefix = {arXiv},
       eprint = {1809.09055},
 primaryClass = {astro-ph.SR},
       adsurl = {https://ui.adsabs.harvard.edu/abs/2019MNRAS.485.5192B}
}

@ARTICLE{Shahbandeh2025ApJ...985..262S,
       author = {{Shahbandeh}, Melissa and {Fox}, Ori D. and {Temim}, Tea and {Dwek}, Eli and {Sarangi}, Arkaprabha and {Smith}, Nathan and {Dessart}, Luc and {Nickson}, Bryony and {Engesser}, Michael and {Filippenko}, Alexei V. and {Brink}, Thomas G. and {Zheng}, WeiKang and {Szalai}, Tam{\'a}s and {Johansson}, Joel and {Rest}, Armin and {Van Dyk}, Schuyler D. and {Andrews}, Jennifer and {Ashall}, Chris and {Clayton}, Geoffrey C. and {De Looze}, Ilse and {DerKacy}, James M. and {Dulude}, Michael and {Foley}, Ryan J. and {Gezari}, Suvi and {Gomez}, Sebastian and {Gonzaga}, Shireen and {Indukuri}, Siva and {Jencson}, Jacob and {Kasliwal}, Mansi and {Lane}, Zachary G. and {Lau}, Ryan and {Law}, David and {Marston}, Anthony and {Milisavljevic}, Dan and {O'Steen}, Richard and {Pierel}, Justin and {Siebert}, Matthew and {Skrutskie}, Michael and {Strolger}, Lou and {Tinyanont}, Samaporn and {Wang}, Qinan and {Williams}, Brian and {Xiao}, Lin and {Yang}, Yi and {Zs{\'\i}ros}, Szanna},
        title = "{JWST/MIRI Observations of Newly Formed Dust in the Cold, Dense Shell of the Type IIn SN 2005ip}",
      journal = {\apj},
         year = 2025,
        month = jun,
       volume = {985},
       number = {2},
          eid = {262},
        pages = {262},
          doi = {10.3847/1538-4357/adce77},
archivePrefix = {arXiv},
       eprint = {2410.09142},
 primaryClass = {astro-ph.HE},
       adsurl = {https://ui.adsabs.harvard.edu/abs/2025ApJ...985..262S}
}

@ARTICLE{Dwek2021ApJ...917...84D,
       author = {{Dwek}, Eli and {Sarangi}, Arkaprabha and {Arendt}, Richard G. and {Kallman}, Timothy and {Kazanas}, Demos and {Fox}, Ori D.},
        title = "{The Infrared Echo of SN2010jl and Its Implications for Shock Breakout Characteristics}",
      journal = {\apj},
         year = 2021,
        month = aug,
       volume = {917},
       number = {2},
          eid = {84},
        pages = {84},
          doi = {10.3847/1538-4357/ac09ea},
archivePrefix = {arXiv},
       eprint = {2106.06531},
 primaryClass = {astro-ph.HE},
       adsurl = {https://ui.adsabs.harvard.edu/abs/2021ApJ...917...84D}
}

@ARTICLE{Bevan2020ApJ...894..111B,
       author = {{Bevan}, A.~M. and {Krafton}, K. and {Wesson}, R. and {Andrews}, J.~E. and {Montiel}, E. and {Niculescu-Duvaz}, M. and {Barlow}, M.~J. and {De Looze}, I. and {Clayton}, G.~C.},
        title = "{Disentangling Dust Components in SN 2010jl: The First 1400 Days}",
      journal = {\apj},
         year = 2020,
        month = may,
       volume = {894},
       number = {2},
          eid = {111},
        pages = {111},
          doi = {10.3847/1538-4357/ab86a2},
archivePrefix = {arXiv},
       eprint = {2004.01503},
 primaryClass = {astro-ph.SR},
       adsurl = {https://ui.adsabs.harvard.edu/abs/2020ApJ...894..111B}
}

@ARTICLE{Gall2014Natur.511..326G,
       author = {{Gall}, Christa and {Hjorth}, Jens and {Watson}, Darach and {Dwek}, Eli and {Maund}, Justyn R. and {Fox}, Ori and {Leloudas}, Giorgos and {Malesani}, Daniele and {Day-Jones}, Avril C.},
        title = "{Rapid formation of large dust grains in the luminous supernova 2010jl}",
      journal = {\nat},
         year = 2014,
        month = jul,
       volume = {511},
       number = {7509},
        pages = {326-329},
          doi = {10.1038/nature13558},
archivePrefix = {arXiv},
       eprint = {1407.4447},
 primaryClass = {astro-ph.SR},
       adsurl = {https://ui.adsabs.harvard.edu/abs/2014Natur.511..326G}
}

@ARTICLE{Maeda2013ApJ...776....5M,
       author = {{Maeda}, K. and {Nozawa}, T. and {Sahu}, D.~K. and {Minowa}, Y. and {Motohara}, K. and {Ueno}, I. and {Folatelli}, G. and {Pyo}, T.-S. and {Kitagawa}, Y. and {Kawabata}, K.~S. and {Anupama}, G.~C. and {Kozasa}, T. and {Moriya}, T.~J. and {Yamanaka}, M. and {Nomoto}, K. and {Bersten}, M. and {Quimby}, R. and {Iye}, M.},
        title = "{Properties of Newly Formed Dust Grains in the Luminous Type IIn Supernova 2010jl}",
      journal = {\apj},
         year = 2013,
        month = oct,
       volume = {776},
       number = {1},
          eid = {5},
        pages = {5},
          doi = {10.1088/0004-637X/776/1/5},
archivePrefix = {arXiv},
       eprint = {1308.0406},
 primaryClass = {astro-ph.SR},
       adsurl = {https://ui.adsabs.harvard.edu/abs/2013ApJ...776....5M}
}

@ARTICLE{Anderson2014MNRAS.441..671A,
       author = {{Anderson}, J.~P. and {Dessart}, L. and {Gutierrez}, C.~P. and {Hamuy}, M. and {Morrell}, N.~I. and {Phillips}, M. and {Folatelli}, G. and {Stritzinger}, M.~D. and {Freedman}, W.~L. and {Gonz{\'a}lez-Gait{\'a}n}, S. and et al.},
        title = "{Analysis of blueshifted emission peaks in Type II supernovae}",
      journal = {\mnras},
         year = 2014,
        month = jun,
       volume = {441},
       number = {1},
        pages = {671-680},
          doi = {10.1093/mnras/stu610},
archivePrefix = {arXiv},
       eprint = {1404.0581},
 primaryClass = {astro-ph.HE},
       adsurl = {https://ui.adsabs.harvard.edu/abs/2014MNRAS.441..671A}
}

@ARTICLE{Thevenot2021TNSAN.212....1T,
       author = {{Th{\'e}venot}, M. and {Kabatnik}, M. and {Gantier}, J.~M.},
        title = "{Supernovae detected in NEOWISE-R 2021}",
      journal = {Transient Name Server AstroNote},
         year = 2021,
        month = aug,
       volume = {212},
        pages = {1-212},
       adsurl = {https://ui.adsabs.harvard.edu/abs/2021TNSAN.212....1T}
}

@ARTICLE{Mainzer2014ApJ...792...30M,
       author = {{Mainzer}, A. and {Bauer}, J. and {Cutri}, R.~M. and {Grav}, T. and {Masiero}, J. and {Beck}, R. and {Clarkson}, P. and {Conrow}, T. and {Dailey}, J. and {Eisenhardt}, P. and {Fabinsky}, B. and {Fajardo-Acosta}, S. and {Fowler}, J. and {Gelino}, C. and {Grillmair}, C. and {Heinrichsen}, I. and {Kendall}, M. and {Kirkpatrick}, J. Davy and {Liu}, F. and {Masci}, F. and {McCallon}, H. and {Nugent}, C.~R. and {Papin}, M. and {Rice}, E. and {Royer}, D. and {Ryan}, T. and {Sevilla}, P. and {Sonnett}, S. and {Stevenson}, R. and {Thompson}, D.~B. and {Wheelock}, S. and {Wiemer}, D. and {Wittman}, M. and {Wright}, E. and {Yan}, L.},
        title = "{Initial Performance of the NEOWISE Reactivation Mission}",
      journal = {\apj},
         year = 2014,
        month = sep,
       volume = {792},
       number = {1},
          eid = {30},
        pages = {30},
          doi = {10.1088/0004-637X/792/1/30},
archivePrefix = {arXiv},
       eprint = {1406.6025},
 primaryClass = {astro-ph.EP},
       adsurl = {https://ui.adsabs.harvard.edu/abs/2014ApJ...792...30M}
}

@ARTICLE{Schulze2018MNRAS.473.1258S,
       author = {{Schulze}, S. and {Kr{\"u}hler}, T. and {Leloudas}, G. and {Gorosabel}, J. and {Mehner}, A. and {Buchner}, J. and {Kim}, S. and {Ibar}, E. and {Amor{\'\i}n}, R. and {Herrero-Illana}, R. and {Anderson}, J.~P. and {Bauer}, F.~E. and {Christensen}, L. and {de Pasquale}, M. and {de Ugarte Postigo}, A. and {Gallazzi}, A. and {Hjorth}, J. and {Morrell}, N. and {Malesani}, D. and {Sparre}, M. and {Stalder}, B. and {Stark}, A.~A. and {Th{\"o}ne}, C.~C. and {Wheeler}, J.~C.},
        title = "{Cosmic evolution and metal aversion in superluminous supernova host galaxies}",
      journal = {\mnras},
         year = 2018,
        month = jan,
       volume = {473},
       number = {1},
        pages = {1258-1285},
          doi = {10.1093/mnras/stx2352},
archivePrefix = {arXiv},
       eprint = {1612.05978},
 primaryClass = {astro-ph.GA},
       adsurl = {https://ui.adsabs.harvard.edu/abs/2018MNRAS.473.1258S}
}

@article{Arnett1989apj,
   author = {Arnett, W. David and Fu, Albert},
    title = "{The Late Behavior of Supernova 1987A. I. The Light Curve}",
  journal = {\apj},
     year = 1989,
    month = may,
   volume = {340},
    pages = {396},
      doi = {10.1086/167402}
}

@ARTICLE{Valerin2025A&A...695A..42V,
       author = {{Valerin}, G. and {Pastorello}, A. and {Reguitti}, A. and {Benetti}, S. and {Cai}, Y. -Z. and {Chen}, T. -W. and {Eappachen}, D. and {Elias-Rosa}, N. and {Fraser}, M. and {Gangopadhyay}, A. and {Hsiao}, E.~Y. and {Howell}, D.~A. and {Inserra}, C. and {Izzo}, L. and {Jencson}, J. and {Kankare}, E. and {Kotak}, R. and {Mazzali}, P.~A. and {Misra}, K. and {Pignata}, G. and {Prentice}, S.~J. and {Sand}, D.~J. and {Smartt}, S.~J. and {Stritzinger}, M.~D. and {Tartaglia}, L. and {Valenti}, S. and {Anderson}, J.~P. and {Andrews}, J.~E. and {Amaro}, R.~C. and {Brennan}, S. and {Bufano}, F. and {Callis}, E. and {Cappellaro}, E. and {Dastidar}, R. and {Della Valle}, M. and {Fiore}, A. and {Fulton}, M.~D. and {Galbany}, L. and {Heikkil{\"a}}, T. and {Hiramatsu}, D. and {Karamehmetoglu}, E. and {Kuncarayakti}, H. and {Leloudas}, G. and {Lundquist}, M. and {McCully}, C. and {M{\"u}ller-Bravo}, T.~E. and {Nicholl}, M. and {Ochner}, P. and {Padilla Gonzalez}, E. and {Paraskeva}, E. and {Pellegrino}, C. and {Rau}, A. and {Reichart}, D.~E. and {Reynolds}, T.~M. and {Roy}, R. and {Salmaso}, I. and {Singh}, M. and {Turatto}, M. and {Tomasella}, L. and {Wyatt}, S. and {Young}, D.~R.},
        title = "{A study in scarlet: I. Photometric properties of a sample of intermediate-luminosity red transients}",
      journal = {\aap},
         year = 2025,
        month = mar,
       volume = {695},
          eid = {A42},
        pages = {A42},
          doi = {10.1051/0004-6361/202451733},
       adsurl = {https://ui.adsabs.harvard.edu/abs/2025A&A...695A..42V}
}

@ARTICLE{Sarangi2025ApJ...993...94S,
       author = {{Sarangi}, Arkaprabha and {Zs{\'\i}ros}, Szanna and {Szalai}, Tam{\'a}s and {Martinez}, Laureano and {Shahbandeh}, Melissa and {Fox}, Ori D. and {Van Dyk}, Schuyler D. and {Filippenko}, Alexei V. and {Bersten}, Melina Cecilia and {De Looze}, Ilse and {Ashall}, Chris and {Temim}, Tea and {Jencson}, Jacob E. and {Rest}, Armin and {Milisavljevic}, Dan and {Dessart}, Luc and {Dwek}, Eli and {Smith}, Nathan and {Tinyanont}, Samaporn and {Brink}, Thomas G. and {Zheng}, WeiKang and {Clayton}, Geoffrey C. and {Andrews}, Jennifer},
        title = "{Two Decades of Dust Evolution in SN 2005af through JWST, Spitzer, and Chemical Modeling}",
      journal = {\apj},
         year = 2025,
        month = nov,
       volume = {993},
       number = {1},
          eid = {94},
        pages = {94},
          doi = {10.3847/1538-4357/ae0645},
archivePrefix = {arXiv},
       eprint = {2504.20574},
 primaryClass = {astro-ph.SR},
       adsurl = {https://ui.adsabs.harvard.edu/abs/2025ApJ...993...94S}
}

@ARTICLE{Fox2013AJ....146....2F,
       author = {{Fox}, Ori D. and {Filippenko}, Alexei V. and {Skrutskie}, Michael F. and {Silverman}, Jeffrey M. and {Ganeshalingam}, Mohan and {Cenko}, S. Bradley and {Clubb}, Kelsey I.},
        title = "{Late-time Circumstellar Interaction in a Spitzer Selected Sample of Type IIn Supernovae}",
      journal = {\aj},
         year = 2013,
        month = jul,
       volume = {146},
       number = {1},
          eid = {2},
        pages = {2},
          doi = {10.1088/0004-6256/146/1/2},
archivePrefix = {arXiv},
       eprint = {1304.0248},
 primaryClass = {astro-ph.SR},
       adsurl = {https://ui.adsabs.harvard.edu/abs/2013AJ....146....2F}
}

@ARTICLE{Szalai2019ApJS..241...38S,
       author = {{Szalai}, Tam{\'a}s and {Zs{\'\i}ros}, Szanna and {Fox}, Ori D. and {Pejcha}, Ond{\v{r}}ej and {M{\"u}ller}, Tom{\'a}s},
        title = "{A Comprehensive Analysis of Spitzer Supernovae}",
      journal = {\apjs},
         year = 2019,
        month = apr,
       volume = {241},
       number = {2},
          eid = {38},
        pages = {38},
          doi = {10.3847/1538-4365/ab10df},
archivePrefix = {arXiv},
       eprint = {1803.02571},
 primaryClass = {astro-ph.HE},
       adsurl = {https://ui.adsabs.harvard.edu/abs/2019ApJS..241...38S}
}

@ARTICLE{Moriya2013MNRAS.428.1020M,
       author = {{Moriya}, Takashi J. and {Blinnikov}, Sergei I. and {Tominaga}, Nozomu and {Yoshida}, Naoki and {Tanaka}, Masaomi and {Maeda}, Keiichi and {Nomoto}, Ken'ichi},
        title = "{Light-curve modelling of superluminous supernova 2006gy: collision between supernova ejecta and a dense circumstellar medium}",
      journal = {\mnras},
         year = 2013,
        month = jan,
       volume = {428},
       number = {2},
        pages = {1020-1035},
          doi = {10.1093/mnras/sts075},
archivePrefix = {arXiv},
       eprint = {1204.6109},
 primaryClass = {astro-ph.HE},
       adsurl = {https://ui.adsabs.harvard.edu/abs/2013MNRAS.428.1020M}
}

@ARTICLE{Dessart2015MNRAS.449.4304D,
       author = {{Dessart}, Luc and {Audit}, Edouard and {Hillier}, D. John},
        title = "{Numerical simulations of superluminous supernovae of type IIn}",
      journal = {\mnras},
         year = 2015,
        month = jun,
       volume = {449},
       number = {4},
        pages = {4304-4325},
          doi = {10.1093/mnras/stv609},
archivePrefix = {arXiv},
       eprint = {1503.05463},
 primaryClass = {astro-ph.SR},
       adsurl = {https://ui.adsabs.harvard.edu/abs/2015MNRAS.449.4304D}
}

@ARTICLE{Ofek2019PASP..131e4204O,
       author = {{Ofek}, E.~O. and {Zackay}, B. and {Gal-Yam}, A. and {Sollerman}, J. and {Fransson}, C. and {Fremling}, C. and {Kulkarni}, S.~R. and {Nugent}, P.~E. and {Yaron}, O. and {Kasliwal}, M.~M. and {Masci}, F. and {Laher}, R.},
        title = "{A Six-year Image-subtraction Light Curve of SN2010jl}",
      journal = {\pasp},
         year = 2019,
        month = may,
       volume = {131},
       number = {999},
        pages = {054204},
          doi = {10.1088/1538-3873/ab0a19},
archivePrefix = {arXiv},
       eprint = {1903.02016},
 primaryClass = {astro-ph.HE},
       adsurl = {https://ui.adsabs.harvard.edu/abs/2019PASP..131e4204O}
}

@ARTICLE{Chugai2018MNRAS.481.3643C,
       author = {{Chugai}, Nikolai N.},
        title = "{Type IIn SN 2010jl: probing dusty line-emitting shell}",
      journal = {\mnras},
         year = 2018,
        month = dec,
       volume = {481},
       number = {3},
        pages = {3643-3650},
          doi = {10.1093/mnras/sty2386},
archivePrefix = {arXiv},
       eprint = {1809.02478},
 primaryClass = {astro-ph.HE},
       adsurl = {https://ui.adsabs.harvard.edu/abs/2018MNRAS.481.3643C}
}

@ARTICLE{Jencson2016MNRAS.456.2622J,
       author = {{Jencson}, J.~E. and {Prieto}, J.~L. and {Kochanek}, C.~S. and {Shappee}, B.~J. and {Stanek}, K.~Z. and {Pogge}, R.~W.},
        title = "{Optical observations of the luminous Type IIn Supernova 2010jl for over 900 d}",
      journal = {\mnras},
         year = 2016,
        month = mar,
       volume = {456},
       number = {3},
        pages = {2622-2635},
          doi = {10.1093/mnras/stv2795},
archivePrefix = {arXiv},
       eprint = {1505.01186},
 primaryClass = {astro-ph.HE},
       adsurl = {https://ui.adsabs.harvard.edu/abs/2016MNRAS.456.2622J}
}

@ARTICLE{Smith2012AJ....143...17S,
       author = {{Smith}, Nathan and {Silverman}, Jeffrey M. and {Filippenko}, Alexei V. and {Cooper}, Michael C. and {Matheson}, Thomas and {Bian}, Fuyan and {Weiner}, Benjamin J. and {Comerford}, Julia M.},
        title = "{Systematic Blueshift of Line Profiles in the Type IIn Supernova 2010jl: Evidence for Post-shock Dust Formation?}",
      journal = {\aj},
         year = 2012,
        month = jan,
       volume = {143},
       number = {1},
          eid = {17},
        pages = {17},
          doi = {10.1088/0004-6256/143/1/17},
archivePrefix = {arXiv},
       eprint = {1108.2869},
 primaryClass = {astro-ph.HE},
       adsurl = {https://ui.adsabs.harvard.edu/abs/2012AJ....143...17S}
}

@ARTICLE{Andrews2011AJ....142...45A,
       author = {{Andrews}, J.~E. and {Clayton}, Geoffrey C. and {Wesson}, R. and {Sugerman}, B.~E.~K. and {Barlow}, M.~J. and {Clem}, J. and {Ercolano}, B. and {Fabbri}, J. and {Gallagher}, J.~S. and {Landolt}, A. and {Meixner}, M. and {Otsuka}, M. and {Riebel}, D. and {Welch}, D.~L.},
        title = "{Evidence for Pre-existing Dust in the Bright Type IIn SN 2010jl}",
      journal = {\aj},
         year = 2011,
        month = aug,
       volume = {142},
       number = {2},
          eid = {45},
        pages = {45},
          doi = {10.1088/0004-6256/142/2/45},
archivePrefix = {arXiv},
       eprint = {1106.0537},
 primaryClass = {astro-ph.CO},
       adsurl = {https://ui.adsabs.harvard.edu/abs/2011AJ....142...45A}
}

@ARTICLE{Smith2011ApJ...732...63S,
       author = {{Smith}, Nathan and {Li}, Weidong and {Miller}, Adam A. and {Silverman}, Jeffrey M. and {Filippenko}, Alexei V. and {Cuillandre}, Jean-Charles and {Cooper}, Michael C. and {Matheson}, Thomas and {Van Dyk}, Schuyler D.},
        title = "{A Massive Progenitor of the Luminous Type IIn Supernova 2010jl}",
      journal = {\apj},
         year = 2011,
        month = may,
       volume = {732},
       number = {2},
          eid = {63},
        pages = {63},
          doi = {10.1088/0004-637X/732/2/63},
archivePrefix = {arXiv},
       eprint = {1011.4150},
 primaryClass = {astro-ph.SR},
       adsurl = {https://ui.adsabs.harvard.edu/abs/2011ApJ...732...63S}
}

@ARTICLE{Sethulakshmi2026ApJ..1001..169S,
       author = {{Sethulakshmi}, V. and {Sutaria}, F.~K. and {Sharma}, R. and {Ray}, A.},
        title = "{SN 2017hcc and SN 2023usc{\textemdash}A Comparative Spectroscopic Study of Type IIn Supernovae}",
      journal = {\apj},
         year = 2026,
        month = apr,
       volume = {1001},
       number = {2},
          eid = {169},
        pages = {169},
          doi = {10.3847/1538-4357/ae47f6},
archivePrefix = {arXiv},
       eprint = {2602.19010},
 primaryClass = {astro-ph.HE},
       adsurl = {https://ui.adsabs.harvard.edu/abs/2026ApJ..1001..169S}
}

@ARTICLE{Smith2020MNRAS.499.3544S,
       author = {{Smith}, Nathan and {Andrews}, Jennifer E.},
        title = "{High-resolution spectroscopy of SN 2017hcc and its blueshifted line profiles from post-shock dust formation}",
      journal = {\mnras},
         year = 2020,
        month = dec,
       volume = {499},
       number = {3},
        pages = {3544-3562},
          doi = {10.1093/mnras/staa3047},
archivePrefix = {arXiv},
       eprint = {2009.14215},
 primaryClass = {astro-ph.HE},
       adsurl = {https://ui.adsabs.harvard.edu/abs/2020MNRAS.499.3544S}
}

@ARTICLE{Prieto2017RNAAS...1...28P,
       author = {{Prieto}, J.~L. and {Chen}, Ping and {Dong}, Subo and {Shappee}, B.~J. and {Seibert}, M. and {Bersier}, D. and {Holoien}, T.~W.-S. and {Kochanek}, C.~S. and {Stanek}, K.~Z. and {Thompson}, T.~A.},
        title = "{The Rise and Peak of the Luminous Type IIn SN 2017hcc/ATLAS17lsn from ASAS-SN and Swift UVOT Data}",
      journal = {Research Notes of the American Astronomical Society},
         year = 2017,
        month = dec,
       volume = {1},
       number = {1},
          eid = {28},
        pages = {28},
          doi = {10.3847/2515-5172/aa9c46},
archivePrefix = {arXiv},
       eprint = {1711.07938},
 primaryClass = {astro-ph.HE},
       adsurl = {https://ui.adsabs.harvard.edu/abs/2017RNAAS...1...28P}
}

@ARTICLE{Smith2011MNRAS.415..773S,
       author = {{Smith}, Nathan and {Li}, Weidong and {Silverman}, Jeffrey M. and {Ganeshalingam}, Mohan and {Filippenko}, Alexei V.},
        title = "{Luminous blue variable eruptions and related transients: diversity of progenitors and outburst properties}",
      journal = {\mnras},
         year = 2011,
        month = jul,
       volume = {415},
       number = {1},
        pages = {773-810},
          doi = {10.1111/j.1365-2966.2011.18763.x},
archivePrefix = {arXiv},
       eprint = {1010.3718},
 primaryClass = {astro-ph.SR},
       adsurl = {https://ui.adsabs.harvard.edu/abs/2011MNRAS.415..773S}
}

@ARTICLE{Jura1990ApJS...73..769J,
       author = {{Jura}, M. and {Kleinmann}, S.~G.},
        title = "{Mass-losing M Supergiants in the Solar Neighborhood}",
      journal = {\apjs},
         year = 1990,
        month = aug,
       volume = {73},
        pages = {769},
          doi = {10.1086/191488},
       adsurl = {https://ui.adsabs.harvard.edu/abs/1990ApJS...73..769J}
}

@ARTICLE{Jacobson2024ApJ...970..189J,
       author = {{Jacobson-Gal{\'a}n}, W.~V. and {Dessart}, L. and {Davis}, K.~W. and {Kilpatrick}, C.~D. and {Margutti}, R. and {Foley}, R.~J. and {Chornock}, R. and {Terreran}, G. and {Hiramatsu}, D. and {Newsome}, M. and {Padilla Gonzalez}, E. and {Pellegrino}, C. and {Howell}, D.~A. and {Filippenko}, A.~V. and {Anderson}, J.~P. and {Angus}, C.~R. and {Auchettl}, K. and {Bostroem}, K.~A. and {Brink}, T.~G. and {Cartier}, R. and {Coulter}, D.~A. and {de Boer}, T. and {Drout}, M.~R. and {Earl}, N. and {Ertini}, K. and {Farah}, J.~R. and {Farias}, D. and {Gall}, C. and {Gao}, H. and {Gerlach}, M.~A. and {Guo}, F. and {Haynie}, A. and {Hosseinzadeh}, G. and {Ibik}, A.~L. and {Jha}, S.~W. and {Jones}, D.~O. and {Langeroodi}, D. and {LeBaron}, N. and {Magnier}, E.~A. and {Piro}, A.~L. and {Raimundo}, S.~I. and {Rest}, A. and {Rest}, S. and {Rich}, R. Michael and {Rojas-Bravo}, C. and {Sears}, H. and {Taggart}, K. and {Villar}, V.~A. and {Wainscoat}, R.~J. and {Wang}, X.-F. and {Wasserman}, A.~R. and {Yan}, S. and {Yang}, Y. and {Zhang}, J. and {Zheng}, W.},
        title = "{Final Moments. II. Observational Properties and Physical Modeling of Circumstellar-material-interacting Type II Supernovae}",
      journal = {\apj},
         year = 2024,
        month = aug,
       volume = {970},
       number = {2},
          eid = {189},
        pages = {189},
          doi = {10.3847/1538-4357/ad4a2a},
archivePrefix = {arXiv},
       eprint = {2403.02382},
 primaryClass = {astro-ph.HE},
       adsurl = {https://ui.adsabs.harvard.edu/abs/2024ApJ...970..189J}
}

@ARTICLE{Hiramatsu2024arXiv241107287H,
       author = {{Hiramatsu}, Daichi and {Berger}, Edo and {Gomez}, Sebastian and {Blanchard}, Peter K. and {Kumar}, Harsh and {Athukoralalage}, Wasundara},
        title = "{Type IIn Supernovae. I. Uniform Light Curve Characterization and a Bimodality in the Radiated Energy Distribution}",
      journal = {arXiv e-prints},
         year = 2024,
        month = nov,
          eid = {arXiv:2411.07287},
        pages = {arXiv:2411.07287},
          doi = {10.48550/arXiv.2411.07287},
archivePrefix = {arXiv},
       eprint = {2411.07287},
 primaryClass = {astro-ph.HE},
       adsurl = {https://ui.adsabs.harvard.edu/abs/2024arXiv241107287H},
      note = {submitted to ApJ}
}

@ARTICLE{Radhakrishnan2024arXiv240717288R,
       author = {{Radhakrishnan Santhakumari}, Kalyan Kumar and {Battaini}, Federico and {Di Filippo}, Simone and {Di Rosa}, Silvio and {Cabona}, Lorenzo and {Claudi}, Riccardo and {Lessio}, Luigi and {Dima}, Marco and {Young}, David and {Landoni}, Marco and {Colapietro}, Mirko and {D'Orsi}, Sergio and {Aliverti}, Matteo and {Genoni}, Matteo and {Munari}, Matteo and {Zanmar Sanchez}, Ricardo and {Vitali}, Fabrizio and {Ricci}, Davide and {Schipani}, Pietro and {Campana}, Sergio and {Achren}, Jani and {Araiza-Duran}, Jose and {Arcavi}, Iair and {Baruffolo}, Andrea and {Ben-Ami}, Sagi and {Bitchkovsky}, Alex and {Brucalassi}, Anna and {Bruch}, Rachel and {Capasso}, Giulio and {Cappellaro}, Enrico and {Cosentino}, Rosario and {D'Alessio}, Francesco and {D'Avanzo}, Paolo and {Della Valle}, Massimo and {Di Benedetto}, Rosario and {Gal-Yam}, Avishay and {Hernandez Diaz}, Marcos and {Hershko}, Ofir and {Kotilainen}, Jari and {Kuncarayakti}, Hanindyo and {Li Causi}, Gianluca and {Marafatto}, Luca and {Martinetti}, Eugenio and {Marty}, Laurent and {Mattila}, Seppo and {Micciche}, Antonio and {Nicotra}, Gaetano and {Oggioni}, Luca and {Perez Ventura}, Hector and {Pariani}, Giorgio and {Pignata}, Giuliano and {Rappaport}, Michael and {Riva}, Marco and {Rubin}, Adam and {Salasnich}, Bernardo and {Savarese}, Salvatore and {Scuderi}, Salvatore and {Smartt}, Steven and {Stritzinger}, Maximilian},
        title = "{What is your favorite transient event? SOXS is almost ready to observe!}",
      journal = {arXiv e-prints},
         year = 2024,
        month = jul,
          eid = {arXiv:2407.17288},
        pages = {arXiv:2407.17288},
          doi = {10.48550/arXiv.2407.17288},
archivePrefix = {arXiv},
       eprint = {2407.17288},
 primaryClass = {astro-ph.IM},
       adsurl = {https://ui.adsabs.harvard.edu/abs/2024arXiv240717288R}
}

@ARTICLE{Anupama2001A&A...367..506A,
       author = {{Anupama}, G.~C. and {Sivarani}, T. and {Pandey}, G.},
        title = "{Early time optical spectroscopy of supernova SN 1998S}",
      journal = {\aap},
         year = 2001,
        month = feb,
       volume = {367},
        pages = {506-512},
          doi = {10.1051/0004-6361:20000427},
       adsurl = {https://ui.adsabs.harvard.edu/abs/2001A&A...367..506A}
}

@ARTICLE{Dessart2024arXiv240504259D,
       author = {{Dessart}, Luc},
        title = "{Interacting supernovae}",
      journal = {arXiv e-prints},
         year = 2024,
        month = may,
          eid = {arXiv:2405.04259},
        pages = {arXiv:2405.04259},
          doi = {10.48550/arXiv.2405.04259},
archivePrefix = {arXiv},
       eprint = {2405.04259},
 primaryClass = {astro-ph.SR},
       adsurl = {https://ui.adsabs.harvard.edu/abs/2024arXiv240504259D}
}

@ARTICLE{Dessart2022A&A...660L...9D,
       author = {{Dessart}, L. and {Hillier}, D. John},
        title = "{Modeling the signatures of interaction in Type II supernovae: UV emission, high-velocity features, broad-boxy profiles}",
      journal = {\aap},
         year = 2022,
        month = apr,
       volume = {660},
          eid = {L9},
        pages = {L9},
          doi = {10.1051/0004-6361/202243372},
archivePrefix = {arXiv},
       eprint = {2204.00446},
 primaryClass = {astro-ph.SR},
       adsurl = {https://ui.adsabs.harvard.edu/abs/2022A&A...660L...9D}
}

@ARTICLE{Gangopadhyay2024arXiv241104107G,
       author = {{Gangopadhyay}, Anjasha},
        title = "{Looking into the world of interacting supernovae}",
      journal = {arXiv e-prints},
         year = 2024,
        month = nov,
          eid = {arXiv:2411.04107},
        pages = {arXiv:2411.04107},
          doi = {10.48550/arXiv.2411.04107},
archivePrefix = {arXiv},
       eprint = {2411.04107},
 primaryClass = {astro-ph.HE},
       adsurl = {https://ui.adsabs.harvard.edu/abs/2024arXiv241104107G}
}

@ARTICLE{SN2014ab_Moriya2020A&A...641A.148M,
       author = {{Moriya}, T.~J. and {Stritzinger}, M.~D. and {Taddia}, F. and {Morrell}, N. and {Suntzeff}, N.~B. and {Contreras}, C. and {Gall}, C. and {Hjorth}, J. and {Ashall}, C. and {Burns}, C.~R. and {Busta}, L. and {Campillay}, A. and {Castell{\'o}n}, S. and {Corco}, C. and {Davis}, S. and {Galbany}, L. and {Gonz{\'a}lez}, C. and {Holmbo}, S. and {Hsiao}, E.~Y. and {Maund}, J.~R. and {Phillips}, M.~M.},
        title = "{The Carnegie Supernova Project II. Observations of SN 2014ab possibly revealing a 2010jl-like SN IIn with pre-existing dust}",
      journal = {\aap},
         year = 2020,
        month = sep,
       volume = {641},
          eid = {A148},
        pages = {A148},
          doi = {10.1051/0004-6361/202038118},
archivePrefix = {arXiv},
       eprint = {2006.10198},
 primaryClass = {astro-ph.HE},
       adsurl = {https://ui.adsabs.harvard.edu/abs/2020A&A...641A.148M}
}

@ARTICLE{Smith2009AJ....137.3558S,
       author = {{Smith}, Nathan and {Hinkle}, Kenneth H. and {Ryde}, Nils},
        title = "{Red Supergiants as Potential Type IIn Supernova Progenitors: Spatially Resolved 4.6 {\ensuremath{\mu}}m CO Emission Around VY CMa and Betelgeuse}",
      journal = {\aj},
         year = 2009,
        month = mar,
       volume = {137},
       number = {3},
        pages = {3558-3573},
          doi = {10.1088/0004-6256/137/3/3558},
archivePrefix = {arXiv},
       eprint = {0811.3037},
 primaryClass = {astro-ph},
       adsurl = {https://ui.adsabs.harvard.edu/abs/2009AJ....137.3558S}
}

@ARTICLE{Huang2018MNRAS.475.1261H,
       author = {{Huang}, Chenliang and {Chevalier}, Roger A.},
        title = "{Electron scattering wings on lines in interacting supernovae}",
      journal = {\mnras},
         year = 2018,
        month = mar,
       volume = {475},
       number = {1},
        pages = {1261-1273},
          doi = {10.1093/mnras/stx3163},
archivePrefix = {arXiv},
       eprint = {1712.01237},
 primaryClass = {astro-ph.HE},
       adsurl = {https://ui.adsabs.harvard.edu/abs/2018MNRAS.475.1261H}
}

@ARTICLE{Sarangi2022ApJ...933...89S,
       author = {{Sarangi}, Arkaprabha and {Slavin}, Jonathan D.},
        title = "{Dust Production in a Thin Dense Shell in Supernovae with Early Circumstellar Interactions}",
      journal = {\apj},
         year = 2022,
        month = jul,
       volume = {933},
       number = {1},
          eid = {89},
        pages = {89},
          doi = {10.3847/1538-4357/ac713d},
archivePrefix = {arXiv},
       eprint = {2205.08352},
 primaryClass = {astro-ph.SR},
       adsurl = {https://ui.adsabs.harvard.edu/abs/2022ApJ...933...89S}
}

@ARTICLE{Dessart2025A&A...698A.293D,
       author = {{Dessart}, Luc and {John Hillier}, D. and {Sarangi}, Arkaprabha},
        title = "{Radiative-transfer models for dusty Type II supernovae}",
      journal = {\aap},
         year = 2025,
        month = jun,
       volume = {698},
          eid = {A293},
        pages = {A293},
          doi = {10.1051/0004-6361/202555161},
archivePrefix = {arXiv},
       eprint = {2504.10928},
 primaryClass = {astro-ph.SR},
       adsurl = {https://ui.adsabs.harvard.edu/abs/2025A&A...698A.293D}
}

@ARTICLE{Mattila2008MNRAS.389..141M,
       author = {{Mattila}, S. and {Meikle}, W.~P.~S. and {Lundqvist}, P. and {Pastorello}, A. and {Kotak}, R. and {Eldridge}, J. and {Smartt}, S. and {Adamson}, A. and {Gerardy}, C.~L. and {Rizzi}, L. and {Stephens}, A.~W. and {van Dyk}, S.~D.},
        title = "{Massive stars exploding in a He-rich circumstellar medium - III. SN 2006jc: infrared echoes from new and old dust in the progenitor CSM}",
      journal = {\mnras},
         year = 2008,
        month = sep,
       volume = {389},
       number = {1},
        pages = {141-155},
          doi = {10.1111/j.1365-2966.2008.13516.x},
archivePrefix = {arXiv},
       eprint = {0803.2145},
 primaryClass = {astro-ph},
       adsurl = {https://ui.adsabs.harvard.edu/abs/2008MNRAS.389..141M}
}

@ARTICLE{Munch1948ApJ...108..116M,
       author = {{M{\"u}nch}, Guido},
        title = "{The Effect of Electron Scattering on the Line Spectrum of High-Temperature Stars.}",
      journal = {\apj},
         year = 1948,
        month = jul,
       volume = {108},
        pages = {116},
          doi = {10.1086/145048},
       adsurl = {https://ui.adsabs.harvard.edu/abs/1948ApJ...108..116M}
}

@ARTICLE{Smith2014ARA&A..52..487S,
       author = {{Smith}, Nathan},
        title = "{Mass Loss: Its Effect on the Evolution and Fate of High-Mass Stars}",
      journal = {\araa},
         year = 2014,
        month = aug,
       volume = {52},
        pages = {487-528},
          doi = {10.1146/annurev-astro-081913-040025},
archivePrefix = {arXiv},
       eprint = {1402.1237},
 primaryClass = {astro-ph.SR},
       adsurl = {https://ui.adsabs.harvard.edu/abs/2014ARA&A..52..487S}
}

@ARTICLE{Gal-Yam2007ApJ...656..372G,
       author = {{Gal-Yam}, Avishay and {Leonard}, D.~C. and {Fox}, D.~B. and {Cenko}, S.~B. and {Soderberg}, A.~M. and {Moon}, D.-S. and {Sand}, D.~J. and {Caltech Core Collapse Program} and {Li}, Weidong and {Filippenko}, Alexei V. and {Aldering}, G. and {Copin}, Y.},
        title = "{On the Progenitor of SN 2005gl and the Nature of Type IIn Supernovae}",
      journal = {\apj},
         year = 2007,
        month = feb,
       volume = {656},
       number = {1},
        pages = {372-381},
          doi = {10.1086/510523},
archivePrefix = {arXiv},
       eprint = {astro-ph/0608029},
 primaryClass = {astro-ph},
       adsurl = {https://ui.adsabs.harvard.edu/abs/2007ApJ...656..372G}
}

@ARTICLE{Smith2009ApJ...697L..49S,
       author = {{Smith}, Nathan and {Ganeshalingam}, Mohan and {Chornock}, Ryan and {Filippenko}, Alexei V. and {Li}, Weidong and {Silverman}, Jeffrey M. and {Steele}, Thea N. and {Griffith}, Christopher V. and {Joubert}, Niels and {Lee}, Nicholas Y. and {Lowe}, Thomas B. and {Mobberley}, Martin P. and {Winslow}, Dustin M.},
        title = "{SN 2008S: A Cool Super-Eddington Wind in a Supernova Impostor}",
      journal = {\apjl},
         year = 2009,
        month = may,
       volume = {697},
       number = {1},
        pages = {L49-L53},
          doi = {10.1088/0004-637X/697/1/L49},
archivePrefix = {arXiv},
       eprint = {0811.3929},
 primaryClass = {astro-ph},
       adsurl = {https://ui.adsabs.harvard.edu/abs/2009ApJ...697L..49S}
}

@ARTICLE{Botticella2009MNRAS.398.1041B,
       author = {{Botticella}, M.~T. and {Pastorello}, A. and {Smartt}, S.~J. and {Meikle}, W.~P.~S. and {Benetti}, S. and {Kotak}, R. and {Cappellaro}, E. and {Crockett}, R.~M. and {Mattila}, S. and {Sereno}, M. and {Patat}, F. and {Tsvetkov}, D. and {van Loon}, J. Th. and {Abraham}, D. and {Agnoletto}, I. and {Arbour}, R. and {Benn}, C. and {di Rico}, G. and {Elias-Rosa}, N. and {Gorshanov}, D.~L. and {Harutyunyan}, A. and {Hunter}, D. and {Lorenzi}, V. and {Keenan}, F.~P. and {Maguire}, K. and {Mendez}, J. and {Mobberley}, M. and {Navasardyan}, H. and {Ries}, C. and {Stanishev}, V. and {Taubenberger}, S. and {Trundle}, C. and {Turatto}, M. and {Volkov}, I.~M.},
        title = "{SN 2008S: an electron-capture SN from a super-AGB progenitor?}",
      journal = {\mnras},
         year = 2009,
        month = sep,
       volume = {398},
       number = {3},
        pages = {1041-1068},
          doi = {10.1111/j.1365-2966.2009.15082.x},
archivePrefix = {arXiv},
       eprint = {0903.1286},
 primaryClass = {astro-ph.SR},
       adsurl = {https://ui.adsabs.harvard.edu/abs/2009MNRAS.398.1041B}
}

@ARTICLE{Moriya2014A&A...569A..57M,
       author = {{Moriya}, Takashi J. and {Tominaga}, Nozomu and {Langer}, Norbert and {Nomoto}, Ken'ichi and {Blinnikov}, Sergei I. and {Sorokina}, Elena I.},
        title = "{Electron-capture supernovae exploding within their progenitor wind}",
      journal = {\aap},
         year = 2014,
        month = sep,
       volume = {569},
          eid = {A57},
        pages = {A57},
          doi = {10.1051/0004-6361/201424264},
archivePrefix = {arXiv},
       eprint = {1407.4563},
 primaryClass = {astro-ph.HE},
       adsurl = {https://ui.adsabs.harvard.edu/abs/2014A&A...569A..57M}
}
\onecolumn 
\begin{appendix}
\nolinenumbers

\section{Observations and data reduction}
\label{appendix:data}
\nolinenumbers
\subsection{Photometric data} \label{appendix:photodata}
Soon after the discovery announcement of SN\,2019vxm, we conducted a comprehensive multiband follow-up campaign in the Johnson-Cousins $UBVRI$ and Sloan $ugriz$ filters. The information on the instruments used are reported in Table~\ref{tab:table_telescope} (Appendix~\ref{appendix:LC_Obs}).

\textit{Swift}/UVOT UV and optical data were retrieved from the NASA \textit{Swift} Data Archive\footnote{\url{https://heasarc.gsfc.nasa.gov/cgi-bin/W3Browse/swift.pl}} and measured with the standard UVOT data-reduction pipeline {\tt HEASoft}\footnote{\url{https://heasarc.gsfc.nasa.gov/lheasoft/download.html}} \citep[version 6.19,][]{HEAsoft2014ascl.soft08004N}. 
The optical photometric data obtained with ground-based telescopes were reduced with the dedicated {\sl ecsnoopy}\footnote{{\sl ecsnoopy} is a package for SN photometry using PSF fitting and/or template subtraction developed by E. Cappellaro. A package description can be found at \url{https://sngroup.oapd.inaf.it/ecsnoopy.html}.} pipeline, following standard procedures as described by \citet[][]{Cai2018MNRAS.480.3424C}. 
We also collected archival data from public surveys, such as ASAS-SN, Asteroid Terrestrial-impact Last Alert System \citep[ATLAS;][]{Tonry2018PASP..130f4505T, Smith2020PASP..132h5002S, Shingles2021TNSAN...7....1S}, and Pan-STARRS \citep[e.g.,][]{Chambers2016arXiv161205560C,Flewelling2020ApJS..251....7F,Magnier2020ApJS..251....3M}. ATLAS \textit{orange} ($o$) and \textit{cyan} ($c$) magnitudes were processed through the ATLAS Forced Photometry service\footnote{\url{https://fallingstar-data.com/forcedphot/}} \citep{Shingles2021TNSAN...7....1S}, while Pan-STARRS1 (PS1) magnitudes were directly adopted from \citet[][]{Lane2026ApJ..1003...19L}. 
We retrieved single-exposure frames in the $W1$ (3.4 $\mu$m) and $W2$ (4.6 $\mu$m) filters from the NASA/IPAC Infrared Science Archive (IRSA), which were observed by the {\it Near-Earth Object WISE (NEOWISE)} Reactivation mission\footnote{While \cite{Thevenot2021TNSAN.212....1T} report that {\it NEOWISE-R} detected SN 2019vxm, we were not able to retrieve their measurements; therefore, we performed our own analysis of the original data.} \citep{Mainzer2014ApJ...792...30M}. During each passage, fields often receive multiple observations over a few days, which we coadded into a single stacked image. Finally, we performed point-spread-function (PSF) fitting photometry on template-subtracted images using pre-explosion archival images obtained in May 2019. The {\it WISE} magnitudes were calibrated against the {\it WISE} All-Sky Data Release catalogue \citep{Cutri2012yCat.2311....0C}. 
The final UV, optical, and MIR magnitudes of SN\,2019vxm are published at the Strasbourg astronomical Data Centre (CDS).

\subsection{Spectroscopic data}  \label{appendix:specdata}

Our spectroscopy of SN\,2019vxm was conducted through multiple telescopes and instruments: 
the 1.82\,m Copernico Telescope (CT) equipped with the Asiago Faint Object Spectrograph and Camera (AFOSC) and Echelle high-resolution spectrograph, located at the Asiago Observatory, Italy; the Beijing-Faint Object Spectrograph and Camera (BFOSC) mounted on the 2.16\,m Xinglong telescope (XLT) at the Xinglong Observatory, China; the Yunnan Faint Object Spectrograph and Camera (YFOSC) on the 2.4\,m Lijiang  telescope (LJT) at the Lijiang Observatory, China; the Kast double spectrograph \citep{Miller1993} mounted on the 3\,m Shane telescope at Lick Observatory, USA; the Astrophysical Research Consortium (ARC) 3.5\,m telescope with the Dual Imaging Spectrograph (DIS) at the Apache Point Observatory, New Mexico, USA; the Device Optimised for the LOw RESolution (DOLORES) on the 3.58\,m Telescopio Nazionale Galileo (TNG), and the 10.4\,m Gran Telescopio Canarias (GTC) with the Optical System for Imaging and low-Intermediate-Resolution Integrated Spectroscopy (OSIRIS); these last two telescopes are located at the Roque de los Muchachos Observatory, La Palma, Spain. In addition, we collected a single-epoch spectrum from the Transient Name Server (TNS\footnote{\url{https://www.wis-tns.org/object/2019vxm}}) captured at its earliest phase on 2019-12-01, which was observed by \citet{Leadbeater2019TNSCR2506....1L}. Basic information for the spectra of SN\,2019vxm is provided in Table~\ref{tab:speclog_2019vxm} (Appendix~\ref{appendix:SpecInfo}). 

The Shane/Kast spectrum was obtained at or near the parallactic angle to minimise slit losses caused by atmospheric dispersion \citep{Filippenko1982PASP...94..715F}. Its data reduction followed standard techniques for CCD frame processing and spectrum extraction \citep{Silverman2012MNRAS.425.1789S} using {\sc iraf} routines and custom \textsc{Python} and IDL codes\footnote{\url{https://github.com/ishivvers/TheKastShiv}}. The Copernico/AFOSC spectra were processed using the dedicated pipeline \texttt{Foscgui}\footnote{\url{https://sngroup.oapd.inaf.it/foscgui.html}}, while spectra collected from other telescopes/instruments were reduced following routine procedures within the \texttt{IRAF} environment. Specifically, raw data were first prereduced through preliminary tasks, including bias subtraction, overscan correction, trimming, and flat-fielding. One-dimensional (1D) spectra were then optimally extracted from the 2D frames. Flux and wavelength calibrations were performed using spectra from spectrophotometric standard stars and comparison lamps, respectively. Subsequently, the strongest telluric absorption bands (e.g.,  $\mathrm{H}_2\mathrm{O}$ and $\mathrm{O}_2$) in the SN spectra were removed using the spectra of standard stars. Finally, the spectral flux was fine-tuned by comparing it with coeval broad-band photometry. 

\onecolumn
\newpage
\section{Photometric facilities}
\label{appendix:LC_Obs}
\nolinenumbers
\vspace{-1.0em}
\begin{table}[htbp]
\centering
\caption{Information on the instrumental setups.}
\label{tab:table_telescope}
\scalebox{0.8}{
\begin{tabular}{@{}llllll@{}}
\hline \hline
Code & Diameter&Telescope & Instrument &Filters& Site \\
                & $\mathrm{(m)}$&            &                   & \\
\hline
Moravian   & 0.67/0.92 & Schmidt Telescope         & Moravian   &$uBVgri$  &Osservatorio Astronomico di Asiago, Asiago, Italy\\
Andor      & 0.80      & Tsinghua-NAOC Telescope (TNT)  & Andor DZ936      &$Bgri$& Xinglong Observatory, Hebei, P.R. China\\
LAIA       & 0.80      & Joan Or\'o Telescope (TJO)& LAIA      &$UBVRI$& Montsec Astronomical Observatory, Catalonia, Spain \\
Nickel     & 1.02      & Nickel Telescope          & Nickel     &$BVRI$& Lick Observatory, California, USA\\
AFOSC      & 1.82      & Copernico 1.82m Telescope & AFOSC      &$BVgriz$&  Osservatorio Astronomico di Asiago, Asiago, Italy\\
IO:O       & 2.00      & Liverpool Telescope (LT)  & IO:O       & $uBVgriz$& Observatorio Roque de Los Muchachos, La Palma, Spain\\
LJT        & 2.40      & Lijiang 2.4\,m Telescope  & YFOSC      &$BVgri$& Gaomeigu site, Lijiang Observatory (LJO), Yunnan, P.R. China\\
ALFOSC     & 2.56      & Nordic Optical Telescope  & ALFOSC     &$BVgriz$&  Observatorio Roque de Los Muchachos, La Palma, Spain\\             
ATLAS      & 0.50      & ATLAS                     & ATLAS       &$co$& Haleakal\={a} Observatory and Mauna Loa Observatory, Hawaii, USA\\
PS1        & 1.80      & Pan-STARRS1               & GPC1        &$i$& Haleakal\={a} Observatory, Maui, Hawaii, USA \\
UVOT       & 0.30      & {\it Swift} Modified Ritchey-Chr\'{e}tien     & UVOT  & $UV$ filters+$ubv$&{\it Neil Gehrels Swift Observatory} \\
{\it WISE}       & 0.40      & {\it Wide-field Infrared Survey Explorer}     &    WISE  &$W1+W2$& {\it Wide-field Infrared Survey Explorer}  \\
\hline \hline
\end{tabular}
}
\vspace{0.3em} 
\begin{minipage}{\columnwidth} 
\end{minipage}
\end{table}
\vspace{-2.5em}

\section{Log of spectroscopic observations}
\label{appendix:SpecInfo}
\nolinenumbers
\begin{table*}[htbp]
\centering
\caption{Log of the spectroscopic observations of SN~2019vxm.}
\label{tab:speclog_2019vxm}
\small
\setlength{\tabcolsep}{5pt}
\begin{tabular}{cccccccc}
\hline\hline
Date & MJD & Phase$^{a}$ & Instrumental setup & Grism/Grating & Spectral range & Exposure time & Resolution \\
     &     & (d) & & & (\AA) & (s) & (\AA) \\
\hline
2019-12-01 & 58818.73 & $+14.73$ & C11+ALPY200 & 200 $\mathrm{mm^{-1}}$ & 3550--7440 & 4200 & $R \approx 130$ \\
2019-12-04 & 58821.68 & $+17.68$ & Copernico+AFOSC & VPH7+6  & 3150--9290 & 1500 & 15 \\
2019-12-05 & 58822.70  & $+18.70$ & Copernico+AFOSC & VPH7+6  & 4990--9290 & 1500 & 15 \\
2019-12-06 & 58823.48  & $+19.48$ & XLT+BFOSC & G4    & 3940--8810 & 2400 & $R \approx 350$ \\
2019-12-06 & 58823.73  & $+19.73$ & Copernico+ECHELLE & 300    & 3840--7150 & 3600 & $R \approx 20,000$ \\
2019-12-08 & 58825.41  & $+21.41$ & XLT+BFOSC & G4    & 3970--8810 & 2400 &  $R \approx 350$ \\
2019-12-10 & 58827.41  & $+23.41$ & XLT+BFOSC &  G4   & 3950--8810 & 3000 &  $R \approx 350$ \\
2019-12-11 & 58828.42  & $+24.42$ & XLT+BFOSC &  G4   & 3970--8810 & 3000 &  $R \approx 350$ \\
2019-12-12 & 58829.43  & $+25.43$ & XLT+BFOSC & G4    & 3980--8810 & 3000 &  $R \approx 350$ \\
2019-12-13 & 58830.43  & $+26.43$ & XLT+BFOSC & G4    & 3970--8810 & 3000 &  $R \approx 350$ \\
2019-12-19 & 58836.44  & $+32.44$ & XLT+BFOSC &  G4   & 3970--8810 & 3000 &  $R \approx 350$ \\
2019-12-22 & 58839.45  & $+35.45$ & XLT+BFOSC & G4    & 3970--8810 & 3000 &  $R \approx 350$ \\
2020-01-01 & 58849.44  & $+45.44$ & XLT+BFOSC & G4    & 3970--8810 & 3000 & $R \approx 350$ \\
2020-01-02 & 58850.72  & $+46.72$ & Copernico+AFOSC & VPH7+6  & 3390--9290 & 1800 &  15 \\
2020-01-05 & 58853.10  & $+49.10$ & Shane+Kast & 4310/7500 & 3620--10700 & 1680 & 9 \\
2020-02-01 & 58880.16  & $+76.16$ & Copernico+AFOSC & VPH7+6  & 3450--8940 & 1800 &  15 \\
2020-04-02 & 58941.09  & $+137.09$ & Copernico+AFOSC & VPH7  & 3420--7280 & 1800 & 14 \\
2020-04-19 & 58958.13  & $+154.13$ & Copernico+AFOSC & VPH7  & 3700--7290 & 600 & 14 \\
2020-04-30 & 58970.20  & $+166.20$ & GTC+OSIRIS & R2500R  & 5590--7680 & 900  & 3.4 \\
2020-05-28 & 58997.09  & $+193.09$ & Copernico+AFOSC & VPH6  & 5000--9300 & 1500 & 15 \\
2020-06-21 & 59021.03  & $+217.03$ & Copernico+AFOSC & VPH7+6  & 3450--7290 & 600 &  15\\
2020-07-09 & 59039.46  & $+235.46$ &  Shane+Kast &  4310/7500  & 3620--10710 & 2760/2700 & 9 \\
2020-07-17 & 59047.10  & $+243.10$ &  TNG+DOLORES  & LRB   & 3300--8000 & 900 & 11 \\
2020-08-11 & 59073.00  & $+269.00$ &  TNG+DOLORES  & LRB+LRR  & 3220--10000 & 1200 & 11 \\
2020-08-21 & 59082.03  & $+278.03$ &  TNG+DOLORES  & LRB   & 3110--7860 & 1200 & 11 \\
2020-08-24 & 59085.38  & $+281.38$ & ARC+DIS &  B400+R300 & 3750--8260 & 3000/3000& 13\\
2020-09-12 & 59104.01  & $+300.01$ & Copernico+AFOSC & VPH7+6  & 3700--9300 & 1200 & 15 \\
2020-10-11 & 59133.52  & $+329.52$ & LJT+YFOSC  & G14  & 3810--7440 & 2100 & 15 \\
2020-11-03 & 59156.59  & $+352.59$ & LJT+YFOSC  & G14  & 3490--8740 & 2100 & 15 \\
2020-11-16 & 59169.81  & $+365.81$ & Copernico+AFOSC & VPH7+6  & 3600--9300 & 1200 & 15 \\
2020-11-17 & 59170.10  & $+366.10$ &  Shane+Kast &  4310/7500  & 3620--10700 & 2460/2400 & 9 \\
2020-11-29 & 59182.76  & $+378.76$ & Copernico+AFOSC & Gr4  & 3400--8200 & 1500 & 14 \\
2021-01-08 & 59222.10  & $+418.10$ &  Shane+Kast &  4310/7500  & 3620--10700 & 2160/2220 & 9 \\
2021-03-19 & 59292.20  & $+488.20$ & Copernico+AFOSC & Gr4  & 3600--9670 & 1200 & 14 \\
2021-04-18 & 59322.50  & $+518.50$ &  Shane+Kast &  4310/7500  & 3630--10700 & 3660/3600 & 9 \\
2021-05-04 & 59339.15  & $+535.15$ &  TNG+DOLORES & LRB+LRR   & 3550--10380 & 1200 & 12 \\
2021-07-06 & 59401.02  & $+597.02$ &  TNG+DOLORES & LRB+LRR   & 3310--10250 & 2700 & 12 \\
2021-09-01 & 59458.86  & $+654.86$ & Copernico+AFOSC & Gr4  & 3580--8200 & 3600 & 14 \\
2021-09-11 & 59468.20  & $+664.20$ &  Shane+Kast & 4310/7500   & 3630--10760 &  3360/3300 & 9 \\
2021-10-14 &  59502.14 & $+698.14$ &  Shane+Kast & 4310/7500  & 3620--10270 & 3660/3600 & 9 \\

\hline\hline
\end{tabular}
\vspace{0.3em}
\begin{minipage}{\linewidth}
\footnotesize
$^{a}$Phases are relative to explosion epoch (MJD = 58804.0) in the observer frame.
\end{minipage}
\end{table*}

\newpage
\section{Supplementary Materials}
\label{sect:SuppMaterial}
\nolinenumbers
\begin{table*}[htbp]
    \centering
    \caption{Basic information of the sample SNe.}
\resizebox{\textwidth}{!}{
    \begin{tabular}{llllllllll}
        \hline\hline
        Object & Explosion Epoch & Peak Epoch & Peak Mag.$^\dagger$&$L_{\rm peak}$& Redshift & Distance & $E(B-V)_\mathrm{MW}$ & $E(B-V)_\mathrm{Host}$ & Source \\
         & [MJD] & [MJD] &[mag]&[erg s$^{-1}$]& $z$ & [Mpc] & [mag] & [mag] & \\
        \hline
        \vspace{0.1em}
        SN~2003ma & 52971 & 52992 & $\sim-22.2$ &$ 5.78\times 10^{43}$& 0.289 & 1534 & 0.348 & $\sim 0$ & 1 \\
        SN~2006gy & 53967 & 54035 & $\sim-22.4$ &$5.94 \times 10^{43}$&0.019 & $76\pm14$ & 0.16 & $\sim 0.54$ & 4,5 \\
        SN~2006tf & -- & 54112 & $\sim-21.1$ &$ 2.39\times 10^{43}$&0.074 & 346 & 0.027 & $\sim 0$ & 6 \\
        SN~2008am & $54438.8\pm1$  &54505& $\sim-21.3$ &$ 9.02\times 10^{43}$& 0.2338 & 1205 & 0.025 & $\sim 0$ & 7 \\
        SN~2010jl & 55470 & 55494 & $\sim-19.4$ &$ 1.39\times 10^{43}$&0.0107 & 49 & 0.027 & $\sim 0$ & 8 \\
        SN 2014ab & $\sim 56625.4$ & $\le 56671.1$ & $\le -19.1$ &$ 1.00\times 10^{43}$& 0.02262 & 94.3 & 0.027 & $\sim 0$ & 9 \\
        SN~2015da & 57030.95 & 57131 & $\sim-20.0$ &$ 2.19\times 10^{43}$&0.006669
 & 32.1 & 0.01 & $0.97 \pm 0.27$ & 10,11 \\
        SN~2017hcc & $58027.4\pm1$  &58084 & $\sim-21.0$ &$ 3.96\times 10^{43}$& 0.0168 & 77.57 & 0.029 & $\lesssim 0.016$ & 12 \\
        SN~2019vxm & 58804 & 58839 & $\sim-20.3$ & $1.59 \times 10^{43}$& $0.019$ & $79.2 \pm 4.2$ & $0.087 \pm 0.002$ & $\sim0$ & This work \\
        \hline
    \end{tabular}}%
    \label{apptab:supm:cmpsinfo}
    \vspace{2pt}
    \scriptsize 
    \raggedright
    
    \textbf{NOTE:} Sources: 
    1 = \cite{SN2003ma_Rest2011ApJ...729...88R}; 
    4 = \cite{SN2006gy_Ofek2007ApJ...659L..13O}; 
    5 = \cite{SN2006gy_Smith2007ApJ...666.1116S}; 
    6 = \cite{SN2006tf_Smith2008ApJ...686..467S}; 
    7 = \cite{SN2008am_Chatzopoulos2011ApJ...729..143C}; 
    8 = \cite{SN2010jl_Ofek2014ApJ...781...42O};
    9 = \cite{SN2014ab_Moriya2020A&A...641A.148M};
    10 = \cite{SN2015da_Smith2024MNRAS.530..405S};
    11 = \cite{SN2015da_Tartaglia2020A&A...635A..39T};
    12 = \cite{SN2017hcc_Moran2023A&A...669A..51M}.
    $\dagger$: The peak magnitudes are obtained from the optical bands with best coverage.
\end{table*}


\begin{table*}[htbp]
\centering

\begin{minipage}[t]{0.34\textwidth}
\centering
\caption{Decline rates (in $\rm mag\, /100\,d$) of the multiband light curves of SN\,2019vxm.}
\label{tab:decline_rate}
\renewcommand{\arraystretch}{1.1}
\setlength{\tabcolsep}{8pt}
\scriptsize
\begin{tabular}{cccc}
\hline\hline
Filter & $\gamma_{0-100}$ & $\gamma_{100-400}$ & $\gamma_{400-700}$\\
\hline
$u$&1.70 $\pm$ 0.03&0.66 $\pm$ 0.02&$-$\\
$U$&1.95 $\pm$ 0.09&$-$&$-$\\
$B$&1.32 $\pm$ 0.03&0.64 $\pm$ 0.02&1.35 $\pm$ 0.05\\
$g$&1.14 $\pm$ 0.04&0.60 $\pm$ 0.01&1.38 $\pm$ 0.03\\
$c$&$-$&0.56 $\pm$ 0.03&1.24 $\pm$ 0.09\\
$V$&1.06 $\pm$ 0.04&0.56 $\pm$ 0.02&1.28 $\pm$ 0.02\\
$r$&0.61 $\pm$ 0.03&0.25 $\pm$ 0.01&0.83 $\pm$ 0.03\\
$o$&0.47 $\pm$ 0.01&0.34 $\pm$ 0.01&1.02 $\pm$ 0.03\\
$i$&0.59 $\pm$ 0.05&0.58 $\pm$ 0.01&1.09 $\pm$ 0.02\\
$z$&0.61 $\pm$ 0.03&0.55 $\pm$ 0.02&1.10 $\pm$ 0.02\\
\hline
Filter & \multicolumn{3}{c}{$\gamma_{600-800}$}\\
\hline
$R$ & \multicolumn{3}{c}{0.83 $\pm$ 0.03}\\
$I$ & \multicolumn{3}{c}{1.21 $\pm$ 0.04}\\
\hline
Filter & \multicolumn{3}{c}{$\gamma_{0-40}$}\\
\hline
$UVW1$ & \multicolumn{3}{c}{3.48 $\pm$ 0.11}\\
$UVM2$ & \multicolumn{3}{c}{4.55 $\pm$ 0.13}\\
$UVW2$ & \multicolumn{3}{c}{4.96 $\pm$ 0.11}\\
\hline
Filter & \multicolumn{3}{c}{$\gamma_{1000-1700}$}\\
\hline
$W1$ & \multicolumn{3}{c}{0.24 $\pm$ 0.03}\\
$W2$ & \multicolumn{3}{c}{0.12 $\pm$ 0.03}\\
\hline\hline
\end{tabular}
\end{minipage}
\hfill
\begin{minipage}[t]{0.62\textwidth}
\centering
\caption{Dust properties at different phases since the explosion epoch.}
\label{tab:dust_properties}
\scriptsize
\setlength{\tabcolsep}{4pt}
\begin{tabular}{l c cc cc cc}
\toprule
\multirow{2}{*}{MJD} & \multirow{2}{*}{Phase} & \multicolumn{2}{c}{Standard Black Body} & \multicolumn{2}{c}{Graphite $a=0.1\,\mu$m} & \multicolumn{2}{c}{Graphite $a=1.0\,\mu$m} \\
\cmidrule(lr){3-4} \cmidrule(lr){5-6} \cmidrule(lr){7-8}
& & $R$ (AU) & $T$ (K) & $M_d$ (M$_\odot$) & $T$ (K) & $M_d$ (M$_\odot$) & $T$ (K) \\
\midrule
59012.58 & 208.6 & 1486 & 1296 & $1.05 \times 10^{-3}$ & 767 & $1.20 \times 10^{-4}$ & 1210 \\
59174.82 & 370.8 & 1926 & 1059 & $1.64 \times 10^{-3}$ & 684 & $1.98 \times 10^{-4}$ & 1003 \\
59379.77 & 575.7 & 1685 & 1447 & $1.42 \times 10^{-3}$ & 812 & $1.55 \times 10^{-4}$ & 1338 \\
59542.04 & 738.0 & 1471 & 1429 & $1.08 \times 10^{-3}$ & 807 & $1.18 \times 10^{-4}$ & 1323 \\
59743.89 & 939.9 & 1518 & 1267 & $1.09 \times 10^{-3}$ & 758 & $1.25 \times 10^{-4}$ & 1185 \\
59906.32 & 1102.3 & 2302 & 935 & $2.26 \times 10^{-3}$ & 632 & $2.81 \times 10^{-4}$ & 892 \\
60105.86 & 1301.8 & 1838 & 950 & $1.44 \times 10^{-3}$ & 639 & $1.79 \times 10^{-4}$ & 906 \\
60273.36 & 1469.3 & 2234 & 817 & $2.06 \times 10^{-3}$ & 578 & $2.63 \times 10^{-4}$ & 785 \\
60475.46 & 1671.4 & 4629 & 594 & $8.54 \times 10^{-3}$ & 458 & $1.12 \times 10^{-3}$ & 577 \\
\midrule
\multirow{2}{*}{MJD} & \multirow{2}{*}{Phase} & \multicolumn{2}{c}{Amorph. Carbon} & \multicolumn{2}{c}{Silicates $a=0.1\,\mu$m} & \multicolumn{2}{c}{Silicates $a=1.0\,\mu$m} \\
\cmidrule(lr){3-4} \cmidrule(lr){5-6} \cmidrule(lr){7-8}
& & $M_d$ (M$_\odot$) & $T$ (K) & $M_d$ (M$_\odot$) & $T$ (K) & $M_d$ (M$_\odot$) & $T$ (K) \\
\midrule
59012.58 & 208.6 & $9.62 \times 10^{-5}$ & 942 & $1.44 \times 10^{-3}$ & 1102 & $9.57 \times 10^{-4}$ & 994 \\
59174.82 & 370.8 & $1.53 \times 10^{-4}$ & 816 & $2.34 \times 10^{-3}$ & 930 & $1.53 \times 10^{-3}$ & 854 \\
59379.77 & 575.7 & $1.28 \times 10^{-4}$ & 1013 & $1.89 \times 10^{-3}$ & 1204 & $1.27 \times 10^{-3}$ & 1074 \\
59542.04 & 738.0 & $9.74 \times 10^{-5}$ & 1004 & $1.44 \times 10^{-3}$ & 1192 & $9.65 \times 10^{-4}$ & 1065 \\
59743.89 & 939.9 & $9.97 \times 10^{-5}$ & 927 & $1.49 \times 10^{-3}$ & 1081 & $9.93 \times 10^{-4}$ & 978 \\
59906.32 & 1102.3 & $2.14 \times 10^{-4}$ & 743 & $3.30 \times 10^{-3}$ & 835 & $2.14 \times 10^{-3}$ & 774 \\
60105.86 & 1301.8 & $1.36 \times 10^{-4}$ & 752 & $2.10 \times 10^{-3}$ & 847 & $1.37 \times 10^{-3}$ & 784 \\
60273.36 & 1469.3 & $1.97 \times 10^{-4}$ & 668 & $3.07 \times 10^{-3}$ & 741 & $1.98 \times 10^{-3}$ & 693 \\
60475.46 & 1671.4 & $8.22 \times 10^{-4}$ & 513 & $1.30 \times 10^{-2}$ & 553 & $8.31 \times 10^{-3}$ & 527 \\
\bottomrule
\end{tabular}
\end{minipage}

\end{table*}

\section{Acknowledgements}
\scriptsize

This work is supported by the National Natural Science Foundation of China (NSFC grants 12303054, 12288102, 12033003, 11633002), the Yunnan Fundamental Research Projects (grants 202401AU070063, 202501AS070078), the National Key Research and Development Program of China (grant 2024YFA1611603), and the International Centre of Supernovae, Yunnan Key Laboratory (grant 202302AN360001). Y.-Z.C., A.R., I.S., and G.V. acknowledge financial support from the SOXS project (PI S. Campana). A.P., A.R., N.E.R., L.T., G.V., and S.B. acknowledge support from the PRIN-INAF 2022, ``Shedding light on the nature of gap transients: from the observations to the models.'' X.-F.W. is also supported by the Tencent Xplorer Prize.
S.M. is funded by Leverhulme Trust grant RPG-2023-240.
J.I. acknowledges funding by Spanish program Unidad de Excelencia Mar\'ia de Maeztu CEX2020-001058-M, PID2023-149918NB-I00, and 2021-SGR-1526 (Generalitat de Catalunya).
P.M. acknowledges financial support from the Spanish MCIU through project PID2022-140871NB-C21 by ``ERDF A way of making Europe,'' and from the Severo Ochoa grant CEX2021-515001131-S funded by MCIN/AEI/10.13039/501100011033.
A.M.G. acknowledges financial support from grant PID2023-152609OA-I00, funded by the Spanish Ministerio de Ciencia, Innovaci\'on y Universidades (MICIU), the Agencia Estatal
de Investigaci\'on (AEI 10.13039/501100011033), and the European Union's European Regional Development Fund (ERDF).
A.V.F.'s research group at UC Berkeley acknowledges financial  assistance from the Christopher R. Redlich Fund, Gary and Cynthia 
Bengier, Clark and Sharon Winslow, Alan Eustace and Kathy Kwan           
(W.Z. is a Bengier-Winslow-Eustace Specialist in Astronomy), 
Timothy and Melissa Draper, Briggs and Kathleen Wood, Ellyn and Alan 
Seelenfreund (T.G.B. is Draper-Wood-Seelenfreund Specialist in Astronomy), and numerous other donors.  

This article is based in part on observations made with the Italian Telescopio Nazionale Galileo (TNG), operated on the island of La Palma by the Fundaci\'on Galileo Galilei of the INAF (Istituto Nazionale di Astrofisica) at the Spanish Observatorio del Roque de los Muchachos of the Instituto de Astrof\'{\i}sica de Canarias, under  program A50TAC\_41 (PI G. Valerin).
Based in part on observations made with the Gran Telescopio Canarias (GTC; Program 98-GTC69/20A) installed at the Spanish Observatorio del Roque de los Muchachos of the Instituto de Astrof\'{\i}sica de Canarias, on the island of La Palma.
Based in part on observations collected at Copernico and Schmidt telescopes (Asiago Mount Ekar, Italy) of the INAF -- Osservatorio Astronomico di Padova.
A major upgrade of the Kast spectrograph on the Shane 3\,m telescope at Lick Observatory, led by Brad Holden, was made possible through gifts from the Heising-Simons Foundation, William and Marina Kast, and the University of California Observatories. Research at Lick Observatory is partially supported by a gift from Google.
We appreciate the expert assistance of the staff at the various observatories where data were obtained.        
We thank T. Zwitter, M. Fiaschi, Z.-H. Chen, F.-Z. Guo, Z.-T. Li, J. Mo, T. Chapman, S. Modak, D. Punjabi, J. Sunseri, K. Bostow, N. Girish, S. Kumar, E. Liu, E. Ma, E. McGinness, A. Parke, and A. Yagubyan for their help with the observations and data reduction for SN\,2019vxm.

\end{appendix}

\end{document}